\documentclass[longauth]{aa}  
\usepackage{graphicx}
\usepackage{txfonts}
\usepackage{amsmath,amsfonts,amssymb} 
\usepackage{natbib}
\usepackage{multicol,caption}
\usepackage{lipsum}
\usepackage{booktabs}
\usepackage{threeparttable}
\usepackage{multirow}
\usepackage{here}
\usepackage{lscape}

\bibpunct{(}{)}{;}{a}{}{,}
\usepackage[colorlinks=true,urlcolor=blue,linkcolor=blue,citecolor=blue]{hyperref}

\newcommand{\planettype}{Earth-sized}

\newcommand{\Atype}{M5V}

\newcommand{\Btype}{M6V}

\newcommand{\ellc}{{\fontfamily{pcr}\selectfont ellc}\,}
\newcommand{\emcee}{{\fontfamily{pcr}\selectfont emcee}\,}
\newcommand{\dynesty}{{\fontfamily{pcr}\selectfont dynesty}\,}

\newcommand{\allesfitter}{{\fontfamily{pcr}\selectfont allesfitter}\,}
\newcommand{\sherlock}{{\fontfamily{pcr}\selectfont SHERLOCK}\,}
\newcommand{\matrixtk}{{\fontfamily{pcr}\selectfont MATRIX}\,}
\newcommand{\wotan}{{\fontfamily{pcr}\selectfont wötan}\,}
\newcommand{\triceratops}{{\fontfamily{pcr}\selectfont TRICERATOPS}\,}

\newcommand{\celerite}{{\fontfamily{pcr}\selectfont celerite}\,}

\usepackage{threeparttable}
\usepackage{booktabs}
\usepackage{longtable}
\usepackage{ragged2e}

\usepackage{tikz,xcolor,hyperref}

\definecolor{lime}{HTML}{A6CE39}
\DeclareRobustCommand{\orcidicon}{%
	\hspace{-1.5mm}
	\begin{tikzpicture}
	\draw[lime, fill=lime] (0,0) 
	circle [radius=0.16] 
	node[white] {{\fontfamily{qag}\selectfont \tiny ID}};
	\draw[white, fill=white] (-0.0625,0.095) 
	circle [radius=0.007];
	\end{tikzpicture}
	\hspace{-2.5mm}
}
\foreach \x in {A, ..., Z}{%
	\expandafter\xdef\csname orcid\x\endcsname{\noexpand\href{https://orcid.org/\csname orcidauthor\x\endcsname}{\noexpand\orcidicon}}
}

\foreach \x in {a, ..., z}{%
	\expandafter\xdef\csname orcid\x\endcsname{\noexpand\href{https://orcid.org/\csname orcidauthor\x\endcsname}{\noexpand\orcidicon}}
}

\begin{document}

\title{Two warm Earth-sized exoplanets and an Earth-sized candidate in the M5V-M6V binary system TOI-2267}
% Define the ORCID iD command for each author separately. 
\newcommand{\orcidauthorA}{0000-0002-9350-830X} 
\newcommand{\orcidauthorB}{0000-0003-1572-7707}
\newcommand{\orcidauthorC}{0000-0002-3481-9052}
\newcommand{\orcidauthorD}{0000-0002-4715-9460}
\newcommand{\orcidauthorE}{0000-0002-0371-1647}
\newcommand{\orcidauthorF}{0000-0002-0149-1302}
\newcommand{\orcidauthorG}{0000-0003-3713-8073}
\newcommand{\orcidauthorH}{0000-0003-0647-6133}
\newcommand{\orcidauthorI}{0000-0002-3627-1676}
\newcommand{\orcidauthorJ}{0009-0006-3846-4558}
\newcommand{\orcidauthorK}{0000-0003-1464-9276}
\newcommand{\orcidauthorL}{0000-0002-4296-2246}
\newcommand{\orcidauthorM}{0000-0002-5370-7494}
\newcommand{\orcidauthorN}{0000-0002-6523-9536}
\newcommand{\orcidauthorO}{0000-0002-9404-6952}
\newcommand{\orcidauthorP}{0000-0002-1787-3444}
\newcommand{\orcidauthorQ}{0000-0002-3164-9086} %M.N. Guenther
\newcommand{\orcidauthorR}{0000-0002-6778-7552}
\newcommand{\orcidauthorS}{0000-0003-2144-4316}
\newcommand{\orcidauthorT}{0000-0002-2214-9258}
\newcommand{\orcidauthorU}{0000-0001-9291-5555}
\newcommand{\orcidauthorV}{0000-0001-5485-4675}
\newcommand{\orcidauthorW}{0000-0002-6780-4252}
\newcommand{\orcidauthorY}{0000-0002-2532-2853}
\newcommand{\orcidauthorX}{0000-0001-7746-5795}
\newcommand{\orcidauthorZ}{0000-0001-9800-6248}
\newcommand{\orcidauthora}{0000-0001-6108-4808}
\newcommand{\orcidauthorb}{0000-0003-1462-7739}
\newcommand{\orcidauthorc}{0000-0003-2415-2191}
\newcommand{\orcidauthord}{0000-0002-8388-6040}
\newcommand{\orcidauthore}{0000-0002-3012-0316}
\newcommand{\orcidauthorf}{0000-0001-6285-9847} %Z. Benkhaldoun
\newcommand{\orcidauthorg}{0000-0002-7486-6726} % Y. Gómez Maqueo Chew
\newcommand{\orcidauthorh}{0000-0002-3937-630X} %G. Dransfield
\newcommand{\orcidauthori}{0009-0000-6625-137X}
\newcommand{\orcidauthorj}{0000-0002-8065-4199} %Sucerquia
\newcommand{\orcidauthorl}{0000-0001-8462-8126} 
\newcommand{\orcidauthorm}{0000-0003-0987-1593}
\newcommand{\orcidauthorn}{0000-0002-0076-6239}
\newcommand{\orcidauthoro}{0000-0002-8039-194X}
\newcommand{\orcidauthorp}{0000-0002-4166-6349}
\newcommand{\orcidauthorq}{0000-0001-9699-1459}
\newcommand{\orcidauthorr}{0000-0001-9892-2406}
\author{
%%%%% Note: Order is not definitive yet
%contributing authors
S.~Z\'u\~niga-Fern\'andez\orcidA{}\inst{\ref{astro_liege}}\thanks{Corresponding authors: \color{blue}sgzuniga@uliege.be; pozuelos@iaa.csic.es} %
\and F.~J.~Pozuelos\orcidB{}\inst{\ref{iaa}}$^\star$ % 
\and M.~D\'evora-Pajares\inst{\ref{ugr},\ref{avanture}} % OK!
\and N.~Cuello\orcidG{}\inst{\ref{ipag}} % planet formation scenario + binary introduction OK!
\and M.~Greklek-McKeon\orcidE{}\inst{\ref{Caltech}} % future observations OK!
% stellar characterization
\and K.~G.~Stassun\orcidC{}\inst{\ref{vanderbilt}} % stellar charact OK!
\and V.~Van Grootel\orcidS\inst{\ref{star_liege}} % stellar charact <Valerie.VanGrootel@uliege.be> OK!
\and B.~Rojas-Ayala\orcidF{}\inst{\ref{IAI-UTA}} % stellar charact OK!
\and J.~Korth\orcidn{}\inst{\ref{geneva},\ref{lund}} % OK!
\and M.~N.~G\"unther\orcidQ{}\inst{\ref{estec}} % legacy work OK!
%spectro
\and A.~J.~Burgasser\orcidN{}\inst{\ref{UCSDiego}} % spectro OK!
\and C.~Hsu\orcidM{}\inst{\ref{CIERA}} % Spectro, chh194@ucsd.edu OK!
\and B.~V.~Rackham\orcidI{}\inst{\ref{MIT},\ref{Kavli_MIT}} % spectro OK!
% Ground-base photometry and data reduction
\and K.~Barkaoui\orcidK{}\inst{\ref{astro_liege},\ref{MIT},\ref{IAC_Laguna}} % OK!
\and M.~Timmermans\inst{\ref{birmingham},\ref{astro_liege}} % OK!
\and C.~Cadieux\orcidU{}\inst{\ref{Univ_Montreal}} %OK!
\and R.~Alonso\orcidl{}\inst{\ref{IAC_Laguna},\ref{Univ_LaLaguna}} % SNO/GTC OK!
%%%% High-resolution imaging
% Lomonosov Moscow
\and I.~A.~Strakhov\orcidH{}\inst{\ref{L_moscow}} % OK!
% Gemini-North
\and S.~B.~Howell\orcidY{}\inst{\ref{Ames_NASA}} % OK!
\and C.~Littlefield\orcidX{}\inst{\ref{Ames_NASA},\ref{Bay_Area}} % OK!
\and E.~Furlan\orcidZ{}\inst{\ref{NASA_caltech}} % OK!
%%%% Comments, feedback or contribution to sections
\and P.~J.~Amado\orcidd{}\inst{\ref{iaa}} %pedro.amado@iaa.csic.es OK!
\and J.~M.~Jenkins\orcidD{}\inst{\ref{Ames_NASA}} %TESS Architects OK!
\and J.~D.~Twicken\orcidR{}\inst{\ref{SETI}} % joseph.twicken@nasa.gov OK!
\and M.~Sucerquia\orcidj\inst{\ref{ipag}} % mario.sucerquia@univ-grenoble-alpes.fr OK!
\and Y.~T.~Davis\orcidi{}\inst{\ref{birmingham}} % OK!
\and N.~Schanche\inst{\ref{maryland},\ref{Goddard_NASA}} % Speculoos OK!
\and K.~A.~Collins\inst{\ref{CfA}} %karen.collins@cfa.harvard.edu (LCO scheduling, data reduction, time contribution) OK!
%%%% SPECULOOS team
\and A.~Burdanov\orcidr{}\inst{\ref{MIT}}% OK!
\and \mbox{F.~Davoudi~\orcidP{}}\inst{\ref{astro_liege}}% OK!
\and B.-O.~Demory\inst{\ref{unibe}} % Saint-Ex OK!
\and L.~Delrez\orcida{}\inst{{\ref{astro_liege},\ref{star_liege}}} % OK!
\and G.~Dransfield\orcidh{}\inst{\ref{oxford},\ref{magdalen},\ref{birmingham}} % OK!
\and E.~Ducrot\inst{\ref{Paris_Region},\ref{cea}} % 
\and L.~J.~Garcia\orcidL{}\inst{\ref{newyork}} % OK!
\and \mbox{M.~Gillon~\orcidb{}}\inst{\ref{astro_liege}}% OK!
\and Y.~G\'omez~Maqueo~Chew\orcidg{}\inst{\ref{unam}} % OK!
\and C.~Jan\'{o}~Mu\~{n}oz\inst{\ref{Cavendish}} % OK!
\and E.~Jehin\inst{\ref{star_liege}} % ejehin@uliege.be OK!
\and C.~A.~Murray\inst{\ref{Colorado}} % Speculoos OK!
\and P.~Niraula\inst{\ref{MIT}} % pniraula@mit.edu  OK!
\and P.~P.~Pedersen\inst{\ref{Cavendish},\ref{ETH_Zur_Queloz}} % OK!
\and D.~Queloz\orcide{}\inst{\ref{Cavendish},\ref{ETH_Zur_Queloz}} % 
\and R.~Rebolo-L\'opez\inst{\ref{IAC_Laguna},\ref{Univ_LaLaguna}} % 
\and M.~G.~Scott\orcidJ{}\inst{\ref{birmingham}} % OK!
\and D.~Sebastian\orcidT{}\inst{\ref{birmingham}} % OK!
\and M.~J.~Hooton\inst{\ref{Cavendish}} % 
\and S.~J.~Thompson\orcido{}\inst{\ref{Cavendish}} % OK! sjt20@cam.ac.uk
\and A.~H.~M.~J.~Triaud\inst{\ref{birmingham}} % OK!
\and J.~de~Wit\orcidc{}\inst{\ref{MIT}} % OK!
%%%% Oukaimeden Observatory
\and M.~Ghachoui\inst{\ref{ouka}} % OK!
\and Z.~Benkhaldoun\orcidf{}\inst{\ref{ouka}} % OK!
%%%% OMM observatory
\and R.~Doyon\orcidV{}\inst{\ref{Univ_Montreal},\ref{OMM}} %(rene.doyon@umontreal.ca) 
\and D.~Lafreni\`ere\orcidW{}\inst{\ref{Univ_Montreal}} %(david.lafreniere@umontreal.ca) OK!
%%%% IAA - OSN observatory
\and V.~Casanova\inst{\ref{iaa}} %casanova@iaa.csic.es OK!
\and A.~Sota\orcidO{}\inst{\ref{iaa}} %sota@iaa.csic.es OK!
%%% Saint-EX
\and I.~Plauchu-Frayn\orcidp{}\inst{\ref{uname}} % (UNAM)
\and A.~Khandelwal\inst{\ref{unam}} % (UNAM)
\and F.~Zong~Lang\inst{\ref{unibe}} % (Bern)
\and U.~Schroffenegger\inst{\ref{unibe}} % (Bern)
\and S.~Wampfler\inst{\ref{unibe}} % (Bern)
\and M.~Lendl\orcidq{}\inst{\ref{geneva}} % (Geneva) monika.lendl@unige.ch
%%%% LCOGT
\and R.~P.~Schwarz\inst{\ref{CfA}} %rpschwarz@comcast.net   (LCO data reduction) OK!
\and F.~Murgas\inst{\ref{IAC_Laguna},\ref{Univ_LaLaguna}} % OK!
\and E.~Palle\orcidm{}\inst{\ref{IAC_Laguna},\ref{Univ_LaLaguna}} %epalle@iac.es   (LCO time contribution from IAC) OK!
\and H.~Parviainen\inst{\ref{Univ_LaLaguna},\ref{IAC_Laguna}} % OK!
%%%% TESS (TESS Publication Policy has just been changed and there is no longer any expectation to invite TESS Architects to be co-authors of papers unless they have contributed directly to the work). 
%\and K.A.~Collins\inst{\ref{Harvard_USA}} %karen.collins@cfa.harvard.edu, LCOGT
%\and G.~Ricker\inst{\ref{Kavli_MIT}} %TESS Architects
%\and R.~Vanderspek\inst{\ref{Kavli_MIT}} %TESS Architects
%\and D.~W.~Latham\inst{\ref{CfA}}  %TESS Architects
%\and S.~Seager\inst{\ref{UCSDiego},\ref{Univ_LaLaguna},\ref{Univ_Maryland}} %TESS Architects
%\and J.~Winn\inst{\ref{Astro_Prin}}   %TESS Architects 
%%%% SPOC representative co-author
%\and J.~D.~Twicken\orcidR{}\inst{\ref{SETI}} % joseph.twicken@nasa.gov OK!
}

\institute{
Astrobiology Research Unit, Universit\'e de Li\`ege, All\'ee du 6 Ao\^ut 19C, B-4000 Li\`ege, Belgium \label{astro_liege}
\and Instituto de Astrof\'isica de Andaluc\'ia (IAA-CSIC), Glorieta de la Astronom\'ia s/n, 18008 Granada, Spain \label{iaa}
\and Dpto. Física Teórica y del Cosmos. Universidad de Granada. 18071. Granada, Spain \label{ugr}
\and Avature Machine Learning, Spain \label{avanture}
\and Univ. Grenoble Alpes, CNRS, IPAG, 38000 Grenoble, France \label{ipag}
\and Division of Geological and Planetary Sciences, California Institute of Technology, Pasadena, CA 91125, USA \label{Caltech}
\and Department of Physics \& Astronomy, Vanderbilt University, 6301 Stevenson Center Ln., Nashville, TN 37235, USA \label{vanderbilt}
\and Space Sciences, Technologies and Astrophysics Research (STAR) Institute, Universit\'e de Li\`ege, All\'ee du 6 Ao\^ut 19C, B-4000 Li\`ege, Belgium \label{star_liege}
\and Instituto de Alta Investigaci\'on, Universidad de Tarapac\'a, Casilla 7D, Arica, Chile \label{IAI-UTA}
\and Observatoire Astronomique de l’Université de Genève, Chemin Pegasi 51, 1290 Versoix, Switzerland \label{geneva}
\and Lund Observatory, Division of Astrophysics, Department of Physics, Lund University, Box 118, 22100 Lund, Sweden \label{lund}
\and European Space Agency (ESA), European Space Research and Technology Centre (ESTEC), Keplerlaan 1, 2201 AZ Noordwijk, The Netherlands \label{estec}
\and Center for Astrophysics and Space Sciences, UC San Diego, UCSD Mail Code 0424, 9500 Gilman Drive, La Jolla, CA 92093-0424, USA \label{UCSDiego}
\and Center for Interdisciplinary Exploration and Research in Astrophysics (CIERA), Northwestern University, 1800 Sherman, Evanston, IL 60201, USA \label{CIERA}
\and Department of Earth, Atmospheric and Planetary Science, Massachusetts Institute of Technology, 77 Massachusetts Avenue, Cambridge, MA 02139, USA \label{MIT}
\and Department of Physics and Kavli Institute for Astrophysics and Space Research, Massachusetts Institute of Technology, Cambridge, MA 02139, USA \label{Kavli_MIT}
\and Instituto de Astrof\'isica de Canarias (IAC), Calle V\'ia L\'actea s/n, 38200, La Laguna, Tenerife, Spain \label{IAC_Laguna}
\and School of Physics \& Astronomy, University of Birmingham, Edgbaston, Birmingham B15 2TT, UK \label{birmingham}
\and Institut Trottier de recherche sur les exoplanètes, Université de Montréal, 1375 Ave Thérèse-Lavoie-Roux, Montréal, QC, H2V 0B3, Canada \label{Univ_Montreal}
\and Departamento de Astrof\'isica, Universidad de La Laguna (ULL), E-38206 La Laguna, Tenerife, Spain \label{Univ_LaLaguna}
\and Lomonosov Moscow State University, Sternberg Astronomical Institute, Universitetskij pr. 13, Moscow 119234, RUSSIA \label{L_moscow}
\and NASA Ames Research Center, Moffett Field, CA 94035, USA \label{Ames_NASA}
\and Bay Area Environmental Research Institute, Moffett Field, CA 94035, USA \label{Bay_Area}
\and NASA Exoplanet Science Institute, Caltech/IPAC, Mail Code 100-22, 1200 E. California Blvd., Pasadena, CA 91125, USA \label{NASA_caltech}
\and Center for Space and Habitability, University of Bern, Gesellschaftsstrasse 6, 3012, Bern, Switzerland \label{unibe}
\and Cavendish Laboratory, JJ Thomson Avenue, Cambridge CB3 0HE, UK \label{Cavendish}
\and Department of Astrophysics, University of Oxford, Denys Wilkinson Building, Keble Road, Oxford OX1 3RH, UK \label{oxford}
\and Magdalen College, University of Oxford, Oxford OX1 4AU, UK \label{magdalen}
\and Paris Region Fellow, Marie Sklodowska-Curie Action \label{Paris_Region}
\and AIM, CEA, CNRS, Universit\'e Paris-Saclay, Universit\'e de Paris, F-91191 Gif-sur-Yvette, France \label{cea}
\and Center for Computational Astrophysics, Flatiron Institute, New York, NY, USA \label{newyork}
\and Universidad Nacional Aut\'onoma de M\'exico, Instituto de Astronom\'ia, AP 70-264, Ciudad de M\'exico, 04510, M\'exico \label{unam}
\and Department of Astrophysical and Planetary Sciences, University of Colorado Boulder, Boulder, CO 80309, USA \label{Colorado}
\and Institute for Particle Physics and Astrophysics, ETH Z\"urich, Wolfgang-Pauli-Strasse 2, 8093 Z\"urich, Switzerland \label{ETH_Zur_Queloz}
\and Department of Astronomy, University of Maryland, College Park, MD  20742, USA \label{maryland}
\and Center for Astrophysics \textbar \ Harvard \& Smithsonian, 60 Garden Street, Cambridge, MA 02138, USA \label{CfA}
\and NASA Goddard Space Flight Center, 8800 Greenbelt Rd, Greenbelt, MD 20771, USA \label{Goddard_NASA}
\and Oukaimeden Observatory, High Energy Physics and Astrophysics Laboratory, Faculty of sciences Semlalia, Cadi Ayyad University, Marrakech, Morocco \label{ouka}
\and Observatoire du Mont-M\'egantic, Universit\'e de Montr\'eal, Montr\'eal H3C 3J7, Canada \label{OMM}
%\and Department of Astronomy, University of Maryland, College Park, College Park, MD 20742 USA \label{Univ_Maryland}
\and SETI Institute, Mountain View, CA 94043 USA/NASA Ames Research Center, Moffett Field, CA 94035 USA \label{SETI}
%\and Department of Astrophysical Sciences, Princeton University, Princeton, NJ 08544, USA \label{Astro_Prin}
\and Universidad Nacional Aut\'onoma de M\'exico, Instituto de Astronom\'ia, AP 106, Ensenada 22800, BC, M\'exico \label{uname}
}
\date{Received/accepted}
\titlerunning{The TOI-2267 planetary system}

\abstract{We report the discovery of two warm exoplanets orbiting the cool binary system TOI-2267, composed of the M5 (TOI-2267A) and M6 (TOI-2267B) stars, whose angular separation is 0.384\,arcsec, corresponding to a projected distance of only about 8\,au at 22\,pc from the Solar System. To confirm the planetary nature of these objects, we combined photometry from the Transiting Exoplanet Survey Satellite (\textit{TESS}) and ground-based facilities together with high-resolution images, archival data, and statistical validation in our analyses. From the current data set, we cannot unambiguously determine which star of the binary the planets orbit. These planets are Earth-sized with radii of 1.00$\pm$0.11 and 1.14$\pm$0.13\,R$_{\oplus}$ for TOI-2267\,b (P = 2.28\,d) and TOI-2267\,c (P = 3.49\,d), respectively, when orbiting TOI-2267A, whereas the radii are of 1.22$\pm$0.29 and 1.36$\pm$0.33\,R$_{\oplus}$ when orbiting TOI-2267B. In addition to the signals attributed to TOI-2267\,b and c, the \textit{TESS} data reveal a third strong signal with a periodicity of 2.03\,d (TOI-2267.02). Although statistical analyses support its planetary nature, ground-based follow-up observations did not detect this signal. Its status therefore remains that of a planetary candidate, with an Earth-size of 0.95$\pm$0.12\,R$_{\oplus}$ or 1.13$\pm$0.30\,R$_{\oplus}$ when orbiting TOI-2267A or B, respectively. If this candidate is confirmed, dynamical analyses indicate that all three planets cannot orbit the same star. The most plausible configurations are b–c or .02–c orbiting the same star, while the .02–b case is unlikely due to strong instabilities. The proximity of b and c to a first-order 3:2 mean motion resonance suggests they likely orbit the same star, with .02 orbiting the other component. This scenario would make TOI-2267 the most compact binary system known to host planets, with both components harbouring transiting worlds, and offer a unique benchmark for studying planet formation and evolution in compact binary environments.}


   \maketitle

%===============================================================================
\section{Introduction}
\label{s:Introduction}
%===============================================================================
The most common type of star in the galaxy are M dwarfs, representing $60-75\%$ of stars within 10\,pc from the Solar System \citep[see, e.g.,][]{Henry2006,Reyle2021}. This prevalence, along with their small sizes, low masses, and cool temperatures, facilitates planetary searches and enables detailed characterisation of the exoplanets orbiting them, especially terrestrial ones. This situation is referred to as the M-dwarf opportunity \citep[][]{Charbonneau2007,triaud2021}, and in recent years it has placed M-dwarfs in the spotlight of many surveys and initiatives, such as SPECULOOS \citep{Gillon2018,Sebastian_2021AA,ZunigaFernandez2025}, CARMENES \citep{ribas2023}, MEarth \citep{MEarth2016ApJ}, TIERRAS \citep{TIERRAS2020}, Project EDEN \citep{Gibbs2020, Dietrich2023}, and Red Dots\footnote{\url{https://www.eso.org/public/announcements/ann17036/}} \citep{anglada2016,dreizler2020}, among others. In synergy with these projects, the Kepler/K2 missions \citep{Borucki2010} and the Transiting Exoplanet Survey Satellite \citep[\textit{TESS};][]{Ricker2014} have also significantly contributed to this endeavour, reaching the current number of $\sim$170 small planets with sizes $\leq1.5\,R_{\oplus}$ found orbiting M dwarfs.\footnote{According to NASA Exoplanet Archive as of January 2025.} 

These efforts have provided a robust framework for advancing the detailed characterisation of Earth-like planets. Some examples are the Rocky Worlds DDT program used with the Hubble Space Telescope (HST) and JWST \citep{redfield2024}, whose main objective is to determine the existence of atmospheres in these planets and establish limits for the cosmic shoreline \citep{zahnle2017}, and the \mbox{TRAPPIST-1} JWST initiative \citep{dewit2024}, which is drawing a road map to efficiently characterise the M8.5 star TRAPPIST-1 and its seven Earth-like planets. The aim of these new scientific initiatives is to elucidate the processes governing the formation and evolution of such planets, thereby laying the foundation for assessing their potential habitability with greater precision, a topic that is actively debated and largely speculated within the community \citep[see, e.g.,][]{dressing2015,shields2016,kalte2019,childs2022}.

Stars typically form as part of multiple stellar systems, with binary systems being the most common configuration \citep{Offner+2023}. Despite this, stellar multiplicity has often been overlooked in statistical studies of planet occurrence rates, even though it plays a pivotal role in shaping the architectures and evolution of planetary systems \citep{Marzari+2019, Bonavita+2020}. The gravitational influence of a nearby stellar companion can significantly alter the physical conditions under which planets form by perturbing the protoplanetary environment. In particular, close companions can truncate circumstellar discs, reduce their masses and lifetimes, and induce misalignments, thereby influencing both the efficiency and the outcome of planet formation processes \citep[see, e.g.,][]{jang2015}.

Recent high-resolution observations from ALMA (Atacama Large Millimeter/submillimeter Array) and the VLT (Very Large Telecope) have markedly expanded our understanding of circumstellar disc evolution in binary systems. For instance, surveys of nearby star-forming regions \citep{Manara+2019, Akeson+2019, zurlo2020} have revealed that circumstellar discs are not only prevalent in many Class II binaries but can also be hosted by both the primary and secondary stars. Notable systems include IRAS04158+2805 \citep{Ragusa+2021}, AS205, EM*SR24, and FUOri \citep{Weber+2023} as well as HBC494 \citep{Nogueira+2023}. These observations support the theoretical expectation that binaries with separations of a few tens to several hundred AU commonly sustain circumstellar discs around each component \citep{Manara+2023}. The diversity in disc morphologies and lifetimes across such systems underscores the complexity of planet formation in a multi-stellar context.

While stellar companions can disrupt disc evolution and trigger dynamical perturbations that influence planetary migration or even ejection \citep{kain2013}, numerical simulations have shown that stable planetary configurations can still exist, especially when planets reside within a few tenths of the binary separation \citep{Holman1999, Quarles+2020}. Observationally, this is supported by the discovery of over 500 exoplanets in multi-stellar systems, many of which orbit main-sequence binaries.\footnote{According to the Encyclopaedia of exoplanetary systems as of June 2025; \url{https://exoplanet.eu/planets_binary/}} However, the presence of a stellar companion introduces additional challenges for detection and characterisation, particularly in transit surveys. Flux contamination from unresolved companions can dilute transit depths, leading to systematic underestimation of planet radii, masses, and bulk densities \citep{Ciardi2015, Furlan2020}, and in some cases, it can hinder the detection of Earth-sized planets altogether \citep{Lester2021}. Moreover, since the habitable zone depends on the luminosity of the host star, identifying whether a transiting planet lies within the habitable zone becomes more complex in multi-stellar systems \citep{Savel2020}.

This study examines the nearby binary system TOI-2267 (see Table~\ref{tab:starlit} for further designations), which consists of an \Atype{} and a \Btype{} star with a projected separation of approximately 8 au. The system hosts at least two warm transiting \planettype{} exoplanets with orbital periods near a 3:2 mean motion resonance. These findings identify TOI-2267 as the coolest binary system with the smallest stellar projected separation known to host planets. The planetary candidate TOI-2267.02 is also assessed, although its planetary nature cannot be fully confirmed at this stage. If future observations validate TOI-2267.02 as a planet, N-body simulations indicate that the three planets cannot orbit the same stellar component. Therefore, TOI-2267.02 would likely orbit a different star than the b-c planets, which would make TOI-2267 the first binary system identified with transiting planets around both stellar components.

This paper is structured as follows: In Section \ref{s:Stellar Characterisation} we describe our stellar characterisation efforts, and in Section \ref{s:Facilities_and_Observations} we summarise all participating facilities and our observations. Next, in Section \ref{s:Validation of the planets}, we check for possible observational bias and false positive scenarios and present the statistical validation of the system. Section \ref{sec:global_fit} encompasses our global analysis of the system and the resulting planetary system parameters. In Section \ref{s:sherlock} we describe the procedure for our independent planetary searches and how we established detection limits. In Section \ref{sec:dyn}, we evaluate the system architecture, and finally, we discuss our findings and draw our conclusions in Sections \ref{s:Discussion} and \ref{s:Conclusion}, respectively.

%===============================================================================
\section{Stellar characterisation}
\label{s:Stellar Characterisation}
%===============================================================================

\subsection{Spectral energy distribution}
\label{SED}
\begin{figure}
    \centering
    \includegraphics[width=0.7\linewidth,trim=70 80 80 90,clip,angle=90]{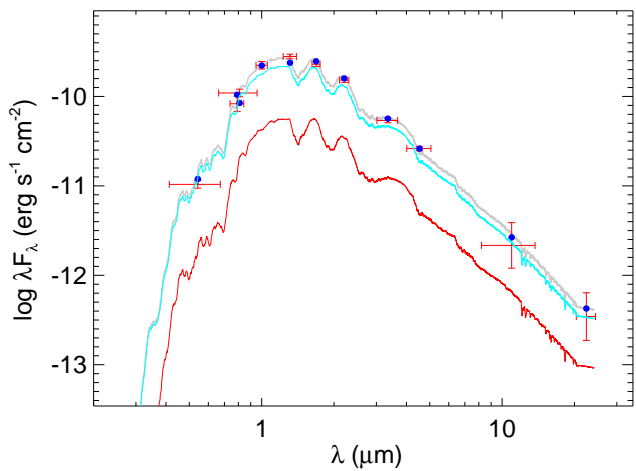}
    \caption{Spectral energy distribution of TOI-2267. Red symbols represent the observed photometric measurements, where the horizontal bars represent the effective width of the passband. Blue symbols are the model fluxes from the best-fit NextGen stellar atmosphere model for the two stellar components (hot component in blue, cool component in red, combined light in black).}
    \label{fig:sed}
\end{figure}

To determine the basic stellar parameters, we performed an analysis of the broadband spectral energy distribution (SED) of the star together with the {\it Gaia\/} DR3 parallax \citep[with no systematic offset applied; see, e.g.,][]{StassunTorres:2021}, to determine an empirical measurement of the stellar radius, following the procedures described in \citet{Stassun:2016,Stassun:2017,Stassun:2018}. We pulled the $JHK_S$ magnitudes from {\it 2MASS}, the W1--W4 magnitudes from {\it WISE}, the $iy$ magnitudes from {\it Pan-STARRS}, and the $G_{\rm BP} G_{\rm RP}$ magnitudes from {\it Gaia}. Together, the available photometry spans the full stellar SED over the wavelength range 0.4--20~$\mu$m (see Fig.~\ref{fig:sed}).  

Due to the presence of the close companion observed in the speckle imaging (see Sect. \ref{sec:HighRes-imaging}), which contributes flux in all of the observed combined-light broadband photometric measurements, we performed a two-component fit following \citet{Stassun:2016}. To this end we used NextGen stellar atmosphere models, with the free parameters being the effective temperatures ($T_{\rm eff}$) and radii ($R_\star$), constrained by the {\it Gaia\/} distance and $I_c$-band flux ratio ($F_I$) measured from the speckle imaging (see Sect. \ref{ss:speckle-SAI}). We set the extinction $A_V \equiv 0$ due to the proximity of the system to Earth. Simple integration of the best-fit SEDs yields the bolometric fluxes at Earth ($F_{\rm bol}$). 

The resulting fit (Fig.~\ref{fig:sed}) has a reduced $\chi^2$ of 1.3 and best fit parameters of $F_{\rm bol,1} = 2.35 \pm 0.19 \times 10^{-10}$ erg s$^{-1}$ cm$^{-2}$, $F_{\rm bol,2} = 4.74 \pm 1.83 \times 10^{-11}$ erg s$^{-1}$ cm$^{-2}$, $R_{\star,1} = 0.2075 \pm 0.0225$~R$_\odot$, $R_{\star,2} = 0.13 \pm 0.03$~R$_\odot$, $T_{\rm eff,1} = 3030 \pm 100$~K, $T_{\rm eff,2} = 2930 \pm 160$~K, $d = 22.54 \pm  0.18$~pc, and $F_I =  0.300 \pm  0.048$. We note that the nature of the fit and the available constraints makes the fitted parameters highly correlated. Hence, for completeness, we provide the full parameter correlation matrix (see Appendix \ref{appendix:SED}).

Using a Monte Carlo approach, we employed the solar metallicity NextGen stellar atmosphere models to calculate synthetic photometry with associated uncertainties for each stellar component. The {\it Species} toolkit \citep{2020A&A...635A.182S} was utilised to interpolate the NextGen grid for the stellar parameters $T_{\rm eff}$, $R_{\star}$, parallax, and $\log{g}$, and to compute synthetic photometry using the filter profiles of the selected magnitudes. The parameter distributions for $T_{\rm eff}$ and $R_{\star}$ were derived from the SED fitting described earlier, while the parallax was taken from Gaia DR3 system values, and $\log{g}$ values were obtained from Table~\ref{table:stellar}. The filter profiles used in this analysis were sourced from the SVO Filter Profile Service \citep{2020sea..confE.182R} and included Gaia DR3, 2MASS, and WISE filters. As with the SED fitting, interstellar extinction was neglected due to the system's proximity. For each stellar component, we performed 1,000 random samples of each parameter assuming Gaussian distributions, interpolated the NextGen grid, and calculated synthetic magnitudes for the relevant filter profiles. Since the upper boundary of the model grid for $\log{g}$ is 5.5 dex, we modified the random sampling process to cap any values exceeding this limit at 5.5 dex. The mean and standard deviation of the resulting magnitude distributions were adopted as the synthetic magnitudes and their uncertainties for each filter profile. The computed synthetic magnitudes and uncertainties are presented in Table~\ref{tab:star_phot}.

\subsection{Spectroscopic analysis}
\label{model_spectra}
To extract the radial velocities (RVs) and projected rotational velocities ($v\sin{i}$) of TOI-2267, we obtained the NIRSPEC high-resolution near-infrared spectra (R$\sim$35,000) on the W. M. Keck II Telescope \citep{McLean:1998aa, McLean:2000aa, Martin:2018aa}. The NIRSPEC spectra were taken on 2022 March 13 (UT) under good seeing ($\sim$0.6$\arcsec$) and $\sim$2\,mm of precipitable water vapour. We chose the ``Kband-new'' filter with the 0$\arcsec$.432$\times$12$\arcsec$ slit, and set the echelle and cross-disperser angles of 62.95$^\circ$ and 35.69$^\circ$, respectively. We then took the A-B nodding sequence for TOI-2267, each with 300~s exposure, with the median signal-to-noise ratios per pixel S/N$\sim$23 and 25, respectively. We note that the double traces were shown due to a focusing issue at a high declination/elevation, not due to the nature of the binarity of TOI-2267. We used the A0V star HD~25175 to serve as our wavelength solution calibrator. We also acquired the associated dark, flat lamp data for our data reduction. Our NIRSPEC data were reduced using a modified version of NIRSPEC Data Reduction Pipeline \citep{Tran:2016aa}, with the changes detailed in \cite{Hsu:2021aa, Theissen:2022aa}. 

Our observed Keck/NIRSPEC spectra and best-fit models are shown in Fig.~\ref{fig:nirspec_spectra} (further details see Appendix \ref{appendix:spectra}). The stellar parameters for both nodes are consistent with the model uncertainties.
The best-fit parameters after combining the results of both spectra are RV = $-$19.0 $\pm$ 0.3 km s$^{-1}$, $v\sin{i}$ = 17.6 $\pm$ 0.3 km s$^{-1}$ (Table~\ref{table:stellar}), effective temperature $T_{\mathrm{eff}}$ = 3167$^{+14}_{-13}$ K, and surface gravity $\log{g}$ = $5.18^{+0.08}_{-0.07}$ cgs dex. We note that our $T_{\mathrm{eff}}$ and $\log{g}$ are merely for reference, as the high-resolution spectra only focused on narrow spectral regions \citep{Hsu:2021aa, Hsu:2024aa}, while the robust and accurate $T_{\mathrm{eff}}$ and $\log{g}$ measurements are determined using the SED and Sect.~\ref{sec:stellar_cha} (Table~\ref{table:stellar}). Finally, we also examined if the secondary component was detected as combined light spectra in our NIRSPEC data. We forward-modelled a binary template, following \cite{Hsu:2023aa}, but we did not find strong evidence of detection of the secondary.

% NIRSPEC spectra
\begin{figure}
    \centering
    \includegraphics[width=\columnwidth]{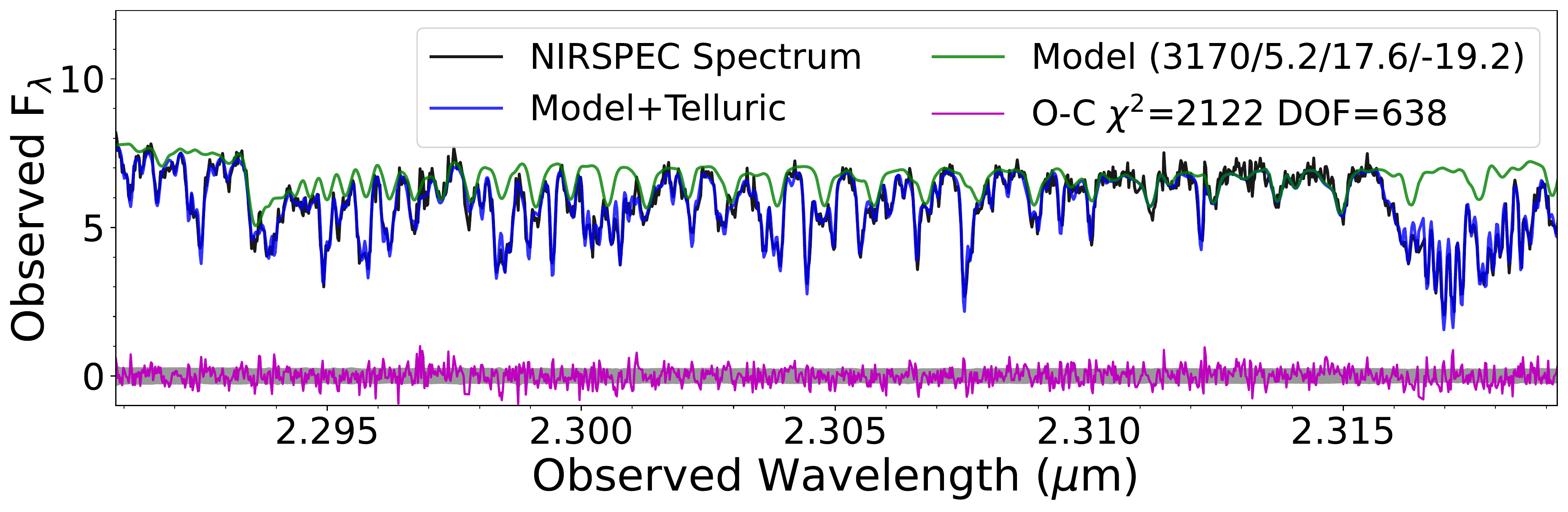}
    \includegraphics[width=\columnwidth]{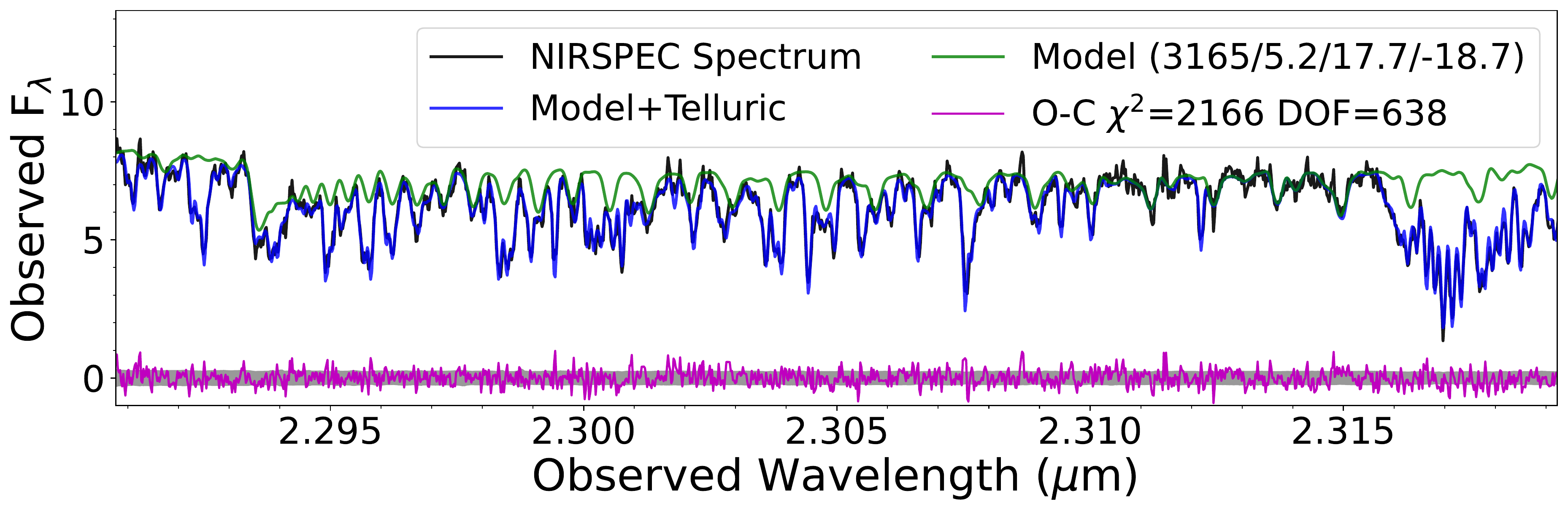}
    \caption{Keck/NIRSPEC \textit{K}-band spectra of TOI-2267 around 2.3~$\mu$m. \textit{Top}: Observed NIRSPEC spectrum in the nodding position A (shown with the black line) with the data noise in the grey-shaded region. The stellar models with and without telluric absorption features are indicated with the blue and green lines, respectively. The residual (data $-$ model) is denoted with the magenta line. \textit{Bottom}: Same as the top panel for NIRSPEC spectra and model fit but in the nodding position B.}
    \label{fig:nirspec_spectra}
\end{figure}

\subsection{Rotational periods from \textit{TESS} photometry}
\label{sec:rot_period}
\begin{figure}
    \centering
\includegraphics[width=0.98\linewidth]{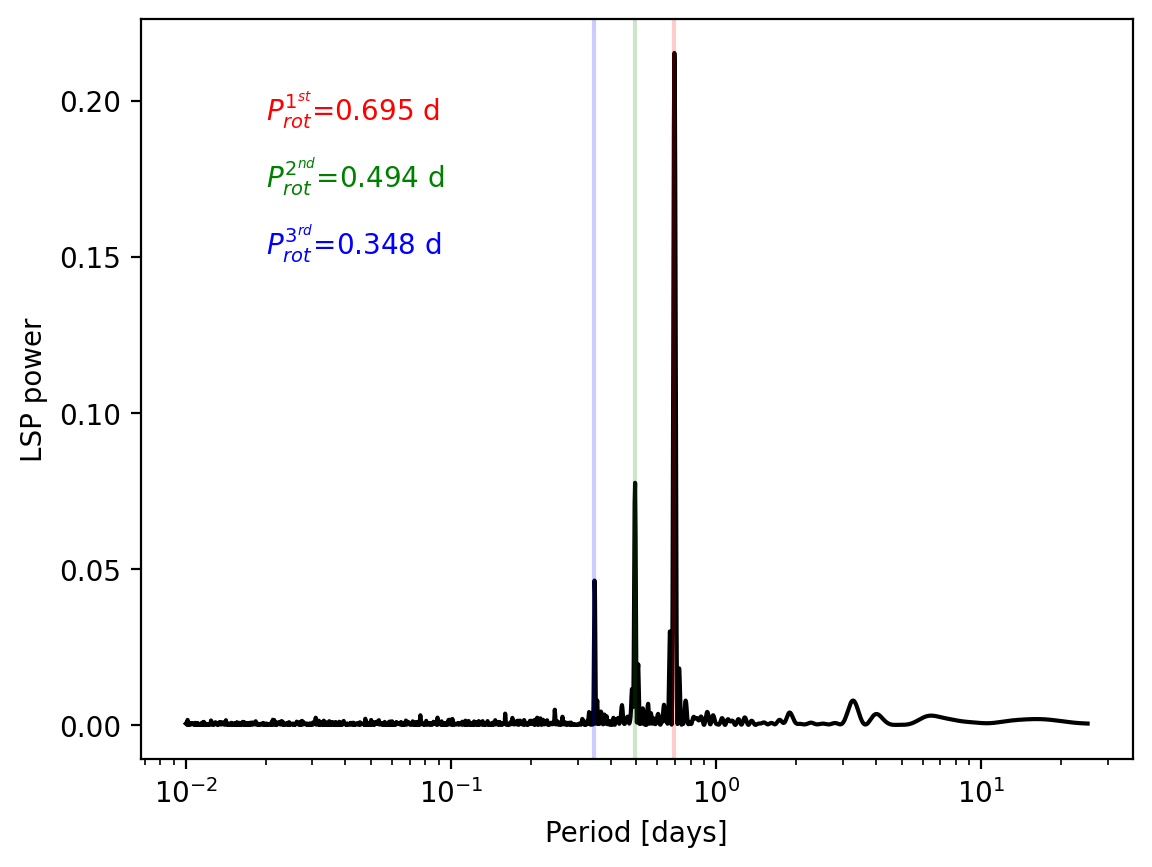}
    \caption{Periodogram obtained with \texttt{TESSExtractor} tool for TOI-2267 from \textit{TESS} Sector 53.}
    \label{fig:Rot_period}
\end{figure}

We performed a Lomb-Scargle periodogram analysis on \textit{TESS} light curves using the \texttt{Lightkurve} package \citep{Lightkurve2018}. We detected three main peaks in the periodogram, with the period at maximum power being 0.6958\,d, consistent with the 0.695\,d rotational period published by \cite{Newton2016b}. Using this period and the Lomb-Scargle model method from \texttt{Lightkurve}, we detrended the light curve and then repeated the periodogram analysis. The period at maximum power, in this case, was 0.4936\,d, which could correspond to the rotational period of the secondary star in our binary system. We repeated the detrending and periodogram procedure as described before and found that the highest peak corresponded to a period of $0.3471$ days, which aligns with the second harmonic of the main rotational period. As an independent approach, we used the \texttt{TESSExtractor} tool\footnote{\url{https://www.tessextractor.app/}} \citep{Brasseur2019,Serna2021}, which performs a correction of systematic effects in the light curve and a periodogram analysis on \textit{TESS} data for a given target and sector. The results that we obtained from \texttt{TESSExtractor} tool are in agreement with the results of our analysis (see Fig. \ref{fig:Rot_period}).

\subsection{Estimated masses and evolutionary modelling}\label{sec:stellar_cha}
We first relied on evolutionary modelling to obtain the stellar masses from the models of very low-mass stars and young brown dwarfs presented in \citet{2019ApJ...879...94F}. We used as constraints the luminosity computed from the bolometric flux $F_{\rm bol}$ and distance $d$ derived from SED fitting (see subsect.~\ref{SED}), the metallicity [Fe/H]= 0.164 $\pm$ 0.11 \citep{MEarth2016ApJ}, and assuming an age of $\gtrsim$ 1 Gyr. Since very low-mass stars evolve so slowly, they keep a constant luminosity once the star has turned on core H-burning and has reached the main sequence. This implies we cannot infer the stellar age from evolutionary modelling. Another consequence is that we could model the two stars separately without considering co-evolution through binary evolution modelling.

We obtained a stellar mass of $M_{\star,1} = 0.1710 \pm 0.0079 M_{\odot}$ for the M5V star and $M_{\star,2} = 0.0989 \pm 0.0130 M_{\odot}$ for the M6V star. The quoted uncertainties reflect the error propagation on the stellar luminosity and metallicity, but also the uncertainty associated with the input physics of the stellar models \citep{2018ApJ...853...30V}. The inferred radii from models are $R_{\star,1}=0.204 \pm 0.021  R_{\odot}$ and $R_{\star,2}=0.125 \pm 0.005 R_{\odot}$, within 1-$\sigma$ agreement with the radii derived from SED fitting  (subsect.~\ref{SED}). Considering these SED radius estimates and the stellar masses inferred from evolutionary modelling, we derived stellar surface gravities of $\log g_{\star,1} = 4.99 \pm 0.05 $ and $\log g_{\star,2} = 5.28 \pm 0.18 $ (cgs), and stellar mean densities of $\rho_{\star,1}=23.0^{+5.3}_{-4.0}$ g cm$^{-3}$ and $\rho_{\star,2}=83^{+113}_{-39}$ g cm$^{-3}$.

\begingroup
\begin{table}
\caption{TOI-2267 stellar designations and astrometric properties.}
\begin{center}
\renewcommand{\arraystretch}{1.15}
\begin{tabular*}{\linewidth}{@{\extracolsep{\fill}}l c c}
\toprule
Parameter & Value & Source \\
\midrule
\multicolumn{3}{c}{\textit{Target designations}} \\
% \midrule
TIC      & 459837008           & 1 \\
2MASS    & J04201254+8454062   & 2 \\
{\it Gaia\/} DR3 & 571488283984760960 & 3 \\
WISE     & J042013.99+845404.0 & 4 \\
USNO-B   & 1749-00009530        & 5 \\
\midrule
\multicolumn{3}{c}{\textit{Astrometry}} \\
RA  (J2000) & 04 20 14.76  & 3 \\
DEC (J2000) & +84 54 02.97  & 3 \\
RA PM (mas/yr)  & 182.388 $\pm$ 0.414 & 3 \\
DEC PM (mas/yr) & -213.503 $\pm$ 0.455 & 3 \\
Parallax (mas) & 44.3496   $\pm$ 0.3556 & 3 \\
\bottomrule
\end{tabular*}
\end{center}
\label{tab:starlit} 
\tablefoot{
1. \citet{stassun2018}, 2. \citet{cutri:2003}, 3. \citet{gaia:2021_EDR3}, 4. \citet{cutri:2014}, 5. \citet{Monet2003}.
}
\end{table}
\endgroup

\begingroup
\begin{table*}
\caption{Derived properties of the primary and secondary stars of the TOI-2267 system.}
\begin{center}
\begin{tabular}{l c c c}
\toprule
Property & \multicolumn{2}{c}{Value} & Source \\
 & Primary & Secondary & \\
\midrule
Sp.\ type & M5V & M6V & SED \& Sect.~\ref{sec:stellar_cha} \\
$\mathrm{T_{eff}}$ (K) & $3030 \pm 100$ & $ 2930 \pm 160$ & SED \\
$\mathrm{M_\star}$ ($\mathrm{M_\odot}$) & $0.1710 \pm 0.0079$ & $0.0989 \pm 0.0130$ & Sect.~\ref{sec:stellar_cha} \\
$\mathrm{R_\star}$ ($\mathrm{R_\odot}$) & $0.2075 \pm 0.0225$ & $0.130 \pm 0.030$ & SED \\
$\mathrm{L_\star}$ ($10^{-3}$ $\mathrm{L_\odot}$) & $3.3 \pm 0.3$ & $1.1 \pm 0.3$ & SED \\
Rotational period (days) & $0.6958$ & $0.4936$ & Sect. \ref{sec:rot_period} \\
$\mathrm{\log\,g}$ & $4.99 \pm 0.05 $ & $5.28 \pm 0.18$ & $\mathrm{M_\star}$, $\mathrm{R_\star}$ \\
$\mathrm{\rho_\star}$ ($\mathrm{g\,cm^{-3}}$) & $23.0^{+5.3}_{-4.0}$ & $83^{+113}_{-39}$ & $\mathrm{M_\star}$, $\mathrm{R_\star}$ \\
$\mathrm{[Fe/H]}$ & \multicolumn{2}{c}{$0.164 \pm 0.11$} & \citet{MEarth2016ApJ} \\
Age (Gyr) & \multicolumn{2}{c}{$\gtrsim 1$}  & Sect.~\ref{sec:stellar_cha} \\
Distance (pc) & \multicolumn{2}{c}{$22.55\pm0.19$} & Parallax$^{a}$ \\
Combined RVs ($\mathrm{km\,s^{-1}}$) & \multicolumn{2}{c}{$-19.0\pm0.3$} & IR spectra \\
Combined vsini ($\mathrm{km\,s^{-1}}$) & \multicolumn{2}{c}{$17.6\pm0.3$} & IR spectra \\
\hline
\end{tabular}
\end{center}
\tablefoot{
$^{a}$ Computed from the parallax provided by \citet{gaia:2021_EDR3}.
}
\label{table:stellar}
\end{table*}
\endgroup

\begin{table*}
\caption{Photometric properties of TOI-2267.}
\centering
\renewcommand{\arraystretch}{1.10}
\begin{tabular*}{\linewidth}{@{\extracolsep{\fill}}l c c c c}
\toprule
\multicolumn{5}{c}{\textit{Photometry}} \\
Parameter & Combined & TOI-2267A & TOI-2267B & Source \\
\midrule
\textit{TESS}   & 12.259 $\pm$ 0.007 & 12.937 $\pm$ 0.338 & 14.264 $\pm$ 0.690 & 1, this work (Sect.~\ref{SED}) \\
B      & 17.031  $\pm$ 0.012  & 16.937 $\pm$ 0.527 & 18.669 $\pm$ 1.045 & 5, this work (Sect.~\ref{SED})\\
V      & 15.412 $\pm$ 0.014 & 15.442 $\pm$ 0.496 & 17.048 $\pm$ 0.972 & 5, this work (Sect.~\ref{SED})\\
Gaia   & 13.762 $\pm$ 0.003 & 14.151 $\pm$ 0.387 & 15.552 $\pm$ 0.755 & 3, this work (Sect.~\ref{SED})\\
J      & 10.353 $\pm$ 0.023   & 10.804 $\pm$ 0.263 & 11.980 $\pm$ 0.576 & 2, this work (Sect.~\ref{SED})\\
H      & 9.765  $\pm$ 0.028   & 10.117  $\pm$ 0.259 & 11.275  $\pm$ 0.570 & 2, this work (Sect.~\ref{SED})\\
K      & 9.459  $\pm$ 0.022   & 9.802  $\pm$ 0.256 & 10.973  $\pm$ 0.569 & 2, this work (Sect.~\ref{SED})\\
WISE 3.4 $\mu$m & 9.268  $\pm$ 0.023 & 9.606  $\pm$ 0.253 & 10.760  $\pm$ 0.566 & 4, this work (Sect.~\ref{SED})\\
WISE 4.6 $\mu$m & 9.073  $\pm$ 0.021 & 9.464  $\pm$ 0.251 & 10.626  $\pm$ 0.563 & 4, this work (Sect.~\ref{SED})\\
WISE 12 $\mu$m  & 8.858  $\pm$ 0.026 & 9.162  $\pm$ 0.245 & 10.284  $\pm$ 0.554 & 4, this work (Sect.~\ref{SED})\\
WISE 22 $\mu$m  & 8.692  $\pm$ 0.314 & 8.829  $\pm$ 0.243 & 9.942  $\pm$ 0.552 & 4, this work (Sect.~\ref{SED})\\
\bottomrule
\end{tabular*}
\tablefoot{
According to:
1.~\citet{stassun2018},
2.~\citet{cutri:2003},
3.~\citet{gaia:2021_EDR3},
4.~\citet{cutri:2014},
5.~\citet{Monet2003}.
The procedure to derive the photometry for each component of this binary system is described in Sect.~\ref{SED}.
}
\label{tab:star_phot}
\end{table*}

%===============================================================================
\section{Photometric observations}
\label{s:Facilities_and_Observations}
This section details all of the observations and facilities used to study the TOI-2267 system. This effort corresponds to time-series photometry measured in nine \textit{TESS} sectors plus observations gathered by four ground-based telescopes used for the photometric follow-up. 
%===============================================================================

\subsection{\textit{TESS}}
\label{ss:TESS}
\textit{TESS} observed TOI-2267 with a 2\,min cadence during the primary mission in  
Sector 19 (midpoint December 2019), 
Sector 20 (midpoint January 2020), 
Sector 25 (midpoint May 2020), and
Sector 26 (midpoint June 2020).
The Science Processing Operations Center (SPOC) pipeline \citep[][]{Jenkins2016} reduced the image data to produce the photometric time series. These time series were used to search for transiting planet signatures with a noise-compensating matched filter \citep{Jenkins2002,Jenkins2010,Jenkins2020}, leading to the detection of a transit signature with an orbital period of $\sim$3.5\,d.
The signal passed all diagnostic tests \citep{Twicken2018,Li2012} and was released as the planetary candidate TOI-2267.01 on September 30, 2020 \citep{Guerrero2021}. No other signal was released at that time. We conducted an independent transit search using our public pipeline \sherlock (see Sect.~\ref{s:sherlock}), which allowed us not only to recover the signal corresponding to TOI-2267.01 but also to find two other low S/N signals with orbital periods of 2.29\,d and 2.03\,d. We then triggered a ground-based follow-up campaign of this system, trying to confirm or refute the three signals detected. We discuss this follow-up effort in the following subsections. During the \textit{TESS} mission extensions, TOI-2267 was re-observed in several sectors. First, it was re-observed in Sector 40 (midpoint July 2021). The reanalysis of all the observations gathered until that date by SPOC drove to the release of the candidate TOI-2267.03 with an orbital period of 2.29\,d on February 28, 2022. Then, \textit{TESS} kept observing the star on   
Sector 52 (midpoint May 2022), 
Sector 53 (midpoint June 2022), 
Sector 59 (midpoint December 2022), and
Sector 60 (midpoint January 2023). 
The reanalysis of these nine sectors together by SPOC led to the detection of a new transiting signal with an orbital period of $\sim$2.03\,d, which was released on August 15, 2023, as TOI-2267.02.
The independent findings by SPOC of TOI-2267.03 and .02, matching our early detections, reinforced their credibility. 
Moreover, \textit{TESS} observed TOI-2267 again in Sector 73 (midpoint December 2023), Sector 79 (midpoint June 2024), and Sector 86 (midpoint December 2024). \\
The field-of-view of these sectors is displayed in Fig.~\ref{fig:fov}, showing the aperture photometry used in each case with red squares and the stars inside it. In all cases, we observed that very few faint stars ($\Delta mag>5$), one or two, are in the aperture photometry, at more than one \textit{TESS} pixel away (>21\,arcsec). Then, to perform our analyses, we retrieved the Presearch Data Conditioning Simple Aperture Photometry (PDCSAP) fluxes, which are corrected from crowding and systematic effects \citep{stumpe2012,stumpe2014,smith2012}. These light curves are displayed in Fig.~\ref{fig:lcs}, with the location of the transits for TOI-2267.01, .02, and .03 candidates highlighted. 

In addition, visual inspection reveals the existence of flares in the data, which, combined with the rapid rotation of both stars, reported in Sec.~\ref{s:Stellar Characterisation}, suggest that TOI-2267 A and B are magnetically active. A comprehensive analysis of activity levels, including flare rates and energy distribution, will be addressed in a subsequent publication.

\subsection{Ground based photometry}
In the following subsubsections, we describe all the ground-based observations. The photometric observations logs and transit detection are also summarised in Table~\ref{tab:GBobservations}.

\subsubsection{SPECULOOS-North/Artemis and SAINT-EX }
We observed a total of 11 transits of TOI-2267.03 and 12 of TOI-2267.01 with SPECULOOS-North/Artemis \citep[SNO/Artemis;][]{Burdanov2022} in the Sloan-$z'$ and -$r'$ filters with an exposure time of 10 and 55\,s, respectively. In addition, we gathered 5 transits of TOI-2267.03 and 1 transit of TOI-2267.01 with SAINT-EX \citep{Demory_AA_SAINTEX_2020} in the $I+z$ and Sloan-$z'$ filters with an exposure time of 10\,s. SNO/Artemis and SAINT-EX are 1.0-m Ritchey–Chrétien telescopes located at the Teide Observatory in the Canary Islands (Spain) and at the Observatorio Astronómico Nacional in San Pedro Mártir (Mexico), respectively. Both are equipped with a thermoelectrically cooled 2K$\times$2K Andor iKon-L BEX2-DD CCD camera with a pixel scale of 0.35$\arcsec$, and a field-of-view (FoV) of 12$\arcmin\times$12$\arcmin$. These facilities are twins of the SPECULOOS-South telescopes \citep{Delrez2018,ZunigaFernandez2024}. Data reduction, calibration, and photometry measurements were performed using the {\tt PROSE}\footnote{{\tt Prose:} \url{https://github.com/lgrcia/prose}} pipeline \citep{prose}.

\subsubsection{TRAPPIST-North}
Two transits of the outer planet TOI-2267.01 were observed with the TRAPPIST-North  \citep{Jehin2011,Gillon2011,Barkaoui2019_TN} telescope. TRAPPIST-North is a 60-cm robotic telescope located at Oukaimeden Observatory in Morocco since 2016. It is equipped with a thermoelectrically cooled 2K$\times$2K Andor iKon-L BEX2-DD CCD camera with a pixel scale of 0.6$\arcsec$, resulting in a fov of 20$\arcmin \times$20$\arcmin$. Both transits were observed in the $I+z'$ filter with an exposure time of 30\,s. The data was processed using the {\tt PROSE} pipeline. 

\subsubsection{LCOGT}
A full transit of the inner planet TOI-2267.03 was observed in the Sloan-$i'$ filter using Las Cumbres Observatory Global Telescope \citep[LCOGT;][]{Brown_2013} 1.0-m at McDonald Observatory. Complementary, three full transits of the outer planet TOI-2267.01 were also observed in the Sloan-$i'$ filter  and one full transit of TOI-2267.02 with the $z'$ filter using LCOGT 1.0-m at Teide Observatory. The science images were calibrated using the standard LCOGT {\tt BANZAI} pipeline \citep{McCully_2018SPIE10707E}, and photometric measurements were extracted using {\tt AstroImageJ}\footnote{{\tt AstroImageJ:}\url{https://www.astro.louisville.edu/software/astroimagej/}} \citep{collins2017} software.

\subsubsection{Sierra Nevada Observatory-T150}
Using the 1.52-m telescope at Sierra Nevada Observatory (OSN/T150), we carried out a full-transit observation of TOI-2267.03 and TOI-2267.01, and two full-transits of TOI-2267.02. We employed the Johnson-Cousin $I$ filter in both observations with an exposure time of 90\,s. The OSN/T150 is a Ritchey-Chr\'etien telescope equipped with a thermoelectrically cooled 2K$\times$2K Andor iKon-L BEX2DD CCD camera with a fov of $7.9'\times7.9'$ and pixel scale of 0.232". The photometric data were extracted using the {\tt AstroImageJ} package.

\subsubsection{Observatoire du Mont-Mégantic}
We observed a full transit of TOI-2267.01 with the PESTO EMCCD camera (1024$\times$1024 pixel) mounted on the 1.6-m telescope at Observatoire du Mont-Mégantic (OMM), Canada. PESTO has an image scale of 0.466$\arcsec$ per pixel, providing an on-sky fov of 7.95$\arcmin$ $\times$ 7.95$\arcmin$. The observation sequence was taken in the $i'$ filter with exposures of 30\,s. The data was processed using the {\tt AstroImageJ} package.

\subsubsection{Gran Telescopio Canarias}
We observed one full transit of TOI-2267.02 with the 10.4-m Gran Telescopio Canarias (GTC), using the Optical System for Imaging and low-Intermediate-Resolution Integrated Spectroscopy \citep[OSIRIS;][]{cepa2000}, a tunable imager and spectrograph installed at the Nasmyth-B focus of the telescope. OSIRIS is equipped with a mosaic of two Marconi CCD42-82 detectors, providing an unvignetted field-of-view of 7.8$\arcmin\times$7.8$\arcmin$ and a pixel scale of 0.127$\arcsec$ per pixel in 1$\times$1 binning mode. The detectors are cooled using a continuous-flow cryostat and offer high quantum efficiency over the 365–1000\,nm wavelength range. For our observations, we used OSIRIS in imaging mode with the Sloan-$i$ filter and an exposure time of 1\,s. Data reduction, calibration, and photometry measurements were performed using the {\tt AstroImage} package.

%===============================================================================
\section{Validation of the planets}
\label{s:Validation of the planets}
%===============================================================================
\subsection{Archival imaging}
\label{sec:archival}
\begin{figure*}
    \centering
    \includegraphics[width=0.99\textwidth]{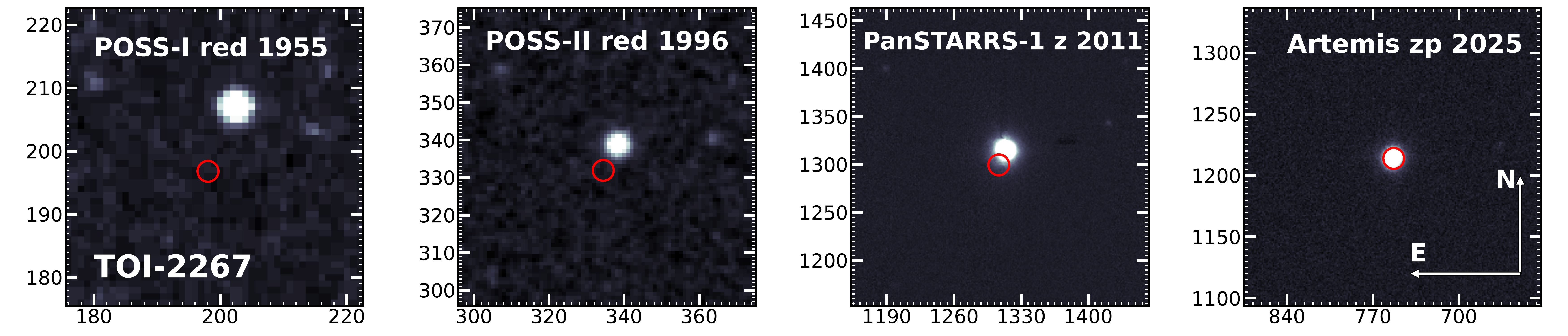}
    \caption{Archival images for TOI-2267. From left to right: 1955 DSS/POSS-I image, 1996 DSS/POSS-II image, 2011 PanSTARRS image, and 2025 Artemis-1m0 image. The red circle corresponds to the target position at 2022 November 09.}
    \label{fig:archive_images}
\end{figure*}

%%% Khalid
We used archival science images of TOI-2267 to exclude background stellar objects that could be blended with the target at its current position. Such an object might introduce the same transit event we observed in our data and skew the system's physical p, which we obtained from the global analysis.
TOI-2267 has a high proper motion of 280$\pm$0.44~$mas/yr$. We used images from DSS/POSS-I \citep{1963POSS-I} in 1955 and DSS/POSS-II 1996 in the red filter, PanSTARRS in 2011 in the $y$ filter, and SNO/Artemis in 2025 in the $z'$ filter, spanning 70 years with our current observations. 
The target has moved by 19.6\,$\arcsec$ from 1995 to 2025. There is no source in the current day positions of TOI-2267; see Fig~\ref{fig:archive_images}.

\subsection{Seeing-limited photometry - SG1}
\label{ss:sg1}
One of the critical steps during the validation process of any TOI is the confirmation of the events in the target star at transit predicted times \citep[see, e.g.,][]{kostov2019,Guenther2019}.
This strategy is motivated by the \textit{TESS}' large pixel size of 21\,arcsec and the associated point-spread function (PSF) that could be as large as 1\,arcmin. These two characteristics of \textit{TESS} observations imply a larger probability of contamination by a nearby eclipsing binary (NEB). Indeed, due to dilution effects, deep eclipses in a faint NEB might mimic a shallow
transit observed on the target star. Hence, it becomes critical to confirm the transits in the target star and, when those are very shallow for ground-based facilities, explore the potential contamination by relatively distant neighbours to rule out the contamination from an NEB. 

On the one hand, for TOI-2267.01 and TOI-2267.03, we confirmed the transits at predicted times using ground-based facilities 
(see Table~\ref{tab:GBobservations} and Fig.~\ref{fig:lc_toi2267b} and ~\ref{fig:lc_toi2267c}), allowing us to rule out the presence of any NEB that might cause these transits and strengthening the planetary interpretation. On the other hand, for TOI-2267.02, we gathered three observations at transit times using the OSN/T150 and the LCO-Teide-1\,m. Unfortunately, due to the shallowness of this transit of $\sim$1.3\,ppt, we could not confirm any event in the target star. We used these observations to search for potential NEBs up to 5\,$\arcmin$ from the target star in the OSN-1.5\,m observations and up to 2.5\,$\arcmin$ in the LCO-Teide-1\,m. None of these observations suggests that an NEB is the origin of the signal. In our effort to confirm this event in the target star, we got a Director's Discretionary Time to use GTC with the Osiris camera (GTC/Osiris). The optimum instrument would have been HiPERCAM\footnote{\url{https://www.gtc.iac.es/instruments/hipercam/hipercam.php}}\citep{GTC_HiPERCAM}, 
which allows for multi-band simultaneous observations using u', g', r', i', and z' bands; unfortunately, the instrument was not mounted when the planet transited. Still, while simultaneous multi-band observations cannot be conducted with GTC/Osiris, its photometric precision should be good enough to detect the transit in the target star. However, an inappropriate election for the exposure time remarkably decreased the photometric quality of the light curve, yielding a non-conclusive observation. We used this observation to search for potential NEBs again in the fov, and we found nothing. 

Therefore, while we have not confirmed the candidate TOI-2267.02 in the target star, we cleaned the fov and ruled out the presence of any NEB that might induce the transit-like signal detected in the \textit{TESS} data. Still, we prudently do not claim this candidate as a validated planet but as pending of a more robust follow-up, which is in our roadmap for this system.

\subsection{High angular resolution imaging}
\label{sec:HighRes-imaging}
Spatially close stellar companions can confound exoplanet discoveries such that the detected transit signal might be a false positive. In addition, even for real planet discoveries, the close companion will yield incorrect stellar and exoplanet parameters if unaccounted for \citep{Ciardi2015,Furlan2017,Furlan2020}.

\subsubsection{Gemini-North-8.0m/‘Alopeke}
\label{ss:speckle-Gemini}

\begin{figure}
    \centering
\includegraphics[width=0.95\linewidth]{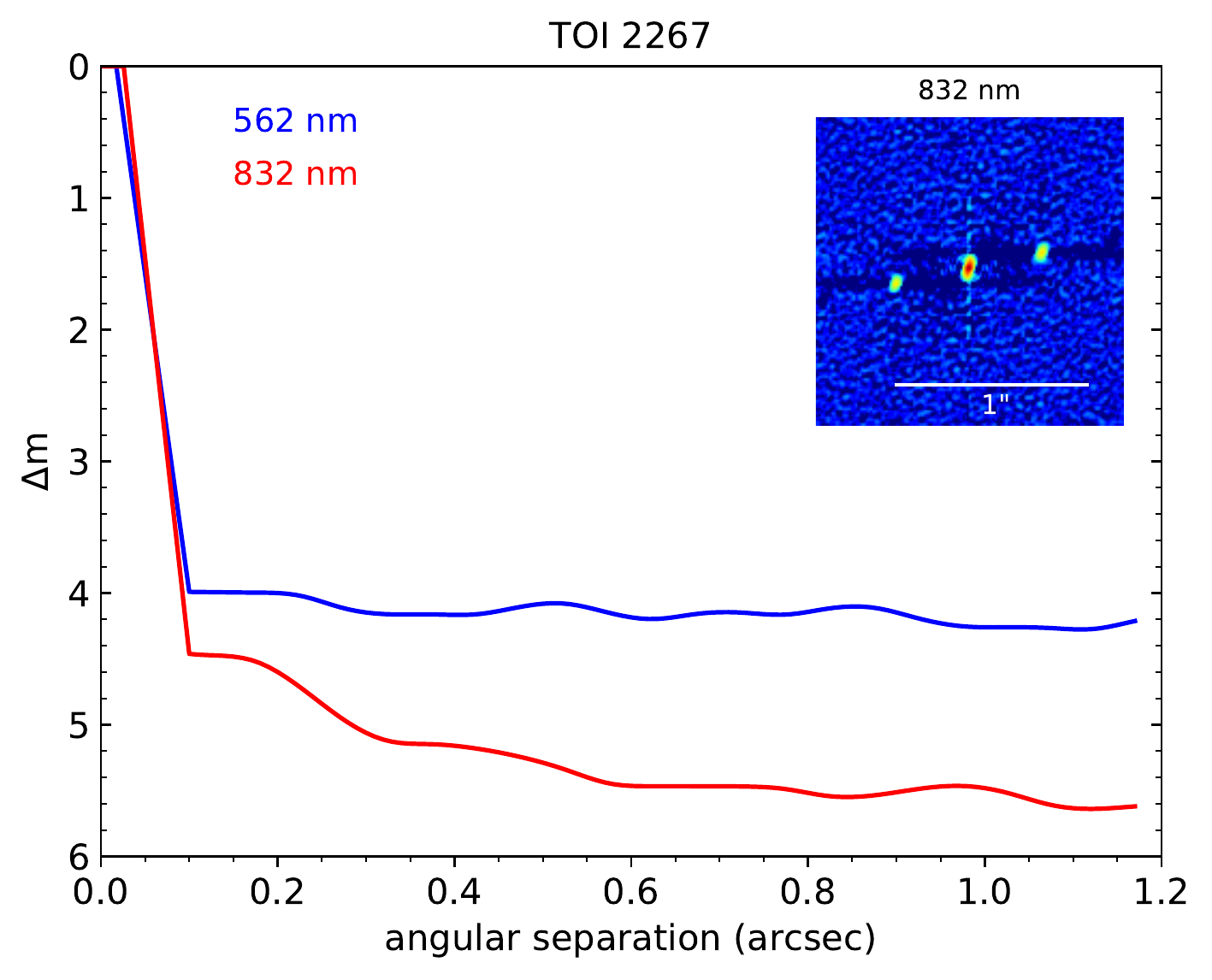}
    \caption{‘Alopeke speckle imaging 5$\sigma$ contrast curves along with the
reconstructed 832-nm image}
    \label{fig:AO-Gemini}
\end{figure}

TOI-2267 was observed on 2021 December 09 UT using the ‘Alopeke speckle instrument on the Gemini North 8-m telescope \citep{Scott2021}.  ‘Alopeke provides simultaneous speckle imaging in two bands (562~nm and 832~nm) with output data products including a reconstructed image with robust contrast limits on companion detections. Eighteen sets of 1000$\times$0.06\,sec images were obtained and processed using our standard reduction pipeline (see \citealt{Howell2011}). Fig.~\ref{fig:AO-Gemini} shows our final contrast curves and the 832 nm reconstructed speckle image. We detect a close companion to TOI-2267 residing at a position angle of 279.7 degrees and a separation of 0.384\,arcsec. The companion star is 1.6 magnitudes fainter than the primary target as measured in the 832 nm filter. No additional close companions brighter than 4-5 magnitudes below that of the target star from the 8-m telescope diffraction limit (20 mas) out to $1.2^{\prime\prime}$ were detected. At the distance of TOI-2267 ($\sim$22.55\,pc), these angular limits correspond to spatial limits of $\sim$0.45 to 27\,AU.

\subsubsection{SAI-2.5m}
\label{ss:speckle-SAI}

TOI-2267 was observed with the speckle polarimeter \citep{Strakhov2023} on the 2.5-m telescope at the Caucasian Observatory of Sternberg Astronomical Institute (SAI) of Lomonosov Moscow State University on three dates, details are given in Table~\ref{tab:stellar_companion}. For observations in 2020 and 2021, EMCCD Andor iXon 897 was used as a detector. The last observation was obtained with low--noise CMOS detector Hamamatsu ORCA--quest. During all observations, the atmospheric dispersion compensator was active, which allowed use of the $I_\mathrm{c}$ band. The respective angular resolution is 83~mas. Exposures and total number of accumulated frames were 60~ms and 3342, 60~ms and 6053, 23~ms and 5222, for three observational epochs, respectively.

\begin{figure}
    \centering
\includegraphics[width=\linewidth]{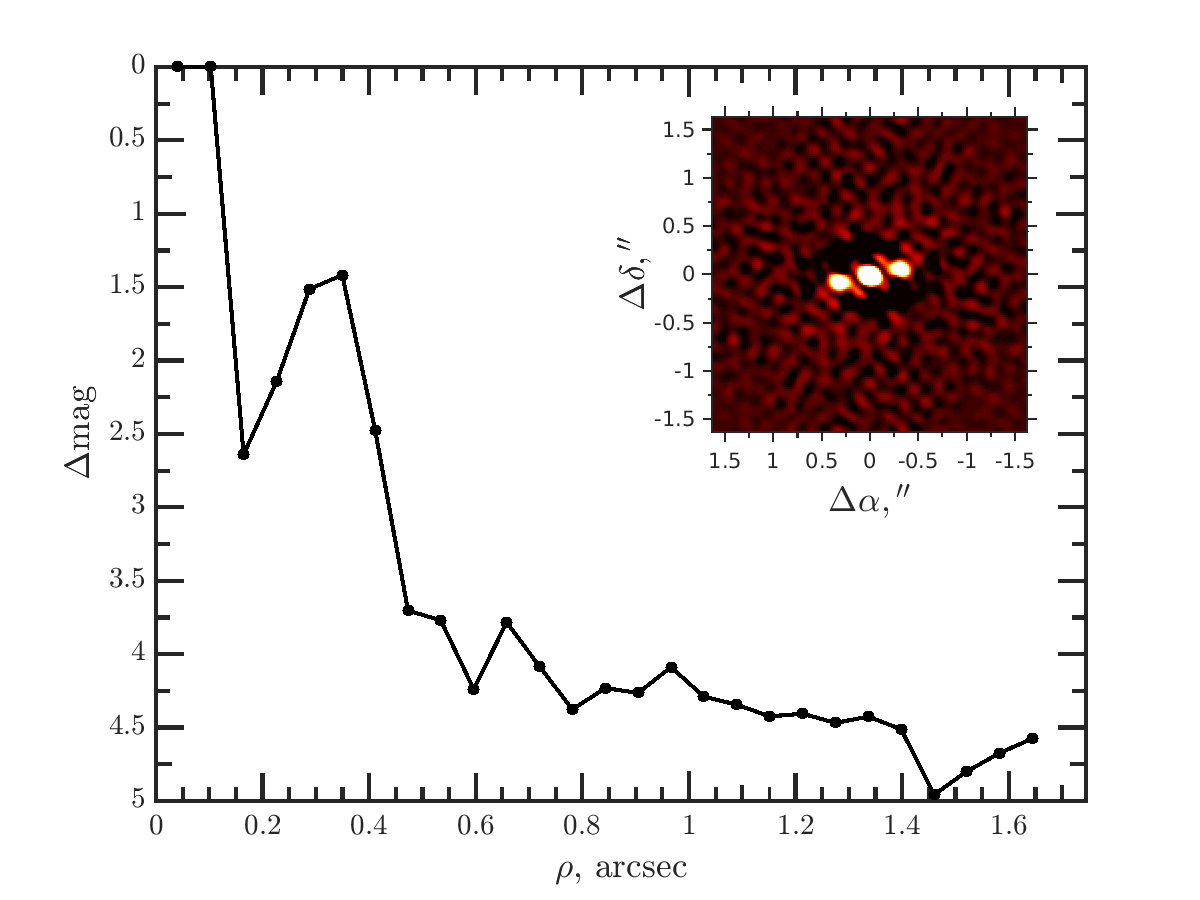}
    \caption{SAI-2.5m contrast curve for the observation obtained on 2024 August 09 UT. The bump at $0.3^{\prime\prime}$ occurs due to the companion. The autocorrelation function is given in the inset.}
    \label{fig:SAI-2.5m}
\end{figure}

In all epochs, a stellar companion was reliably detected (Table~\ref{tab:stellar_companion}). Binary parameters were determined using an approximation of the average power spectrum by the method described in \citep{Strakhov2023}. The $180^{\circ}$ ambiguity in position angle characteristic for speckle interferometry was resolved by use of the bispectrum. While the flux ratio demonstrates consistency between observations, separation clearly decreased during the past 3.8~yr. It is interesting that at the same time, the position angle of the binary changed only a little, indicating almost edge-on orientation of the binary orbit. Further high angular resolution and radial velocity observations of the object in the coming years will allow us to constrain the orbit. The epoch with the best seeing conditions (2024 August 09) was used to put limits on additional companions: $\Delta I_\mathrm{c}=2.9^m$ and $4.2^m$ at distances $0.25$ and $1.0^{\prime\prime}$ from the main star, respectively (see Fig.~\ref{fig:SAI-2.5m}).

\begin{table}
\caption{Speckle interferometric observations of TOI-2267 at SAI-2.5m.}
\label{tab:stellar_companion}
\centering
\resizebox{0.48\textwidth}{!}{
\begin{tabular}{lccccr}
\toprule
Date, UT & $\beta, ^{\prime\prime}$ & sep, mas & P.A.,$^{\circ}$ & flux ratio & SNR  \\
\midrule
2020-10-25 & 0.96 & $408\pm5$ & $283.8\pm0.4$ & $0.31\pm0.02$ & 30.97 \\
2021-10-22 & 1.12 & $393\pm2$ & $283.4\pm0.5$ & $0.27\pm0.05$ & 53.10 \\
2024-08-09 & 0.89 & $324\pm3$ & $282.3\pm0.5$ & $0.31\pm0.02$ & 24.16 \\
\bottomrule
\end{tabular}}
\tablefoot{
$\beta$ corresponds to the long-exposure seeing.
}
\end{table}

\subsection{Statistical validation}
The final step of our validation procedure was statistically validating the three planetary candidates. To this end, we used the \triceratops package \citep{giacalone2021}, which is widely used to validate \textit{TESS} planets statistically. \triceratops
uses the folded light curve of the candidates at their given epochs and periods and runs a Bayesian fit of several different possible astrophysical scenarios \citep[see Table 1 and Sec. 2.2 from][]{giacalone2021}, obtaining for each one the corresponding marginal likelihoods. These likelihoods are normalised once all the scenarios are examined so their sum equals one. Under this assumption, \triceratops computes two parameters to elucidate if a given candidate is a validated planet, likely a planet, or a nearby false positive. These parameters are the False Positive Probability (FPP) and the Nearby False Positive Probability (NFPP) with the following thresholds for each case: 1) FPP < 0.015 and NFPP < 0.001 for validated planets, 2) FPP < 0.5 and NFPP < 0.001 for likely planetary signals, and 3) NFPP > 0.1 for likely nearby false positives. 

In our case, using the high-resolution images described in 
Sec.~\ref {sec:HighRes-imaging} we learned that the system is composed of two stars, both carefully characterised in Sec.~\ref{s:Stellar Characterisation}. Since this information was not in GAIA, the default information imported by \triceratops was incomplete. Fortunately, the package allows for updating the stellar information, adding new stars when needed, and appropriately considering them when computing the marginal likelihoods for each astrophysical scenario. Hence, using our results presented in Tables~\ref{table:stellar} and \ref{tab:star_phot}, we included in our validation process the two stars that compound the TOI-2267 system.  

Moreover, \triceratops allows for adding extra constraints depending on the follow-up observations gathered. In particular, it allows for dropping those astrophysical scenarios that are known to not be at work; for example, the scenarios TP, EB, and EBx2P correspond to the case where the target star does not have an unresolved companion; what we know is false in our case. Then, we dropped them. Moreover, by examining the archival images presented in Sect.~\ref{sec:archival}, we know that there are no stars in the background at the current location of TOI-2267 that might be inducing the transit signals; hence, these scenarios can also be dropped; they are: DTP, DEB, DEBx2P, BTP, BEB, BEBx2P. 

It is important to note that in this study, we are unable to determine unambiguously around which star each planetary candidate orbits (see Sec.~\ref{sec:global_fit} and Sec.\ref{s:Conclusion}). Thus, for the purposes of validating the planetary nature of the signals in this binary system, we adapted the original \triceratops methodology \citep[see equations 4 and 5 from][]{giacalone2021} by adding to the FPP computation the probabilities corresponding to transiting planets around the bound companion scenarios (STP and NTP), and correspondingly removing the NTP scenario from the NFPP calculation. Hence, our adjusted FPP and NFPP values accurately reflect the probabilities relevant to planetary validation in this particular binary system configuration. 

Taking all these considerations into account, we obtained for TOI-2267.01 values of FPP=$(5.6 \pm 3.4) \times 10^{-5}$ and NFPP=$(4.8 \pm 3.1) \times 10^{-5}$. For TOI-2267.02, we derived FPP=$(1.6\pm 0.6) \times 10^{-4}$ and NFPP=$(1.5\pm 0.6) \times 10^{-4}$. Finally, for TOI-2267.03, the calculated values were FPP=$(1.2\pm 0.5) \times 10^{-5}$ and NFPP=$(9.9 \pm 4.6) \times 10^{-6}$.
These computed values for FPP and NFPP are notably low, clearly satisfying the established validation criteria for planetary candidates with high statistical confidence. Specifically, their corresponding false-positive probabilities translate to confidence levels exceeding 99.99\% regarding their planetary nature.

%===========================================================================
\section{Global transit analysis}

%===========================================================================

\subsection{Derivation of the system parameters}
\label{sec:global_fit}

We carried out our light-curve analyses using the \allesfitter package \citep{allesfitter-code,allesfitter-paper}, which allowed us to model planetary transits using the \ellc package \citep{maxted2016} while accounting for other phenomena such as stellar flares, spots, and variability.
\allesfitter also allows several ways to model the correlated noise, including polynomials, splines, and Gaussian processes \citep[GPs;][]{Rasmussen2004}, which are implemented through the \celerite package \citep{Foreman-Mackey2017,Foreman-Mackey2017b}. The parameters of interest are retrieved using a Bayesian approach implementing a Markov chain Monte Carlo method \citep[see, e.g.,][]{hastings1970,ford2005} using the \emcee package \citep{foreman2013}, or the Nested Sampling inference algorithm \citep[see, e.g.,][]{feroz2009,feroz2019} using the \dynesty package \citep{speagle2020}. We used the Dynamic Nested Sampling algorithm to estimate the Bayesian evidence in this study directly. This strategy allows us to compare a diverse set of orbital configurations robustly.

For each planet, we fitted the ratio of planetary radius over stellar radius ($R_{p}/R_{\star}$), the sum of the stellar and planetary radius scaled to the orbital semi-major axis ($(R_{p}+R_{\star})/a$), the cosine of the orbital inclination ($\cos~i_{p}$), the mid-transit time ($T_{0}$), the orbital period ($P$), and, when considering eccentric scenarios, jointly the eccentricity and the argument of pericenter ($\sqrt{e}\cos \omega$ and $\sqrt{e}\sin \omega$). Given that the host star is part of a stellar binary system, the photometric data sets are contaminated by the companion star; in that regard, we included a dilution factor ($D_0$). The dilution factor prior distributions were calculated from the flux ratio at each bandpass. The dilution factor for the $I_c$ filter was obtained directly from high angular resolution observation (see Table \ref{tab:stellar_companion}), and we leave it fixed as a reference. The flux ratios for the rest of the bandpasses were estimated from the SED model. We report the flux-ratio values and the corresponding dilution factor priors in Table \ref{tab:dilution_factor}. 

\begin{table}[hbt!]
\caption{Flux ratios and dilution factors for TOI-2267.}
\label{tab:dilution_factor}
\centering
\resizebox{0.35\textwidth}{!}{%
\begin{tabular}{lcc}
\toprule
\multirow{2}{*}{Bandpass} & Flux ratio & Dilution factor \\
 &  &  \\
\midrule
$i'$ & $0.2894\pm0.05$ & $0.2244\pm0.05$ \\
$z'$ & $0.3342\pm0.05$ & $0.2505\pm0.05$ \\
$Ic$ & $0.3000\pm0.05$ & $0.230769$ (fixed)\\
\textit{TESS} & $0.3045\pm0.05$ & $0.2334\pm0.05$ \\
\bottomrule
\end{tabular}
}
\tablefoot{
Flux ratio values are estimated from the SED model (see Sect.~\ref{SED}), except for the $I_c$ filter which was obtained from high-resolution imaging observations (see Sect.~\ref{sec:HighRes-imaging}). The corresponding dilution factor values were used as Gaussian prior distributions in the global fit analysis.
}
\end{table}

We used a hybrid spline to model the baseline for our ground-base observations. For more complex systematics, present in the data obtained by \textit{TESS}, we used GPs to model the correlated noise using the stochastically driven damped harmonic oscillator (SHO) kernel, described by its three hyper-parameters; the frequency of the undamped oscillator ($\omega_0$), the quality factor of the oscillator ($Q$) and $S_0$, which is related to the amplitude of the variability \citep[further detail see,][]{Foreman-Mackey2017}. \allesfitter also fits an error scaling term to account for white noise in the data.

To reduce the number of free parameters in the models, we fixed the quadratic limb-darkening (LD) coefficients during the fitting process \citep[see, e.g.,][]{kipping2017,Guenther2019}. We computed the values of the LD coefficients in the physical $\mu$-space, $u_1$ and $u_{2}$, for each bandpass used in the data interpolating from the tables of \cite{Claret2012}. For the non-standard $I+z$ filter, we took the averages of the values for the standard filters $Ic$ and Sloan-z$'$. Then, following the relations of \cite{kipping2013}\footnote{$q_{1}=(u_{1}+u_{2})^{2}$ and $q_{2}=0.5u_{1}/(u_{1}+u_{2})$}, we converted $u_1$ and $u_{2}$ to the transformed $q$-space, $q_{1}$ and $q_{2}$ used in \allesfitter. We report these values in Tables \ref{table:LD_A} and \ref{table:LD_B}.

Before performing a global model accounting for all the available data, we analysed each light curve independently to estimate the GPs hyper-parameters and the white noise parameter \citep[see for detailed description,][]{Guenther2019,pozuelos2023}. For the \textit{TESS} data, we used the orbital periods and transit times obtained from the SPOC pipeline to refine the transit locations by performing a preliminary fit across all \textit{TESS} sectors, applying wide, uniform priors. Next, we masked eight-hour windows around each identified transit midpoint and, using out-of-transit data, fit for noise and GPs hyper-parameters with wide uniform priors. Finally, we refined the planetary and orbital parameters by propagating the out-of-transit GPs posteriors as priors into a fit of the full \textit{TESS} dataset, applying Gaussian distributions. This allowed us to sample the planetary and orbital parameters from wide uniform priors. We tested the configuration with eccentricity set to zero and as a free parameter in the \textit{TESS} dataset fit. In both cases, we obtained practically the same results and no significant preference from the Bayesian evidence. Then, to reduce the number of free parameters, we fixed $\sqrt{e}\cos \omega$ and $\sqrt{e}\sin \omega$ to zero during the fitting process. In the context of planets in multiple-star systems, recent population analyses indicate that orbital eccentricities primarily correlate with the ratio between the star-star projected separation and the planetary semi-major axis, s/a, rather than with the absolute separation \citep{GonzalezPayo2024}. For TOI-2267 planets, the s/a ratios range from 300 to 600, placing the system in the high s/a regime where no eccentricity enhancement is expected, consistent with our near-circular solutions. The presence of Earth-sized planets in this close binary therefore adds to the small but growing set of compact low-eccentricity architectures known in multiple systems.

The follow-up observations could present high levels of red noise. To determine which observations could effectively refine the transit parameters and which were dominated by red noise at scales larger than the transit signals, we independently fit each dataset using a pure-noise model and a transit-and-noise model, recording the Bayesian evidence for each. For the pure-noise model, we fit the light curves by setting the planetary radius ratio ($R_{p}/R_{\star}$) to zero (indicating no transit) and use a hybrid spline to model the red noise. In the transit-and-noise model, we fit the planetary and orbital parameters by sampling the posteriors from the \textit{TESS}-only model using uniform distributions. We then computed the Bayes factor for each follow-up observation as $\Delta \ln Z = \ln Z_\mathrm{transit} - \ln Z_\mathrm{noise}$. Strong evidence for a transit signal was identified in the data when $\Delta \ln Z > 5$ \citep{Trotta2008}. Ground-based observations that met this criterion were directly included in the global analysis. The complete list of follow-up observations, along with their corresponding $\Delta \ln Z$ values, is presented in Table~\ref{tab:GBobservations}.

During the modelling procedure, \allesfitter uses the stellar mass and radius obtained from our stellar characterisation (see Sect. \ref{sec:stellar_cha}) to compute a normal prior on the stellar density (see Table \ref{table:stellar}). During the fitting process, the stellar density is calculated at each Nested Sampling step from the fitted parameters via $\rho_{\star}\approx\frac{3\pi}{GP^{2}}(\frac{a}{R_{\star}})^{3}$ \citep{seager2003} and compared with the value provided in the prior. The fits are penalised when these two values disagree.

Given the particular architecture of this system, before performing the global model combining \textit{TESS} and ground-based observation, we masked each transit in \textit{TESS} data to fit each candidate independently. Each of these single planets \textit{TESS}-only fits were tested with the primary and the secondary star as a host. For each candidate fit, we computed the Bayes factor as $\Delta \ln Z_\mathrm{host} = \ln Z_\mathrm{primary} - \ln Z_\mathrm{secondary}$. The TOI-2267.01 and TOI-2267.03 candidates showed a Bayes factor slightly favourable with the primary star as a host, while TOI-2267.02 showed a Bayes factor favourable with the secondary star. Despite these results, the Bayesian evidence was not strong enough to confirm the host star for each candidate ($-1 < \Delta \ln Z_\mathrm{host}< 1$). As we describe in Sec.~\ref{sec:dyn}, the three planets orbiting the same star would lead to a highly unstable architecture. Given the close 3:2 mean motion resonance for TOI-2267.01 and TOI-2267.03, and the lack of ground-based confirmation for TOI-2267.02, we decided to joint-fit TOI-2267.01 and TOI-2267.03, including the ground-based observation. Then, the orbital fit of TOI-2267.02 was performed isolated from the other candidates using just \textit{TESS} data.

\begin{figure*}[htb!]
    \centering
    \includegraphics[width=0.95\textwidth]{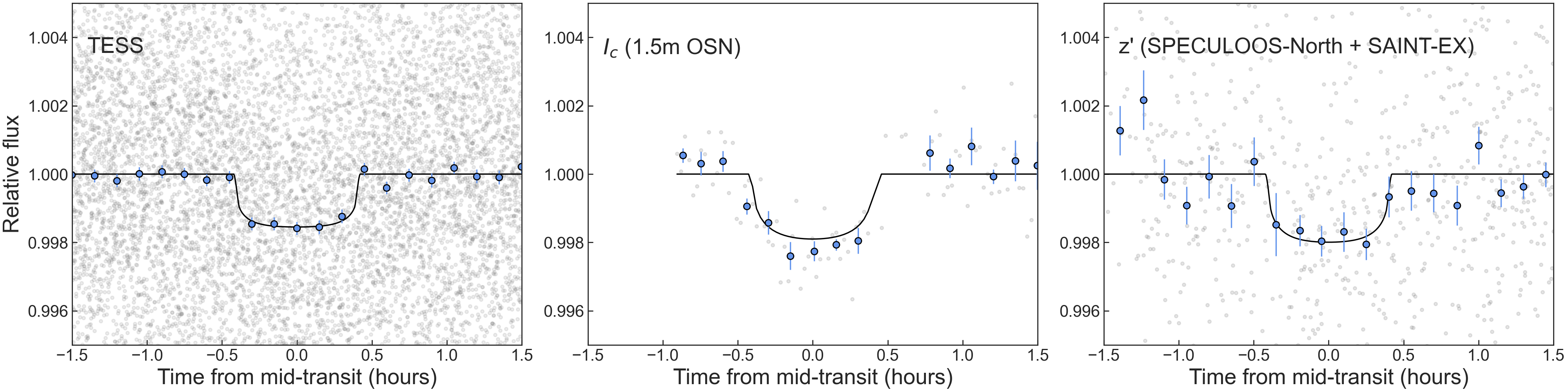}
    \caption{Phase-folded and detrended photometry of TOI-2267\,b transits along with the best-fit transit model (solid black line). The unbinned data points are shown in grey, while the blue circles with error bars correspond to 9-min bins.}
    \label{fig:lc_toi2267b}
\end{figure*}

\begin{figure*}[htb]
    \centering
    \includegraphics[width=0.95\textwidth]{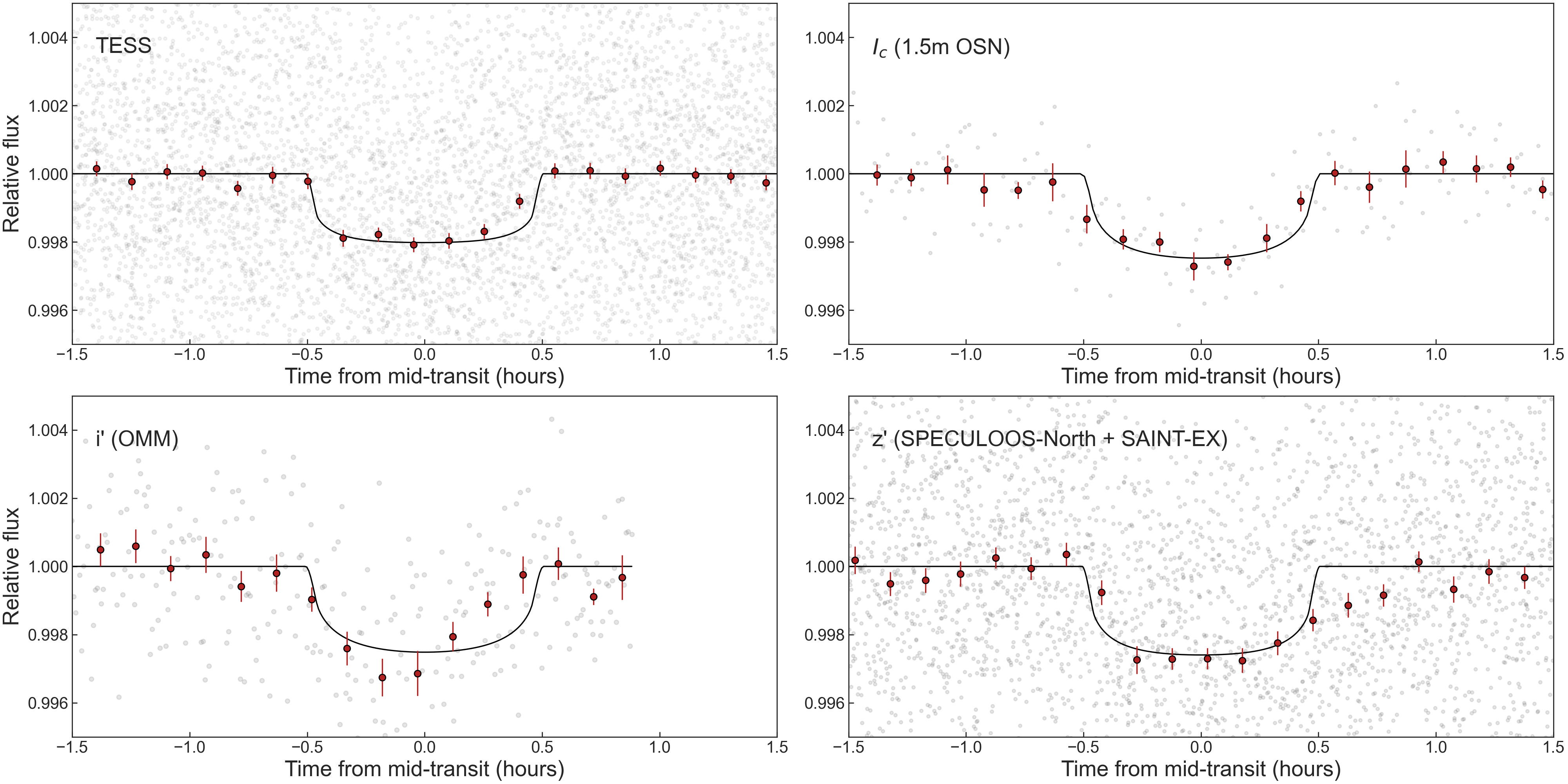}
    \caption{Phase-folded and detrended photometry of TOI-2267\,c transits along with the best-fit transit model (solid black line). The unbinned data points are shown in grey, while the red circles with error bars correspond to 9-min bins.}
    \label{fig:lc_toi2267c}
\end{figure*}

\begin{figure}[htb]
    \centering
    \includegraphics[width=0.95\linewidth]{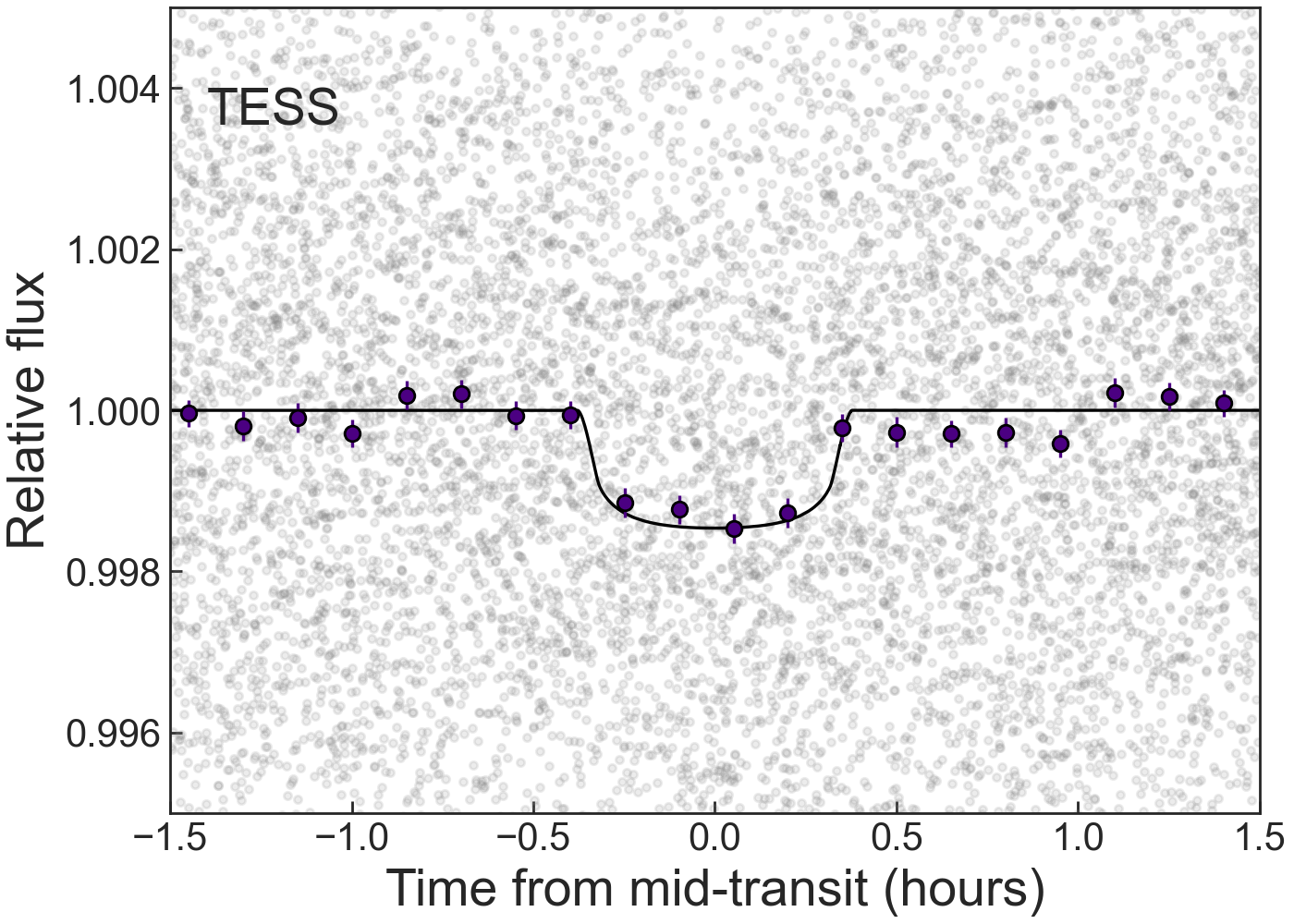}
    \caption{Phase-folded and detrended photometry of TOI-2267.02 transit along with the best-fit transit model (solid black line, secondary star host). The unbinned data points are shown in grey, while the dark blue circles with error bars correspond to 9-min bins.}
    \label{fig:lc_toi2267_02}
\end{figure}

The result of the global fit shows a preference for the configuration where planets TOI-2267.01 and .03 orbit the primary star, with a $\Delta \ln Z_\mathrm{host} \sim 0.5$ favourable to primary star host scenario and host density from orbit in agreement at 1$\sigma$ level with the host density from stellar characterisation (see Fig. \ref{fig:bc_density} and \ref{fig:d_density}). These results reveal a tentative configuration of the planetary system where TOI-2267.01 and .03 are genuine planets orbiting the primary star of the binary, TOI-2267A. Hence, hereafter, we denoted these planets as TOI-2267\,b and TOI-2267\,c. However, since this result is based on the limited Bayes evidence and 1$\sigma$ agreement host density from orbit but not directly observed, future follow-up studies will focus on confirming this hypothesis. In Fig.~\ref{fig:lc_toi2267b} and ~\ref{fig:lc_toi2267c}, we display our best-fitting model using the \textit{TESS} and the ground-based data. In Fig.~\ref{fig:lc_toi2267_02}, we show our best-fitting model using \textit{TESS} for TOI-2267.02, which remains a planetary candidate due to the lack of independent confirmation using ground-based observations. The fitted and derived physical parameters for primary and secondary host cases for planets b, c, and the candidate TOI-2267.02 are reported in Table ~\ref{table:planet_params_bc} and \ref{table:planet_params_d}.

\subsection{Check for transit chromaticity}
\label{sec:chromaticity}

\begin{figure}[htb]
    \centering
    \includegraphics[width=0.9\linewidth]{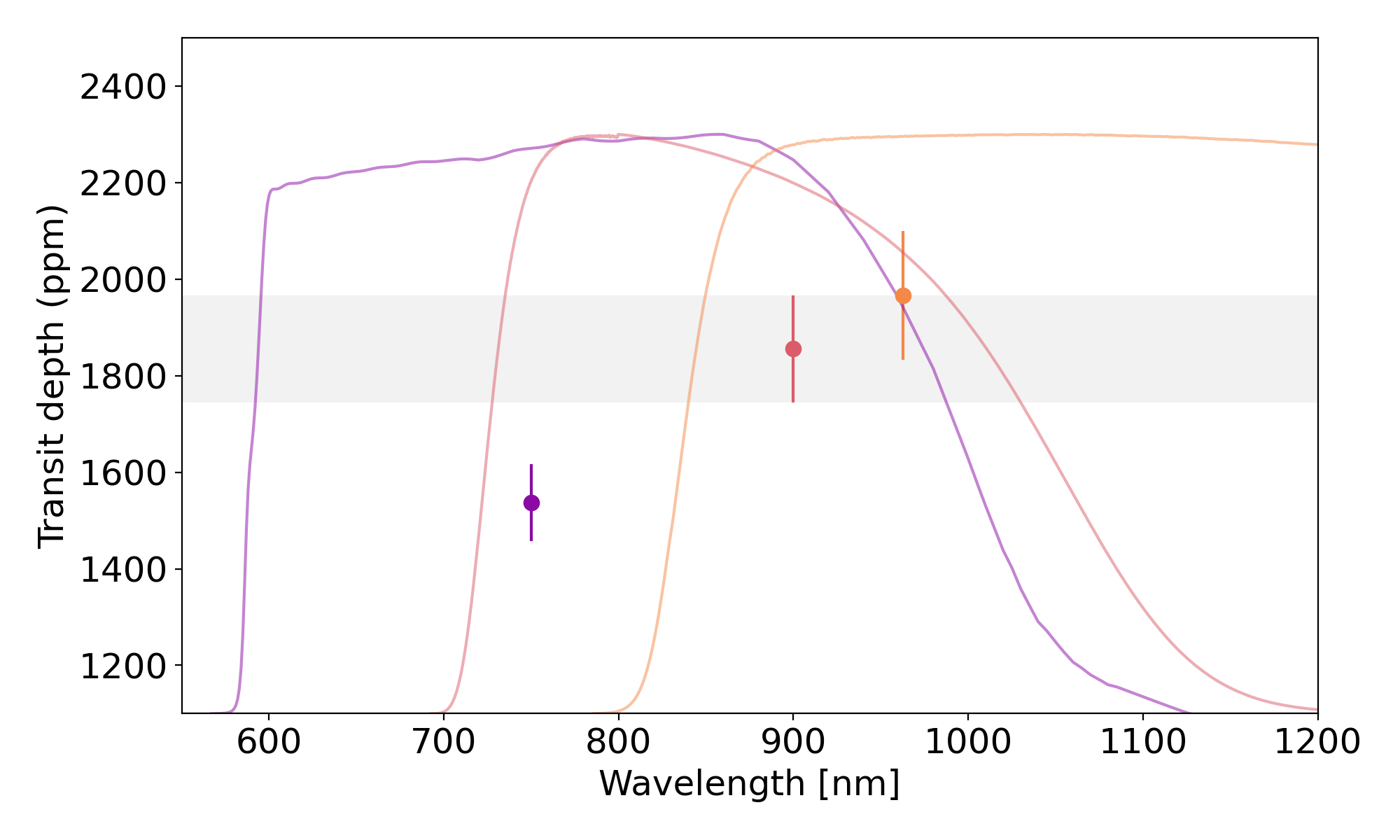}
    \includegraphics[width=0.9\linewidth]{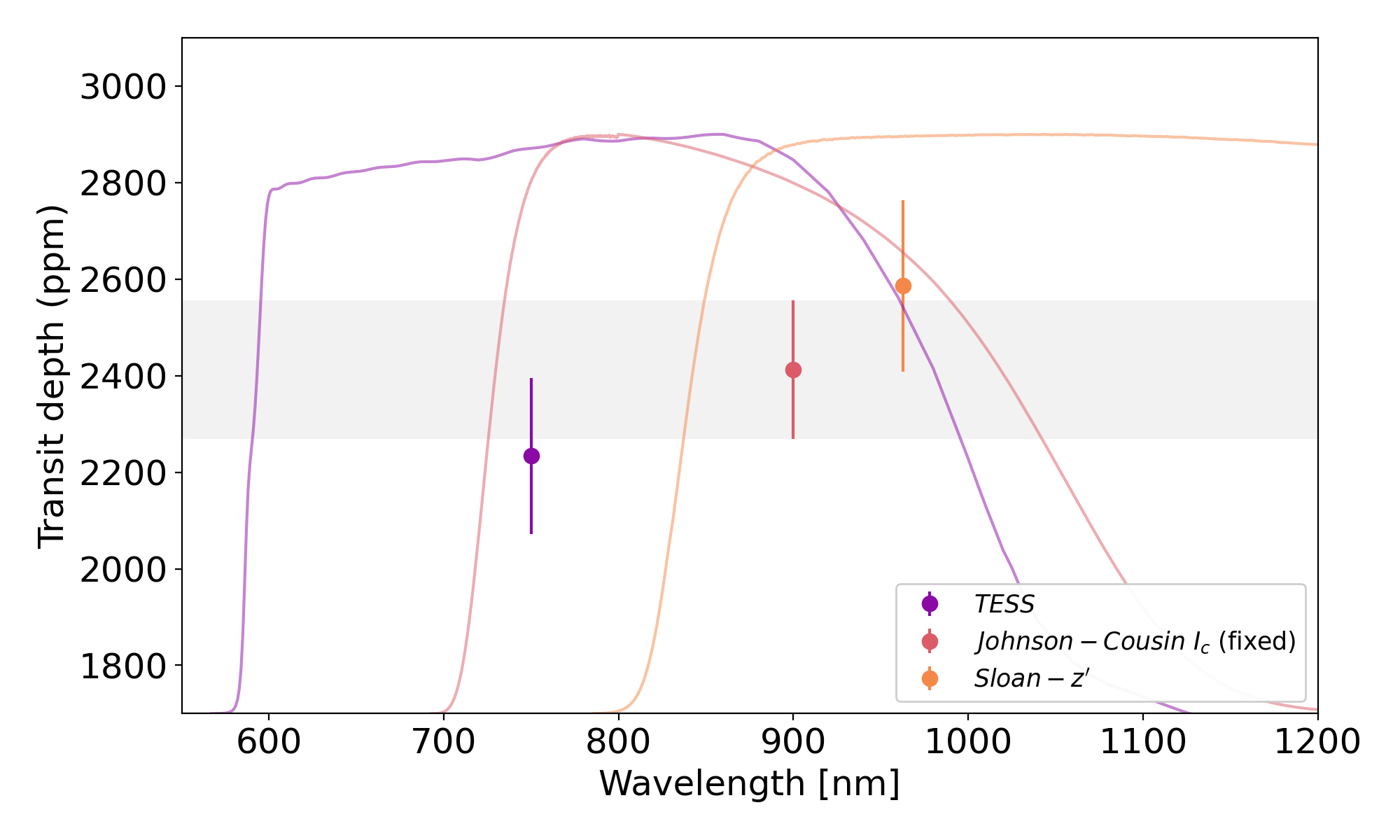}
    \caption{Measured transit depth vs wavelength for TOI-2267A b. The dark grey horizontal line indicates the depth of the Johnson-Cousin $I_c$ filter (dilution factor fixed in the global fit, see Sect. \ref{sec:global_fit}); other filters are highlighted by circles for comparison. \textit{Top}: Diluted transit depths. \textit{Bottom}: Undiluted transit depths corrected by the dilution factor.}
    \label{fig:transit_dephts_b}
\end{figure}

\begin{figure}[htb]
    \centering
    \includegraphics[width=0.9\linewidth]{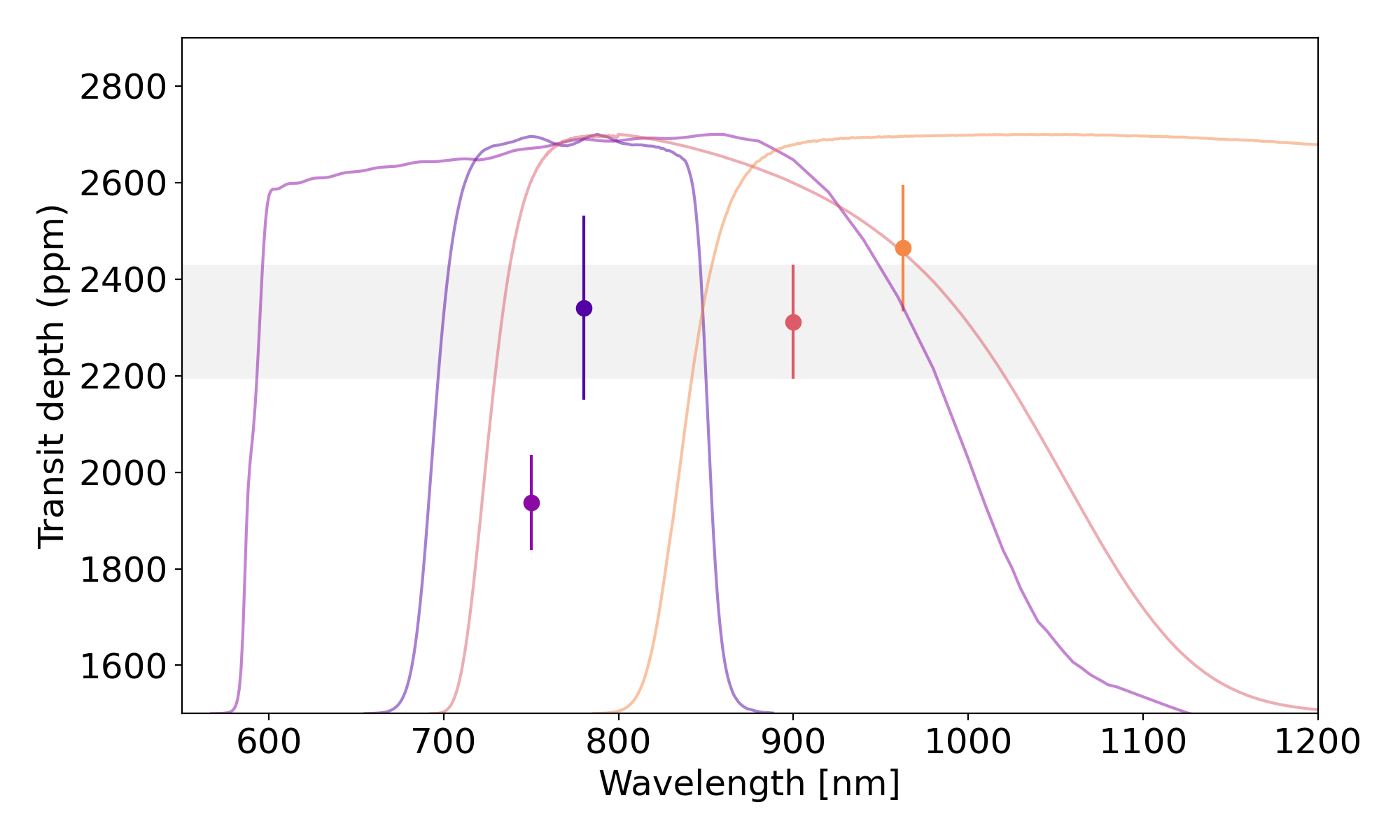}
    \includegraphics[width=0.9\linewidth]{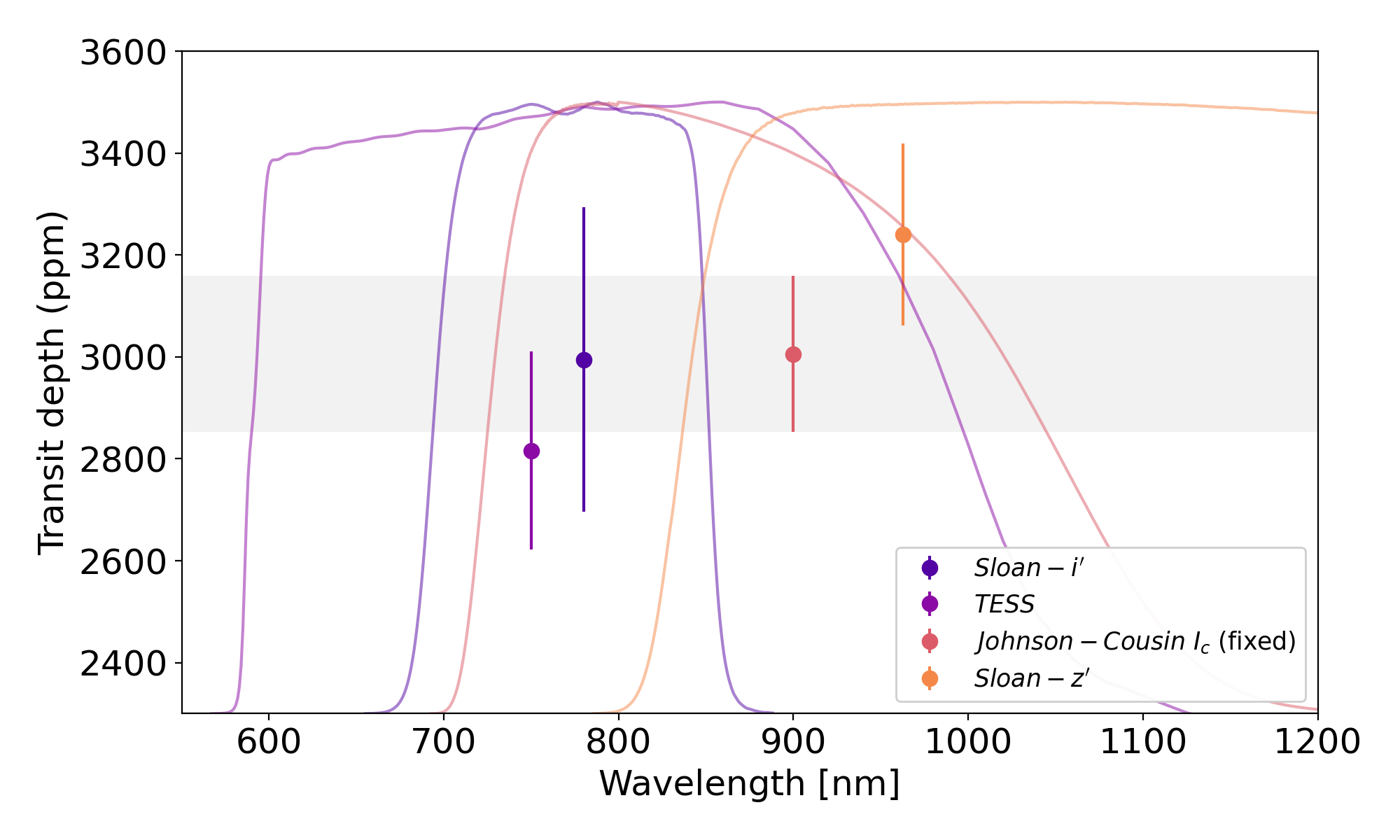}
    \caption{Measured transit depth vs wavelength for TOI-2267A c. The dark grey horizontal line indicates the depth of the Johnson-Cousin $I_c$ filter (dilution factor fixed in the global fit, see Sect. \ref{sec:global_fit}); other filters are highlighted by circles for comparison. \textit{Top}: Diluted transit depths. \textit{Bottom}: Undiluted transit depths corrected by the dilution factor.}
    \label{fig:transit_dephts_c}
\end{figure}

The early analyses of ground-based observations showed no evidence for wavelength-dependent transit depths. To verify this behaviour, we checked the diluted and undiluted transit depth values from our global analysis of the TOI-2267\,b, and c planets. We compared the posterior distributions of the derived transit depths corresponding to the ground-based observations and \textit{TESS} (see Tables~\ref{tab:transit_depths} and \ref{tab:transit_depths_b}), finding that all undiluted depths, after dilution factor correction, agree at the 1$\sigma$ level (see Fig.~\ref{fig:transit_dephts_b} and \ref{fig:transit_dephts_c} for primary host, and Fig. \ref{fig:B_transit_dephts_b} and \ref{fig:B_transit_dephts_c} for secondary host). Hence, we confirm that the transits for both planets do not show any chromatic dependence, thus reaffirming the results and conclusions presented in the previous section.

\renewcommand{\arraystretch}{1.2} % más espacio entre filas
\begin{table}[htb!]
\caption{Measured transit depths at each bandpass for the global fit assuming the primary star as the host.}
\centering
\resizebox{0.35\textwidth}{!}{%
\begin{threeparttable}
\begin{tabular}{lcc}
\toprule
Transit depth, $\delta$ & Diluted (ppm) & Undiluted (ppm)\\
& & \\
\midrule
Bandpass & \multicolumn{2}{c}{TOI-2267A b}\\
\midrule
\textit{TESS} & $1537\pm80$ & $2230_{-150}^{+160}$ \\
$Ic$ & $1854_{-83}^{+110}$ & $2410_{-120}^{+150}$ \\
$z'$ & $1970\pm130$ & $2630\pm160$ \\
\midrule
 & \multicolumn{2}{c}{TOI-2267A c}\\
\midrule
$i'$ & $2340_{-190}^{+180}$ & $2990_{-280}^{+300}$ \\
\textit{TESS} & $1937_{-88}^{+99}$ & $2820_{-180}^{+190}$ \\
$Ic$ & $2312_{-120}^{+93}$ & $3010_{-150}^{+120}$ \\
$z'$ & $2470_{-140}^{+130}$ & $3290_{-170}^{+180}$ \\
\bottomrule
\end{tabular}
\tablefoot{
Undiluted transit depths are corrected by the dilution factor (see Sect.~\ref{sec:global_fit}).
}
\end{threeparttable}
}
\label{tab:transit_depths}
\end{table}
\renewcommand{\arraystretch}{1.0} % restaurar valor por defecto

%===============================================================================
\section{Planet searches and detection limits from the \textit{TESS} photometry}
\label{s:sherlock}
%===============================================================================

As discussed in Section~\ref{ss:TESS}, we executed our pipeline \sherlock when only the candidate TOI-2267.01 had been identified by the SPOC pipeline. \sherlock is a dedicated Python package designed to explore space-based photometric data in search of planetary signals, originally introduced in \cite{pozuelos2020} and \cite{Demory_AA_SAINTEX_2020}. We refer the reader to \cite{devora2024} for the latest version of the pipeline and a comprehensive overview of its functionalities.

During our first data exploration using Sectors 19, 20, 25, and 26, \sherlock identified a strong sinusoidal signal attributable to a fast rotator using a similar procedure as described in Section~\ref{sec:rot_period}, with a maximum peak at 0.695\,d (see Fig.~\ref{fig:Rot_period}). Such variability might hinder any planetary detection; hence, to remove it previously to the planetary search, \sherlock automatically used the \texttt{cosine} function provided by the \wotan package \citep{hippke2019}. Once this correction was applied, \sherlock properly identified a signal with an orbital period of 3.5\,d. This detection corresponds to the released TOI-2267.01, allowing us to recover this candidate independently. In the subsequent runs, we identified two unknown extra signals: one with an orbital period of 2.28\,d and the other with an orbital period of 2.03\,d. We executed our vetting module and found no evidence of any potential false positive origin. Each time a new sector became available, we executed \sherlock to check if more signals appeared; however, until our last execution, including the twelve sectors currently available with a 2\,min cadence, only the three original signals were detected.

The apparent absence of additional candidates using \textit{TESS} data, whether analysed by \sherlock or SPOC, could be attributed to one of the following scenarios \cite[see, e.g.,][]{wells2021,Schanche2021,pozuelos2023}: (1) no other planets exist in the system; (2) other planets exist but do not transit; (3) other planets exist and transit, but their orbital periods are longer than those explored in this study; or (4) other planets exist, and transit, but the photometric precision of the data is insufficient to detect them. Scenarios (1) and (2) could be further investigated through high-precision radial velocity follow-up, which is out of the scope of this study. Scenario (3) can be tested by extending the observational baseline. To evaluate the final scenario, we performed injection-and-recovery experiments using the \matrixtk  code\footnote{{The \texttt{MATRIX} (\textbf{M}ulti-ph\textbf{A}se \textbf{T}ransits \textbf{R}ecovery from \textbf{I}njected e\textbf{X}oplanets) code is open access on GitHub: \url{https://github.com/PlanetHunters/tkmatrix}}} \citep{devora2022}.

\begin{figure}[hbt]
    \centering
    \includegraphics[width=\columnwidth]{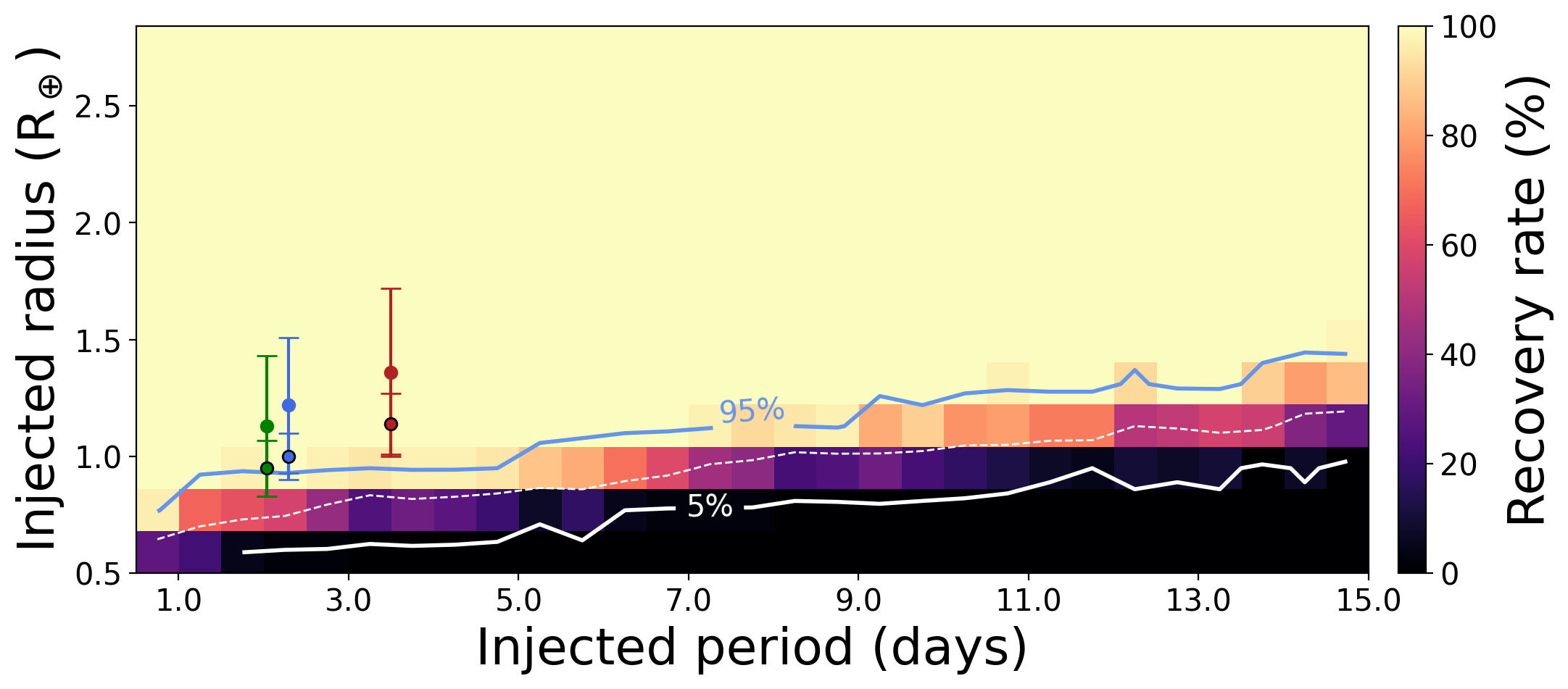}
    \caption{\matrixtk injection-and-recovery experiment conducted to establish the detection limits using the ten \textit{TESS} sectors described in Sect.~\ref{ss:TESS}. We explored a total of 18000 different scenarios, where each pixel evaluated 40 of them. Colours indicate the recovery rate: Bright yellow is for a high recovery, and dark purple and black are for a low recovery. The solid blue line indicates the 95\% recovery contour, the dashed white line marks the 50\% recovery contour, and the solid white line corresponds to the 5\% recovery contour. Blue, red, and green points show planets b and c and candidate .02, respectively. Black-edged circles denote planets orbiting the primary star; edge-free circles indicate the secondary star.}
    \label{fig:ir_results}
\end{figure}

This code conducts a three-dimensional parameter space exploration, generating a grid of orbital periods, planetary radii, and transit epochs. Each of these sets of parameters generates a synthetic scenario that is injected into the original light curve; in our case,  we set up a grid of 60 periods (from 0.5 to 15\,d), 30 radii (from 0.5 to 3 $R_\oplus$), and 10 different epochs, making a total of 18000 scenarios. We note that the upper limit of 15 days was chosen not only to ensure computational feasibility but also because, under the assumption of a co-planar system, planets with longer periods would likely not transit \citep[see, e.g.,][]{jenkins2019}. 
For each of these scenarios, the rotational signal at 0.70\,d was corrected as we did with \sherlock, and then we applied a detrend with a bi-weight filter using a window size of 0.25\,d, which was found to be the optimal length during the \sherlock exploration. Then, the light curves are processed in the search for planets using the same algorithm as \sherlock. A synthetic planet is considered as retrieved when its found period and epoch differ by at most 1\% and up to 1 hour from the injected values, respectively. 
The execution results are shown in Fig.~\ref{fig:ir_results}. On the one hand, the smallest detectable planets have sizes of 0.6\,$R_\oplus$ and would be placed in the innermost orbits below $\sim$1\,d, as expected. That is, any planet smaller than that value would be undetectable for any orbital period. On the other hand, for any transiting planets with sizes larger than 1.5\,$R_{\oplus}$ with orbital periods up to 15\,d, we got 100\% of recovery rate; hence, their existence can be discarded. Above an orbital period of 8\,d, transiting Earth-sized planets begin to be missed at some epochs, becoming undetectable with very low recovery rates at periods $\sim$11\,d and above.

%===============================================================================
\section{System architecture}
\label{sec:dyn}
%===============================================================================
\subsection{Instability of the three-planet solution}
\label{sec:3planet}
Throughout the previous sections, we fully validated the planetary nature of the candidates TOI-2267\,b, and c, but the candidate TOI-2267.02, while statistically validated, still remains elusive from a ground-based confirmation. According to our results, if this candidate becomes a validated planet, its orbital period would be 2.03\,d, dangerously close to TOI-2267\,b, whose orbital period is 2.28\,d. Then, we decided to test the system's stability, considering that three planets orbit the same star using the nominal values presented in Tables~\ref{table:planet_params_bc} and \ref{table:planet_params_d}, and circular orbits for the three bodies. The planets' masses are derived from the mass-radius relationship obtained using the \texttt{Springht} package \citep{spright2024}. Since we do not know the actual physical distance between the two stellar components of TOI-2267, only the projected distance between them, we conducted two analyses with single-star scenarios. In the first case, the three planets orbit the primary star, and in the second, they orbit the secondary. For stability tests, we used the most recent version of the Stability Orbital Configuration Klassifier\footnote{SPOCK:  \url{https://github.com/dtamayo/spock}} \citep[{\tt SPOCK 2.0.0};][]{spock2020}, a machine-learning model capable of classifying the stability of compact 3+ planetary systems over $10^{9}$ orbits of the innermost planet, which in this case translates into $\sim$10$^{7}$\,yr \citep[see, e.g.,][]{pozuelos2023,dreizler2024}. To further enhance the robustness of our study, we performed a Monte Carlo analysis using 1000 realisations of the planetary masses drawn from their \texttt{Spright}-derived distributions. We then computed the corresponding stability probabilities with \texttt{SPOCK}. We found that for the scenario in which the three planets orbit the primary star, none of the simulations resulted in a stability probability greater than 0.9, and only 3\% exceeded a threshold of 0.5. In the scenario where the planets orbit the secondary star (M2), this percentage increased slightly to 6.5\%, but none reached a stability greater than 0.9. Hence, for both scenarios, we found highly unstable architectures, hinting that the three-planet configurations orbiting the same star are highly disfavoured. We foresee two potential solutions: the TOI-2267.02 is not a genuine planet, or, if real, the three planets do not orbit the same star.

\subsection{Stability of the two-planet solution}
\label{sec:megno}

Due to the high instability of the three-planet configuration orbiting a single star, this subsection examines the dynamical stability of two-planet combinations: TOI-2267.02 and b, TOI-2267.02 and c, and TOI-2267\,b and c. Each scenario considers the two planets orbiting either the primary or secondary stellar component. As seen in Sec.\ref{sec:3planet}, due to the lack of well-constrained separation between the two stars, each case is modelled as a single-star system hosting two planets.

To this end, we employ the Mean Exponential Growth factor of Nearby Orbits, $Y(t)$ \citep[MEGNO;][]{cincottasimo1999,cincottasimo2000,cincotta2003}, a well-established diagnostic for assessing the dynamical stability of planetary systems \citep[see, e.g.,][]{hinse2015,jenkins2019,delrez2021}. The MEGNO parameter was computed using the {\scshape rebound} $N$-body integrator \citep{rein2012}, with the Wisdom-Holman WHfast symplectic algorithm \citep{rein2015}. The time-averaged value, $\langle Y(t) \rangle$, enhances stochastic variations in the orbital evolution, allowing a clear distinction between quasi-periodic motion (when $\langle Y(t \rightarrow \infty) \rangle \approx 2$) and chaotic trajectories (when $\langle Y(t \rightarrow \infty) \rangle \rightarrow \infty$).

For each two-planet configuration, 1000 realisations were generated by sampling orbital parameters from the posterior distributions in Tables~\ref{table:planet_params_bc} and \ref{table:planet_params_d}. Planetary masses were sampled from the distributions provided by the SPRIGHT package. Two eccentricity regimes were considered: strictly circular orbits (e = 0) and orbits with eccentricities up to e $\leq$ 0.025. Each configuration was integrated for 10$^{7}$ orbital periods of the outermost planet, using a timestep equal to 5\% of the innermost planet’s orbital period. A realisation was classified as dynamically unstable if its MEGNO value exceeded 2.1.
 
Our MEGNO analysis shows that configurations including planets .02 and b orbiting the same star are highly unstable across the explored parameter space. For strictly circular orbits, $\sim$78\% of the realisations are dynamically unstable, while allowing eccentricities up to $e \leq 0.025$ increases the fraction to $\sim$95\%. In contrast, configurations pairing planet .02 with c, or b with c, are found to be highly stable. In both cases, the fraction of unstable realisations is 0\% in the circular and eccentric regimes, indicating that these architectures remain stable for the entire integration timespan. These results strongly suggest that the actual system architecture consists of one of these two pairs orbiting the same stellar component, while the remaining planet (b or .02, respectively) orbits the other star.

The orbital periods of planets b and c lie close to a 3:2 mean-motion commensurability, whereas the .02–c pair is far from any low-order resonance. Proximity to such resonances can be the outcome of the system’s evolutionary history, as first-order resonances are frequently observed in compact multi-planet systems and are often interpreted as the signature of convergent migration within a protoplanetary disk \citep[see, e.g.,][]{bryden2000,ramos2017,huang2023,wong2024}. In this scenario, planets exchange angular momentum with the surrounding disk material, and the smooth inward migration facilitates resonant capture when the orbits converge \citep[e.g.][]{lee2002,batygin2015}. Resonant locking can then act as a stabilising mechanism, protecting the planets from close encounters despite their tight orbital spacing.

Given the stability of the b–c configuration, we extended our simulations to explore higher eccentricity regimes before investigating the possible resonant behaviour. Following \citet{Demory_AA_SAINTEX_2020}, we computed a MEGNO stability map of $150\times150$ pixels in the $e_b$--$e_c$ parameter space, ranging from 0 to 0.3. The resulting stability map shows that the mutual eccentricities must satisfy $e_{b} \leq 0.65 - 1.10\,e_{c}$ for long-term stability, implying that low eccentricities are strongly favoured. In particular, maximum mutual values of about $e_b \simeq e_c \simeq 0.025$ are allowed.

\begin{figure}
    \centering
\includegraphics[width=0.99\columnwidth]{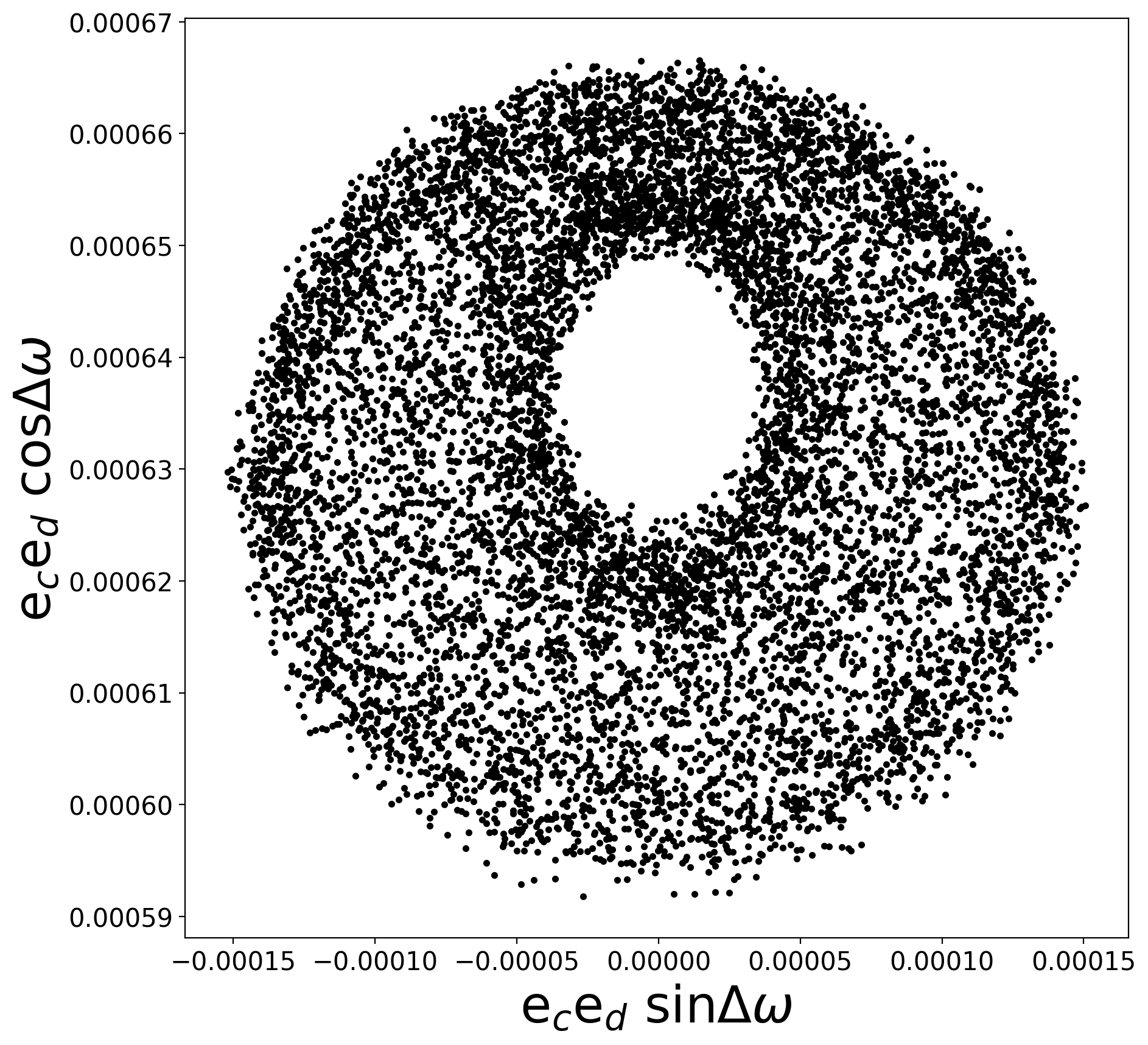}
    \caption{Polar plot of the apsidal trajectory ($e_b e_c$ vs $\Delta \omega$) for the TOI-2267\,b and c planets over a period of 1\,Myr. In this particular scenario, the planetary eccentricities were chosen as 0.025. The figure shows that the apsidal modes are librating around the 3:2 MMR (see Sect.~\ref{sec:megno}).}
    \label{fig:polar}
\end{figure}

Using these constraints, we then examined the dynamical behaviour of the b–c system to test whether it is indeed in 3:2 resonance. We computed the apsidal evolution over 1\,Myr for eccentricities between 0.001 and 0.025. Apsidal motion in interacting planetary systems can manifest as libration or circulation, separated by a secular separatrix \citep[see, e.g.,][]{barnes2006a,barnes2006b,barnes2006c,kane2014,kane2019}. In all simulations, the b–c system exhibited apsidal libration around a fixed point offset from the origin (Fig.~\ref{fig:polar}), with trajectories lying entirely in the positive $e_b e_c \cos \Delta\omega$ direction, indicative of an aligned configuration. This behaviour confirms that planets b and c are likely locked in a stable, aligned libration around the 3:2 mean motion resonance (MMR), consistent with an evolutionary pathway involving resonant capture during disk-driven migration.

Although the specific stellar component hosting each planet cannot be determined unambiguously, the dynamical analysis supports a configuration in which planets b and c orbit the same star in a first-order mean-motion resonance, with planet .02 orbiting the companion star. This arrangement prevents rapid dynamical instability and aligns with a formation pathway involving convergent migration within a protoplanetary disk. However, the presence of two low-mass stars may increase the complexity of the system’s formation and subsequent dynamical evolution.

\subsection{Transit timing variations}

As we have described previously, TOI-2267\,b and c are close to the first order 3:2 MMR, an orbital architecture that might produce significant TTVs \citep[see, e.g.,][]{lithwick2012} allowing us to measure planetary masses, which is especially critical to deepening our understanding for a given planetary system when other techniques such as radial velocities are challenging or impracticable \citep[][]{agol2021}. To assess whether a dedicated photometric campaign could yield precise masses, we generated 10,000 synthetic system realisations following the approach of \citet{pozuelos2023} for TOI‑2096.  Orbital periods and mid‑transit epochs were drawn from Gaussian distributions centred on the values in Table~\ref{table:planet_params_bc}, planetary masses were sampled from the mass–radius priors produced by the \texttt{Spright} package, and eccentricities uniformly from 0 to 0.025 to ensure dynamical stability.
For simplicity, and because the third candidate, TOI‑2267.02, remains not fully validated and, if real, may orbit the companion star instead, we restrict our analysis to two two‑planet hypotheses: one in which planets b and c orbit the primary component, and another in which they orbit the secondary.

\begin{figure}
    \centering
\includegraphics[width=0.99\linewidth]{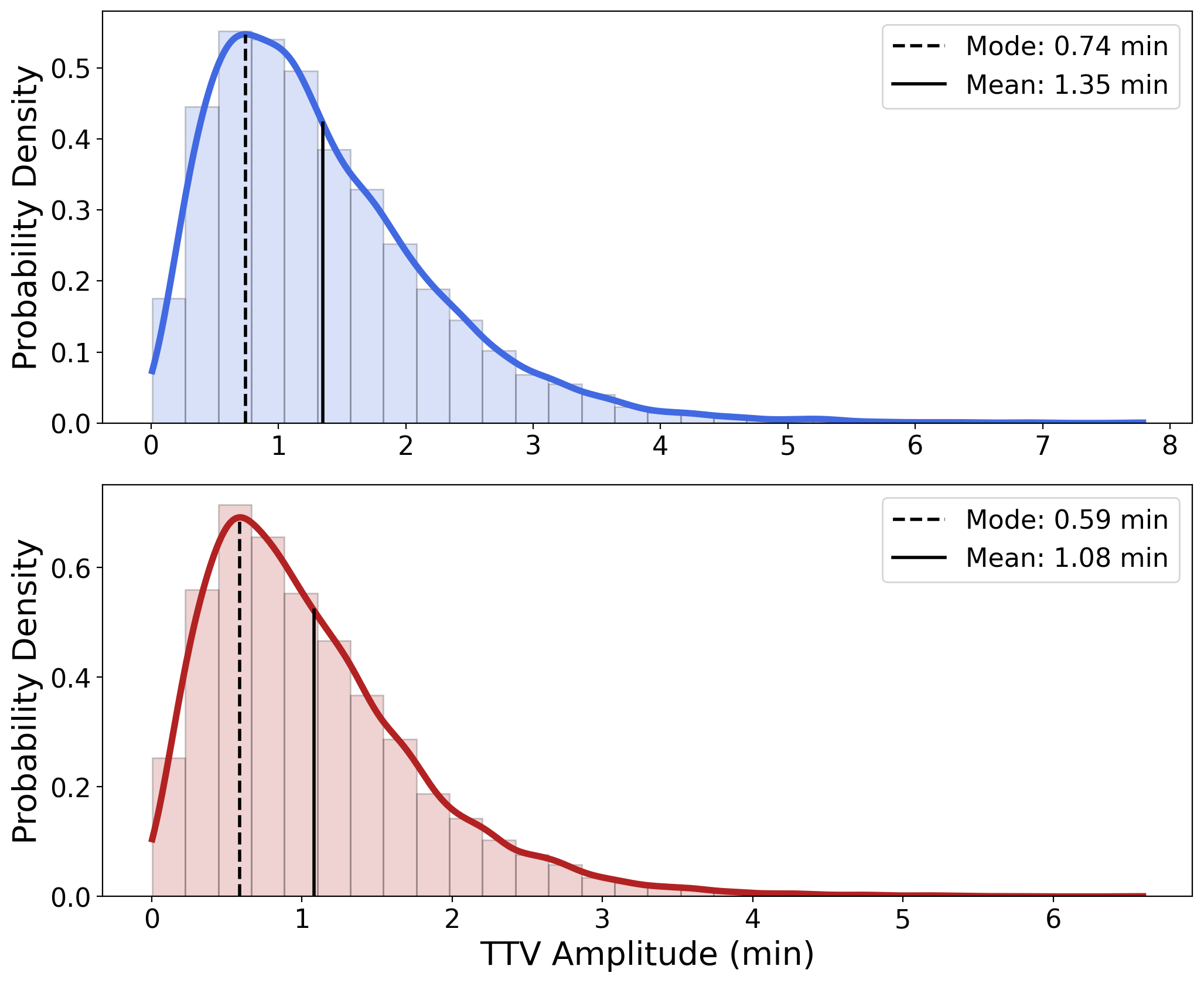}
    \caption{Expected TTV amplitudes for planets TOI‑2267\,b (upper panel) and TOI‑2267\,c (lower panel), assuming both planets orbit the primary star. Dashed and solid vertical lines correspond to the mode and mean of the PDFs, respectively.}
    \label{fig:ttvpre}
\end{figure}

We observed that the TTVs for each planet do not follow a symmetric distribution. Hence, we employed a 
Gaussian Kernel Density Estimation (KDE) to characterise these distributions using the
\texttt{gaussian-kde} function from the \texttt{scipy} package \citep{scipy}. This non-parametric
approach allowed us to accurately estimate the probability density functions (PDFs) of the TTVs without
assuming an underlying distribution. When assuming that planets b and c orbit the primary star, the KDE modes occur at $\sim$0.7 min and $\sim$0.6\,min, respectively, with PDF means of $\sim$1.4\,min (b) and $\sim$1.0\,min (c), and maximum TTV excursions of $\sim$4\,min (see Fig.~\ref{fig:ttvpre}). By contrast, under the alternative hypothesis that both planets orbit the secondary component, we obtain slightly larger amplitudes: KDE modes of 0.83 min (b) and 0.75 min (c), and means of 2.53\,min (b) and 2.26\,min (c), and maximum TTV excursions of $\sim$10 min.

Based on these results, we conclude that measuring the planetary masses of TOI-2267\,b and c using TTVs might be very challenging due to it requiring extremely high precision
photometry with mid-transit times of about 0.5\,min, which is out of reach of small- and mid-sized ground-based telescopes. In a recent work by \cite{gillon2024}, the
authors reported the discovery of SPECULOOS-3b, an Earth-sized planet orbiting an M6 star that is a
similar star-planet system to TOI-2267. 
In such a study, the HiPERCAM mounted on the GTC was used, which yielded a mid-transit precision of $\sim$5.2\,sec, well below the required precision that we would need for TOI-2267\,b and c. Hence, we conclude that GTC/HiPERCAM would be an ideal instrument for accurately measuring the TTVs and deriving the planetary masses, setting the stage for further studies using space-based facilities such as the James Webb Space Telescope.

%===============================================================================
\section{Discussion and prospects}
\label{s:Discussion}
%===============================================================================

\subsection{TOI-2267 in the context of the binary systems with planets}
\label{sec:binaries}
\begin{figure}
    \centering
\includegraphics[width=0.98\linewidth]{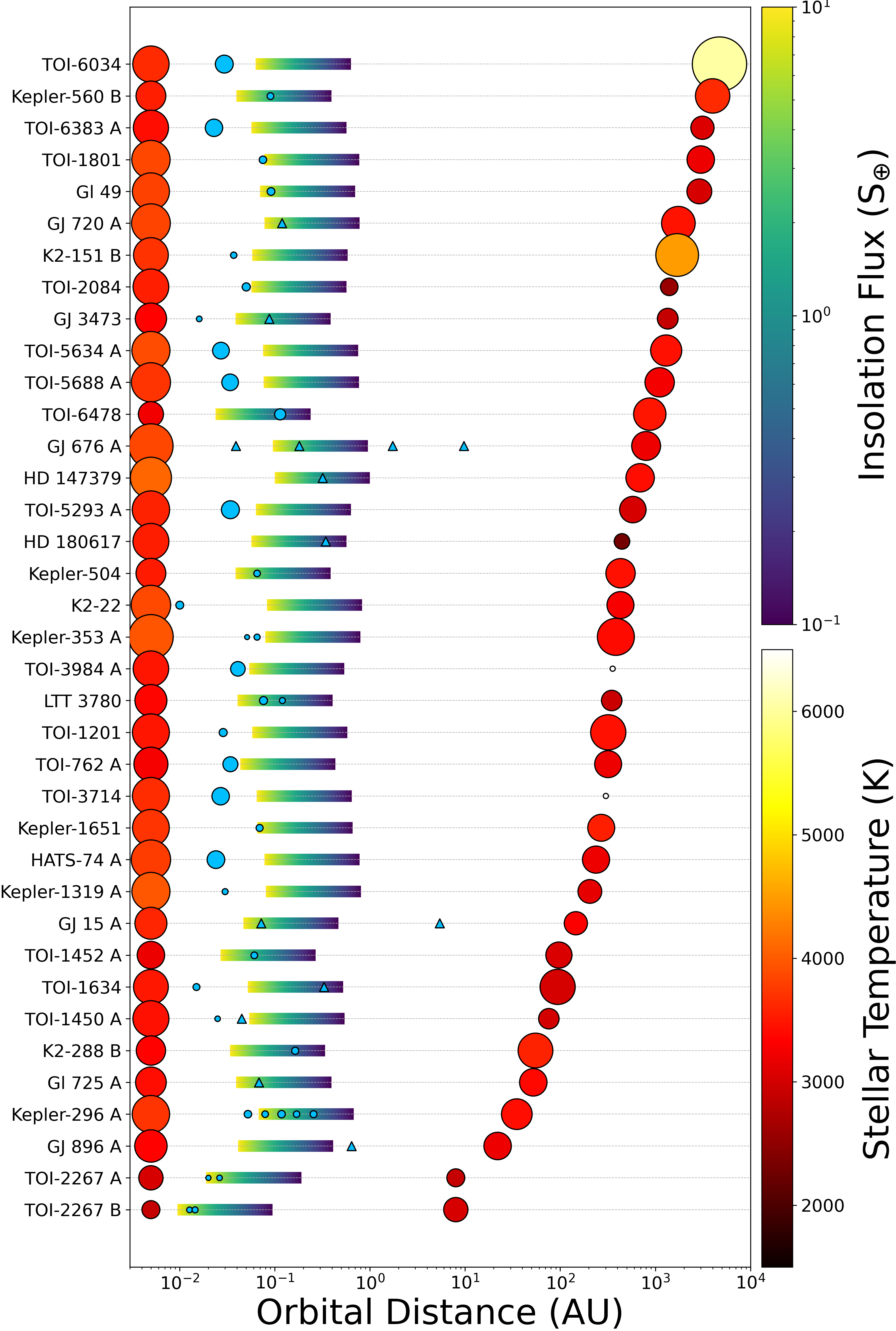}
    \caption{List of binary systems where the stellar component hosting planets is a low-mass star with $T_{\text{eff}} < 4000\,K$. The systems are arranged according to the projected separation between the two stellar components of the binary, with TOI-2267 standing out as the closest binary system found to harbour planets. The sizes of the stars are to scale relative to each other, and their colours represent their stellar temperatures. Blue circles indicate transiting planets, which are also scaled relative to each other. In contrast, blue triangles denote non-transiting planets (detected by radial velocity only), all set to an arbitrary size. For each system, the orbital distances from the host star where the insolation flux ranges from 10 to 0.1 $S_{\oplus}$ are highlighted.}
    \label{fig:binarios}
\end{figure}

Stellar multiplicity is a common outcome of star formation, with multiplicity fractions ranging from 40–50\% for FGK stars and 20–30\% for M-dwarfs \citep[e.g.][]{duchene2013,winters2019,clark2022}. However, among stars hosting planets, the occurrence of close stellar companions (projected separations $<$100\,au) appears strongly suppressed. For instance, the POKEMON survey \citep{pokemon1,pokemon2,pokemon3} found that M-dwarf binaries hosting planets tend to have projected separations peaking around 200\,au, while those without known planets peak at $\sim$6\,au. Similar trends have been observed for solar-type stars \citep{hirsch2021}.
These findings suggest that close stellar companions may hinder planet formation or survival.
Currently, approximately 8\% of confirmed exoplanets reside in binary systems, with 50 such planets orbiting M-dwarfs, spanning 37 distinct systems\footnote{According to the NASA Exoplanet Archive consulted in July 2025}. Figure~\ref{fig:binarios} shows these systems sorted by projected separation, using complementary estimates from \citet{pecaut2013} when some parameters of the secondary component were unavailable. Two systems, Kepler-779 and LHS 1678, were excluded from our final list due to insufficient information regarding the secondary component.  

 In this context, TOI‑2267 is remarkable for its exceptionally tight configuration: a projected separation of only 8\,au between its components, the smallest in our compiled sample. The only possible rival, LHS~1678, has an uncertain companion whose nature remains debated (planet vs. brown dwarf; \citealt{silverstein2022,silverstein2024}). Thus, TOI‑2267 stands as the clearest example to date of an extremely compact low-mass binary system hosting planets, challenging the current view that such architectures inhibit planet formation or long-term stability.

\subsection{Planet formation scenario}
\label{sec:formation}
At the dynamical level, circumstellar discs in binary systems exhibit significant differences from those around single stars \citep{Cuello+2025}. Most notably, they tend to dissipate more rapidly, thereby shortening the timescale for planet formation to occur \citep{Cieza+2009, Kraus+2012}. In addition, solid material in these discs drifts inward more quickly, potentially leading to substantial mass loss unless mitigated by pressure traps or instabilities \citep{Rosotti+2018, Zagaria+2023}. Binarity also leads to disc truncation, yielding smaller and less massive discs \citep{Pichardo+2005}. Consequently, if planet formation proceeds similarly in all environments, planets forming in binaries are expected to be more compact and less massive than those around single stars \citep{Akeson+2019, Quarles+2020}. For instance, under plausible assumptions about disc properties, sufficient material may have existed to form several rocky planets in the Alpha Centauri system \citep{Cuello+2024}. These findings suggest that planet formation in binaries may be less inhibited than previously thought, and that rocky planets in such systems could be common.

Overall, the study of S-type exoplanets (those orbiting one star in a binary) is still maturing. Comprehensive reviews by \cite{Marzari+2019} and \cite{Bonavita+2020} highlight the progress and limitations in this field. Current observational techniques remain biased, particularly in their ability to detect rocky planets in binaries, which hinders a comprehensive understanding of this population. Nonetheless, the potential for discovering multi-planet systems in binaries is growing thanks to an increase in sensitivity and longer monitoring campaigns. Our results on the TOI-2267 binary system, with at least one transiting planet candidate around each component, mark a significant step forward. This demonstrates the feasibility of detecting such systems and offers new opportunities to test planet formation models. Enhancing detection capabilities and correcting for survey incompleteness will be crucial for developing a more comprehensive view of planet occurrence in binary systems.

\subsection{Disentangling the planetary hosts}

As described in Section \ref{sec:global_fit}, a comparison of Bayesian evidence from transit fits accounting for the binary’s blended light dilution factor and the implied host star densities suggests that TOI-2267\,b and c orbit the primary star. However, this indirect statistical evidence cannot unambiguously assign the planets to their host. One potential method to assign the TOI-2267 planets to a host star is by directly detecting the transits through high-precision time-series photometry with simultaneous extremely high spatial resolution ($< 0.4''$) that resolves the binary components. This method has been demonstrated previously  for the Kepler-13 AB system \citep{Howell2019}, which hosts much brighter stars and has deeper transits. We scaled the photometric precision obtained for Kepler-13 with the 'Alopeke instrument on Gemini-N, and found that it would be impossible to resolve the binary and directly detect the shallow TOI-2267 transits using this technique for even the largest current ground-based telescopes. Instead, this direct transit detection could be made from space, as HST/WFC3 or JWST/NIRCam have sufficient technical capabilities. Both instruments have a sufficiently wide FOV and angular resolution to resolve the full PSFs of TOI-2267A and B, and a high enough photometric precision to detect the transits. Any space-based transit observations of the TOI-2267 planets would also provide exquisite transit timing measurements that could be combined with ground-based follow-up to better constrain the planetary masses through TTVs.

If TOI-2267.02 is confirmed to be a genuine planet, observations to assign the stellar host could reveal the first-ever double-transiting binary system architecture. There are a few known binary star systems with planets orbiting both components \citep{GonzalezPayo2024}, but none in which both components have transiting planets assigned. Hence, the TOI-2267 system may present the only feasible opportunity to characterise a double transiting binary system, with significant implications for studying planet formation scenarios in close binary systems.

%===============================================================================
\section{Conclusion}
\label{s:Conclusion}
%===============================================================================
Through this work, we have identified, validated, and provided a preliminary characterisation of the nearby system TOI-2267, which is composed of two very low-mass stars with at least two warm Earth-sized exoplanets. 
We first conducted a detailed characterisation of this peculiar cool binary. To do so, we leveraged SED fitting, spectroscopic analyses, stellar atmosphere modelling, and evolutionary models. These analyses revealed that the binary is composed of M5V and M6V stars, with effective temperatures of about 3030 and 2930 K, respectively. The system has a projected separation of around 8 au. This makes it the coolest binary with the smallest stellar projected separation known to host planets.     

We then validated the planetary nature of two of the three planetary candidates independently found by our team and \textit{TESS}/SPOC. To achieve this, we combined photometry from ground-based facilities, data from 12 \textit{TESS} sectors, archival information, high angular resolution imaging, and statistical considerations. As a result, we validated two planets, TOI-2267\,b and TOI-2267\,c, though we cannot yet unambiguously determine which star they orbit. For the third Earth-sized candidate, TOI-2267.02, our comprehensive validation analysis strongly supports its planetary nature, but it was not detected using ground-based observations. Despite favourable statistical evidence, we prudently keep it as a planetary candidate pending further confirmation.

When studying the system's architecture, we found an interesting result. If TOI-2267.02 is a genuine planet, placing all three planets around the same star becomes rapidly unstable. In addition, planets b and .02 cannot orbit the same star. The most stable solutions are the pairs b-c or .02-c orbiting the same star, while the other planet orbits the other component. Since planets TOI-2267 b and c are close to the 3:2 first-order MMR, the scenario with b-c orbiting one star and .02 orbiting the other is the most plausible.

This hypothesis makes the TOI-2267 system a benchmark case for further studies, as it would represent the first detection of transiting planets orbiting both components of an extremely compact binary. This unique configuration provides an opportunity to investigate how planetary formation and orbital dynamics operate under the influence of two close stellar hosts.

%===============================================================================
% Bibliography
%===============================================================================
\bibliographystyle{aa.bst} % style aa.bst
\bibliography{references.bib} % your references Yourfile.bib

\clearpage

%\onecolumn
\begin{appendix}
\section{}
%===============================================================================
% Acknowledgements
%===============================================================================
\begin{acknowledgements}\\
We thank the anonymous referee for providing constructive comments and suggestions, which helped improve the clarity and quality of this manuscript.
We sincerely thank Steven Giacalone for his helpful advice and guidance on properly using TRICERATOPS.
%Personal
F.J.P and P.J.A acknowledge financial support from the Severo Ochoa grant CEX2021-001131-S funded by MCIN/AEI/10.13039/501100011033 and Ministerio de Ciencia e Innovación through the project PID2022-137241NB-C43. Partially based on observations made at the Observatorio de Sierra Nevada (OSN), operated by the Instituto de Astrofísica de Andalucía (IAA-CSIC).
%% M. Gillon
 M.G. is F.R.S.-FNRS Research Director.
% J. Korth
J.K. acknowledge support from the Swiss NCCR PlanetS and the Swiss National Science Foundation. This work has been carried out within the framework of the NCCR PlanetS supported by the Swiss National Science Foundation under grants 51NF40182901 and 51NF40205606. J.K. gratefully acknowledges the support of the Swedish National Space Agency (SNSA; DNR 2020-00104) and of the Swedish Research Council (Project Grant 2017-04945 and 2022-04043) and of the Swiss National Science Foundation under grant number TMSGI2\_211697.
%% M.Gunther
M.N.G. acknowledges support from the European Space Agency (ESA) as an ESA Research Fellow. 
%% K.Barkaoui
The postdoctoral fellowship of KB is funded by F.R.S.-FNRS grant T.0109.20 and by the Francqui Foundation.
%% N. Cuello, M. Sucerquia
N.C. and M.S. acknowledge funding support from the European Research Council (ERC) under the European Union Horizon Europe program (grant agreement No. 101042275, project Stellar-MADE).
%% B.R-A
B.R-A acknowledges funding support from the ANID Basal project FB210003.
%% N. Schanche
The material is based upon work supported by NASA under award number 80GSFC24M0006.
%% B.V. Rackham
This material is based upon work supported by the National Aeronautics and Space Administration under Agreement No.\ 80NSSC21K0593 for the program ``Alien Earths''.
The results reported herein benefited from collaborations and/or information exchange within NASA’s Nexus for Exoplanet System Science (NExSS) research coordination network sponsored by NASA’s Science Mission Directorate.
%%%% Y. Gómez Maqueo Chew
YGMC acknowledges support from UNAM-PAPIIT-101224.
%% E. Palle
We acknowledge financial support from the Agencia Estatal de Investigaci\'on of the Ministerio de Ciencia e Innovaci\'on MCIN/AEI/10.13039/501100011033 and the ERDF “A way of making Europe” through project PID2021-125627OB-C32, and from the Centre of Excellence “Severo Ochoa” award to the Instituto de Astrofisica de Canarias.
%% Felipe Murgas
F. M. acknowledges financial support from the Agencia Estatal de Investigaci\'{o}n del Ministerio de Ciencia, Innovaci\'{o}n y Universidades (MCIU/AEI) through grant PID2023-152906NA-I00.
%% Julien de Wit and Artem Burdanov
J.d.W. and MIT gratefully acknowledge financial support from the Heising-Simons Foundation, Dr. and Mrs. Colin Masson and Dr. Peter A. Gilman for Artemis, the first telescope of the SPECULOOS network situated in Tenerife, Spain.
%% H. Parviainen
H.P. acknowledges support from the Spanish Ministry of Science and Innovation with the Ramon y Cajal fellowship number RYC2021-031798-I, and funding from the University of La Laguna and the Spanish Ministry of Universities.
%Facilities
%\textit{TESS}
Funding for the \textit{TESS} mission is provided by NASA's Science Mission Directorate. We acknowledge the use of public \textit{TESS} data from pipelines at the \textit{TESS} Science Office and at the \textit{TESS} Science Processing Operations Center. This research has made use of the Exoplanet Follow-up Observation Program website, which is operated by the California Institute of Technology, under contract with the National Aeronautics and Space Administration under the Exoplanet Exploration Program. This paper includes data collected by the \textit{TESS} mission that are publicly available from the Mikulski Archive for Space Telescopes (MAST).
%SPECULOOS 
Based on data collected by the SPECULOOS-North Observatory at the Observatorio del Teide in Canary Islands, Spain. The ULiege's contribution to SPECULOOS has received funding from the European Research Council under the European Union's Seventh Framework Programme (FP/2007-2013) (grant Agreement n$^\circ$ 336480/SPECULOOS), from the Balzan Prize and Francqui Foundations, from the Belgian Scientific Research Foundation (F.R.S.-FNRS; grant n$^\circ$ T.0109.20), from the University of Liege, and from the ARC grant for Concerted Research Actions financed by the Wallonia-Brussels Federation. This work is supported by a grant from the Simons Foundation (PI Queloz, grant number 327127). This research is in part funded by the European Union's Horizon 2020 research and innovation program (grants agreements n$^{\circ}$ 803193/BEBOP), and from the Science and Technology Facilities Council (STFC; grant n$^\circ$ ST/S00193X/1, and ST/W000385/1).
%%% SAINT-EX
This work is based on observations taken with the SAINT-EX telescope at the Observatorio
Astron\'omico Nacional on the Sierra de San Pedro M\'artir (OAN-SPM), Baja California, M\'exico, that is supported by the Swiss National Science Foundation (SNSF; PP00P2-163967 and PP00P2-190080, SPIRIT-216537), the Centre for Space and Habitability (CSH) of the University of Bern, the National Centre for Competence in Research PlanetS, supported by the SNSF, and UNAM PAPIIT-IG101224.
%%% For TRAPPIST-S/N
The research leading to these results has received funding from the ARC grant for Concerted Research Actions, financed by the Wallonia-Brussels Federation. TRAPPIST is funded by the Belgian Fund for Scientific Research (Fond National de la Recherche Scientifique, FNRS) under the grant PDR T.0120.21. TRAPPIST-North is a project funded by the University of Liege (Belgium), in collaboration with Cadi Ayyad University of Marrakech (Morocco).
%%% OMM
Based on observations obtained at the Observatoire du Mont-Mégantic, financed by Université de Montréal, Université Laval, the Canada Economic Development program and the Ministère de l'Économie et de l'Innovation du Québec. This work benefited from support of the Fonds de recherche du Québec – Nature et technologies (FRQNT), through the Center for Research in Astrophysics of Quebec.
%%% Gemini North
Some of the observations in this paper made use of the High-Resolution Imaging instrument ‘Alopeke and were obtained under Gemini LLP Proposal Number: GN/S-2021A-LP-105. ‘Alopeke was funded by the NASA Exoplanet Exploration Program and built at the NASA Ames Research Center by Steve B. Howell, Nic Scott, Elliott P. Horch, and Emmett Quigley. ‘Alopeke was mounted on the Gemini North telescope of the international Gemini Observatory, a program of NSF’s OIR Lab, which is managed by the Association of Universities for Research in Astronomy (AURA) under a cooperative agreement with the National Science Foundation. on behalf of the Gemini partnership: the National Science Foundation (United States), National Research Council (Canada), Agencia Nacional de Investigación y Desarrollo (Chile), Ministerio de Ciencia, Tecnología e Innovación (Argentina), Ministério da Ciência, Tecnologia, Inovações e Comunicações (Brazil), and Korea Astronomy and Space Science Institute (Republic of Korea).
%%%% GTC
Based on observations made with the GTC telescope, in the Spanish Observatorio del Roque de los Muchachos of the Instituto de Astrofísica de Canarias, under Director’s Discretionary Time (GTC10-23BDDT, IP F.J.P).
%%% IAS
IAS acknowledges the support of M.V. Lomonosov Moscow State University Program of Development.
%%%
% Software, data archives, cluster, etc
This research made use of Lightkurve, a Python package for Kepler, and \textit{TESS} data analysis \citep{2018ascl.soft12013L}. This work made use of Astropy:\footnote{http://www.astropy.org}, a community-developed core Python package and an ecosystem of tools and resources for astronomy \citep{astropy:2013, astropy:2018, astropy:2022}. This work made use of \texttt{allesfitter}, a public and user-friendly astronomy software package for modelling photometric and radial velocity data \citep{allesfitter-paper,allesfitter-code}.
This research has made use of the NASA Exoplanet Archive, which is operated by the California Institute of Technology, under contract with the National Aeronautics and Space Administration under the Exoplanet Exploration Program. Resources supporting this work were provided by the NASA High-End Computing (HEC) Program through the NASA Advanced Supercomputing (NAS) Division at Ames Research Center for the production of the SPOC data products.
Computational resources have been provided by the Consortium des Équipements de Calcul Intensif (CÉCI), funded by the Fonds de la Recherche Scientifique de Belgique (F.R.S.-FNRS) under Grant No. 2.5020.11 and by the Walloon Region. This research has made use of data obtained from or tools provided by the portal exoplanet.eu of The Extrasolar Planets Encyclopaedia.

\end{acknowledgements}

\newpage

\section{SED modelling}
\label{appendix:SED}

\renewcommand{\arraystretch}{1.1} % más espacio entre filas
\begin{table}[hbt!]
\caption{Fitted parameters correlation matrix.}
\centering
\resizebox{0.45\textwidth}{!}{%
\begin{threeparttable}
\begin{tabular}{lcccccc}
\toprule
 & $R_{\star,1}$ & $R_{\star,2}$ & $T_{\mathrm{eff},1}$ & $T_{\mathrm{eff},2}$ & $d$ & $F_I$ \\
\midrule
$R_{\star,1}$       &  1.00  & -0.89 & -0.90 &  0.83 &  0.15 & -0.08 \\
$R_{\star,2}$       & -0.89  &  1.00 &  0.65 & -0.98 & -0.07 &  0.10 \\
$T_{\mathrm{eff},1}$ & -0.90  &  0.65 &  1.00 & -0.57 &  0.08 & -0.12 \\
$T_{\mathrm{eff},2}$ &  0.83  & -0.98 & -0.57 &  1.00 &  0.08 & -0.03 \\
$d$                 &  0.15  & -0.07 &  0.08 &  0.08 &  1.00 & -0.31 \\
$F_I$               & -0.08  &  0.10 & -0.12 & -0.03 & -0.31 &  1.00 \\
\bottomrule
\end{tabular}
\tablefoot{
$R_{\star,1}$ and $R_{\star,2}$ are the radii of the primary and secondary stars, respectively; 
$T_{\mathrm{eff},1}$ and $T_{\mathrm{eff},2}$ are their effective temperatures; 
$d$ is the distance; 
$F_I$ is the flux in the $I$-band.
}
\end{threeparttable}
}
\label{tab:correlation_matrix}
\end{table}
\renewcommand{\arraystretch}{1.0} % restaurar valor por defecto

\section{NIRSPEC spectra modelling}
\label{appendix:spectra}
We briefly summarise our modelling method, following the analysis detailed in \cite{Hsu:2021aa}. We employed the forward-modelling method \citep{Blake:2010aa, Burgasser:2015ac} that fits our spectra with a single stellar template and telluric absorption simultaneously with the \texttt{SMART (Spectral Modelling Analysis and RV Tool)} package \citep{Hsu:2021ab}.
We chose the BT-Settl models \citep{Allard:2012ab} and the ESO earth atmosphere models \citep{Moehler:2014aa} to fit our NIRSPEC spectra.
We focused our analysis on order 33 (2.29–2.34 $\mu$m) because this order has the CO (2--0) rotational-vibrational bandhead, ideal for measuring precise RVs and $v\sin{i}$s values. 
We used the Markov chain Monte Carlo sampling method to derive our best-fit parameters using the \texttt{emcee} package \citep{foreman2013}, for a total of 9 parameters, 50 chains, and 600 steps, with the first 200 steps as burn-ins. 

\newpage
\onecolumn

\section{Ground-based photometric observations}
\begin{table*}[hbt!]
\caption{Ground-based time-series photometric observations logs of TOI-2267.}
\begin{center}
\begin{tabular}{l l c c c c}
\toprule
Candidate & Date (UT) & Telescope \& size & Bandpass & Exp. time (s) & $\Delta$ln$Z$ \\
\midrule
\multirow{16}{*}{TOI-2267.01 / TOI-2267\,c} & 14 Oct 2020 & TRAPPIST-North-0.6m & $I+z$ & 10 & <2.3 \\
  & 18 Nov 2020 & TRAPPIST-North-0.6m & $I+z$ & 10 & <2.3  \\
  & 25 Nov 2020 & SAINT-EX-1.0m & $z'$ & 10 & 8.95$^{\dag}$\\
  & 10 Dec 2020 & LCO-McD-1.0m & $i'$ & 82 & <2.3  \\
  & 17 Mar 2021 & SNO/Artemis-1.0m & $z'$ & 10 & 6.25$^{\dag}$\\
  & 18 Sep 2021 & SNO/Artemis-1.0m & $z'$ & 10 & 5.1$^{\dag}$\\
  & 25 Sep 2021 & SNO/Artemis-1.0m & $z'$ & 10 & <0 \\
  & 13 Nov 2021 & SNO/Artemis-1.0m & $z'$ & 10 & 6.34$^{\dag}$\\
  & 04 Dec 2021 & SNO/Artemis-1.0m & $z'$ & 10 & 11.3$^{\dag}$\\
  & 06 Nov 2021 & SNO/Artemis-1.0m & $z'$ & 10 & 23.5$^{\dag}$\\
  & 11 Dec 2021 & SNO/Artemis-1.0m & $z'$ & 10 & <0 \\
  & 18 Dec 2021 & SNO/Artemis-1.0m & $z'$ & 10 & <0  \\
  & 05 Mar 2022 & SNO/Artemis-1.0m & $r'$ & 55 & <2.3\\ 
  & 12 Mar 2022 & SNO/Artemis-1.0m & $z'$ & 10 & 8.77$^{\dag}$\\
  & 19 Mar 2022 & SNO/Artemis-1.0m & $z'$ & 10 & <0  \\
  & 08 Nov 2022 & SNO/Artemis-1.0m & $z'$ & 10 & 26.9$^{\dag}$ \\
  & 09 Nov 2022 & OMM-1.6m & $i'$ & 30 & 7.71$^{\dag}$ \\ 
  & 07 Oct 2023 & OSN/T150-1.52m & $I_c$ & 90 & 20.4$^{\dag}$ \\
\midrule
\multirow{4}{*}{TOI-2267.02 / TOI-2267\,d} & 20 Sept 2023 & OSN/T150-1.52m & $I_c$ & 90 & No detection\\ 
  & 25 Sept 2023 & LCO-Teid-1.0m & $i'$ & 82 & No detection \\
  & 18 Nov 2023 & OSN/T150-1.52m & $I_c$ & 90 & No detection \\
  & 7 Jan 2024 & GTC/10.4m & $ i'$ & 1 & No detection \\
\midrule 
\multirow{21}{*}{TOI-2267.03 / TOI-2267\,b} & 25 Nov 2020 & SAINT-EX-1.0m & $z'$ & 10 & 8.95$^{\dag}$\\
  & 04 Dec 2020 & SAINT-EX-1.0m & $z'$ & 10 & 3.8 \\
  & 22 Mar 2021 & SNO/Artemis-1.0m & $I+z$ & 10 & <2.3\\
  & 31 Mar 2021 & SAINT-EX-1.0m & $z'$ & 10 & <0 \\
  & 05 Sep 2021 & SNO/Artemis-1.0m & $z'$ & 10 & <0 \\
  & 21 Sep 2021 & SNO/Artemis-1.0m & $z'$ & 10 & <0 \\
  & 30 Oct 2021 & SNO/Artemis-1.0m & $z'$ & 10 & <0 \\
  & 06 Nov 2021 & SNO/Artemis-1.0m & $z'$ & 10 & 23.5$^{\dag}$\\
  & 15 Nov 2021 & SNO/Artemis-1.0m & $z'$ & 10 & <0 \\
  & 26 Nov 2021 & SAINT-EX-1.0m & $I+z$ & 10 & <2.3\\
  & 10 Dec 2021 & SNO/Artemis-1.0m & $z'$ & 10 & <0 \\
  & 10 Dec 2021 & SAINT-EX-1.0m & $I+z$ & 10 & <2.3\\
  & 17 Dec 2021 & SNO/Artemis-1.0m & $z'$ & 10 & <0 \\
  & 05 Mar 2022 & SNO/Artemis-1.0m & $r'$ & 55 & <2.3\\ 
  & 05 Mar 2022 & LCO-Teid-1.0m & $i'$ & 82 & <2.3\\
  & 21 Mar 2022 & SNO/Artemis-1.0m & $z'$ & 10 & <0 \\
  & 29 Aug 2022 & LCO-Teid-1.0m & $i'$ & 82 & <0 \\
  & 07 Oct 2022 & LCO-Teid-1.0m & $zs$ & 120 & <2.3 \\  
  & 07 Nov 2022 & LCO-Teid-1.0m & $i'$ & 85 & <2.3\\
  & 07 Nov 2022 & SNO/Artemis-1.0m & $z'$ & 10 & <0 \\
  & 20 Nov 2023 & OSN/T150-1.52m & $I_c$ & 90 & 23.9$^{\dag}$ \\
\bottomrule
\end{tabular}
\tablefoot{
Only those with Bayes factor $\Delta$ ln Z$>5$ (highlighted with a $\dag$ symbol) are selected for use in the global analysis (see Sect.~\ref{sec:global_fit} for details).
}
\label{tab:GBobservations}
\end{center}
\end{table*}

\newpage
\twocolumn
\section{Quadratic limb darkening coefficients}

\renewcommand{\arraystretch}{1.1} % más espacio entre filas
\begin{table}[hbt!]
\caption{Quadratic limb-darkening coefficients for TOI-2267A in each bandpass reported in Table~\ref{tab:GBobservations}.}
\centering
\resizebox{\columnwidth}{!}{%
\begin{threeparttable}
\begin{tabular}{l c c c c}
\toprule
bandpass & $u_{1}$ & $u_{2}$ & $q_{1}$ & $q_{2}$ \\
\midrule
$z'$ & $0.2719\pm0.05$ & $0.5352\pm0.05$ & $0.6514\pm0.04$ & $0.1684\pm0.06$ \\
$i'$ & $0.4323\pm0.07$ & $0.4699\pm0.07$ & $0.8139\pm0.06$ & $0.2395\pm0.07$ \\
$I_{c}$ & $0.4188\pm0.08$ & $0.4877\pm0.06$ & $0.8217\pm0.06$ & $0.2309\pm0.08$ \\
$I+z'$ & $0.3453\pm0.07$ & $0.5114\pm0.05$ & $0.7341\pm0.04$ & $0.2015\pm0.07$ \\
$r'$ & $1.0553\pm0.07$ & $-0.1423\pm0.04$ & $0.8335\pm0.04$ & $0.5779\pm0.06$ \\
\textit{TESS} & $0.1693\pm0.06$ & $0.5023\pm0.04$ & $0.4510\pm0.04$ & $0.1260\pm0.06$ \\
\bottomrule
\end{tabular}
\tablefoot{
The $u_{1}$ and $u_{2}$ values are from the theoretical tabulations of \citet{Claret2012,Claret2017}, and $q_{1}$ and $q_{2}$ are computed using the parametrisation of \citet{kipping2013}.
}
\end{threeparttable}
}
\label{table:LD_A}
\end{table}
\renewcommand{\arraystretch}{1.0} % restaurar valor por defecto

\renewcommand{\arraystretch}{1.1} % más espacio entre filas
\begin{table}[hbt!]
\caption{Quadratic limb-darkening coefficients for TOI-2267B in each bandpass reported in Table~\ref{tab:GBobservations}.}
\centering
\resizebox{\columnwidth}{!}{%
\begin{threeparttable}
\begin{tabular}{l c c c c}
\toprule
bandpass & $u_{1}$ & $u_{2}$ & $q_{1}$ & $q_{2}$ \\
\midrule
$z'$ & $0.3988\pm0.05$ & $0.4626\pm0.05$ & $0.7420\pm0.04$ & $0.2314\pm0.06$ \\
$i'$ & $0.6505\pm0.07$ & $0.2843\pm0.07$ & $0.8738\pm0.06$ & $0.3479\pm0.07$ \\
$I_{c}$ & $0.6135\pm0.08$ & $0.3195\pm0.06$ & $0.8704\pm0.06$ & $0.3287\pm0.08$ \\
$I+z'$ & $0.5061\pm0.07$ & $0.3910\pm0.05$ & $0.8049\pm0.04$ & $0.2820\pm0.07$ \\
$r'$ & $0.9784\pm0.07$ & $-0.0886\pm0.04$ & $0.7917\pm0.04$ & $0.5497\pm0.06$ \\
\textit{TESS} & $0.2275\pm0.06$ & $0.5520\pm0.04$ & $0.6076\pm0.04$ & $0.1459\pm0.06$ \\
\bottomrule
\end{tabular}
\tablefoot{
The $u_{1}$ and $u_{2}$ values are from the theoretical tabulations of \citet{Claret2012,Claret2017}, and $q_{1}$ and $q_{2}$ are computed using the parametrisation of \citet{kipping2013}.
}
\end{threeparttable}
}
\label{table:LD_B}
\end{table}
\renewcommand{\arraystretch}{1.0} % restaurar valor por defecto

\newpage
\onecolumn

\section{TESS PCDSAP light curves}

\begin{figure*}[ht] 
    \includegraphics[width=0.96\linewidth]{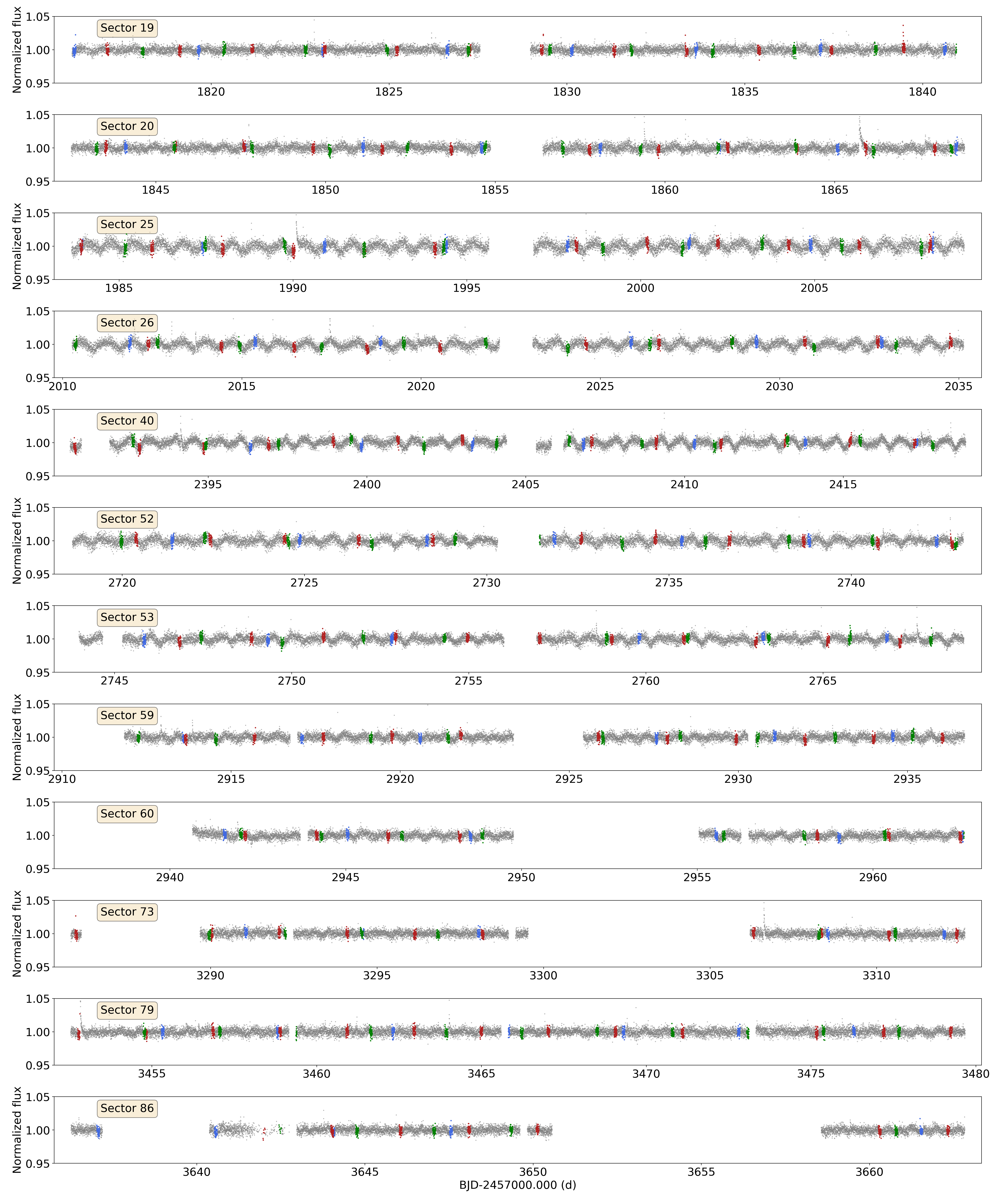}
    \caption{\textit{\textit{TESS}} photometric time series of TOI-2267 obtained for sectors 19, 20, 25, 26, 40, 52, 53, 59, 60, 73, 79, and 86. In all cases, the grey points correspond to the PDCSAP fluxes obtained from the SPOC pipeline. The blue, red and green points correspond to the location of the transits for the candidates TOI-2267.01, TOI-2267.02 and TOI-2267.03, respectively.}
    \label{fig:lcs}
\end{figure*}

\newpage
\section{\textit{TESS} field of view}
\label{appendix:TES_FoV}

\begin{figure*}[ht]
\includegraphics[width=0.99\textwidth]{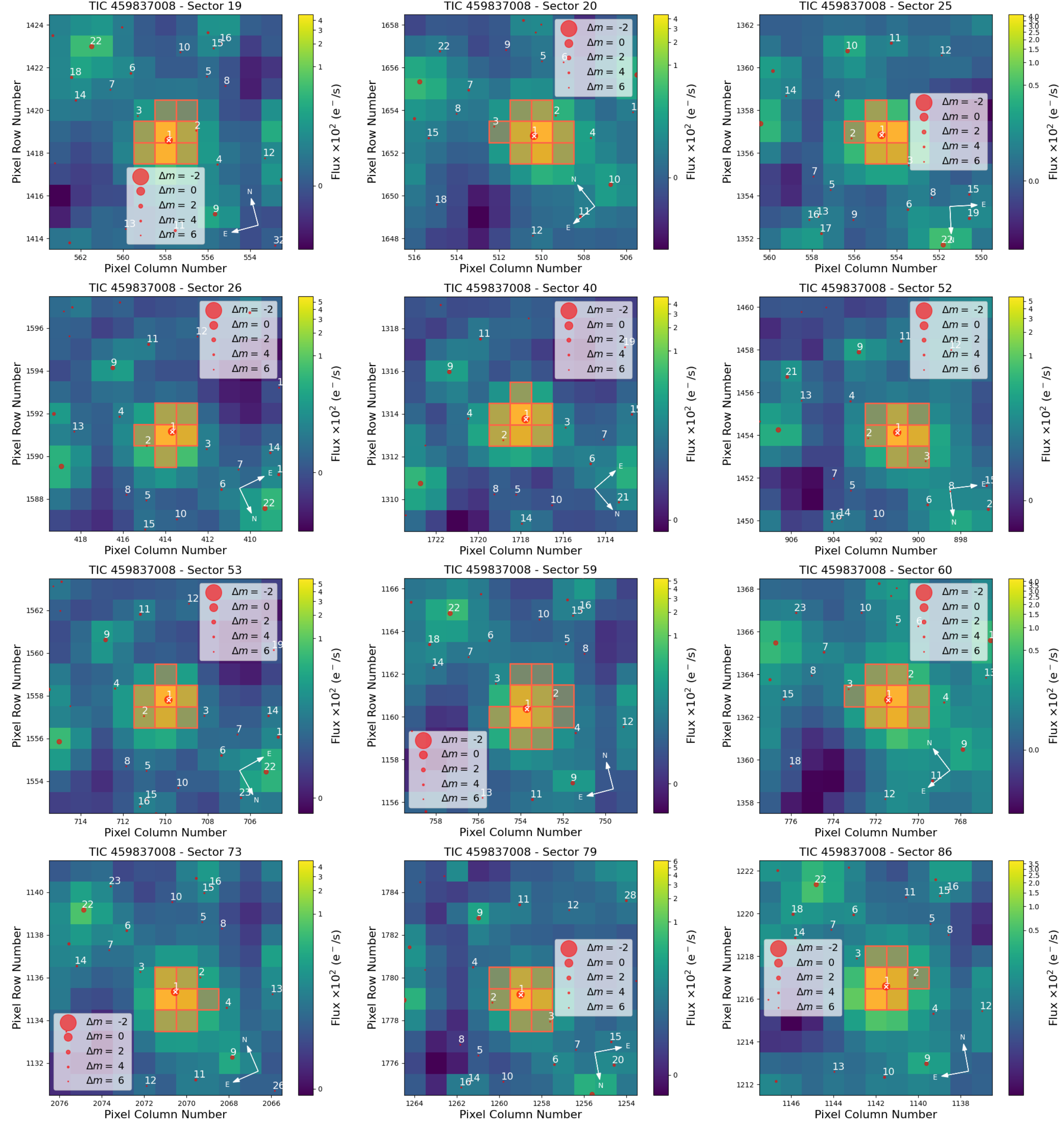}
\caption{\textit{\textit{TESS}} target pixel files (TPFs) showing the field of view of the 12 sectors used in this study, generated employing \texttt{tpfplotter} \citep{aller:2020}. The photometric apertures used to extract the light curves in each case are shown as red-shaded regions. The {\it Gaia\/} DR2 catalogue is overplotted, with all sources up to six magnitudes in contrast with TOI-2267 shown as red circles. We note that the symbol size scales with the magnitude contrast. The target star, TOI-2267, is highlighted with a white cross and numbered as 1 in all cases.}
\label{fig:fov}
\end{figure*}

\newpage
\section{Global fit}
\label{appendix:global_fit}

\renewcommand{\arraystretch}{1.2} % aumenta el alto de las filas
\begin{table*}[ht!]
\caption{Parameters for the TOI-2267\,b and TOI-2267\,c planets.}
\label{table:planet_params_bc}
\centering
\resizebox{0.9\textwidth}{!}{%
\begin{threeparttable}
\begin{tabular}{l c c r}
\toprule
Parameter & Unit & Primary host & Secondary host \\ 
\midrule
\multicolumn{2}{c}{\it{Fitted Parameters}} & \multicolumn{2}{c}{TOI-2267\,b / TOI-2267.03}  \\
\midrule
$R_p / R_\star$ &  &  $0.0441_{-0.0011}^{+0.0013}$ & $0.0859\pm0.0040$ \\
$(R_\star + R_p)/a_p$ & & $0.0486_{-0.0019}^{+0.0033}$ & $0.0505_{-0.0022}^{+0.0047}$ \\
$\cos{i_p}$ & & $-0.006_{-0.014}^{+0.020}$ & $-0.001\pm0.017$ \\
Orbital period,  $P$ & days &  $2.2890900_{-0.0000011}^{+0.0000012}$ & $2.2890896_{-0.0000017}^{+0.0000015}$ \\
Mid-transit time, $T_0$ & BJD$_{TDB}$-2457000 & $2525.40333_{-0.00035}^{+0.00028}$ & $2525.40311\pm0.00041$  \\
\midrule
\multicolumn{2}{c}{\it{Derived Parameters}} & & \\
\midrule
Planet Radius, $R_p$ & $R_\oplus$ & $1.00\pm0.11$ & $1.22\pm0.29$ \\
Semimajor axis, $a$ & AU & $0.0205\pm0.0025$ & $0.0127\pm0.0032$ \\
Inclination, $i$ & $^{\circ}$ &  $90.35_{-1.2}^{+0.85}$ &  $90.0\pm1.0$ \\
Transit duration, $T_{1-4}$ & hrs &  $0.815\pm0.017$ & $0.855_{-0.024}^{+0.028}$ \\
Impact parameter, $b$ & & $-0.13_{-0.29}^{+0.44}$ &  $-0.02\pm0.37$ \\
$^{a}$ Equilibrium Temperature, $T_{eq}$ & K & $424_{-17}^{+19}$ & $411_{-25}^{+29}$ \\
Insolation Flux, $S$ & S$_{\oplus}$ & $8.4_{-2.6}^{+3.7}$ & $7.6_{-4.0}^{+8.0}$  \\
Host density from orbit, $\rho_\mathrm{\star}$ & $g\,cm^{-3}$ & $35.7_{-6.4}^{+4.7}$ & $35.9_{-8.4}^{+5.2}$ \\  
\midrule 
\multicolumn{2}{c}{\it{Fitted Parameters}} & \multicolumn{2}{c}{TOI-2267\,c / TOI-2267.01}\\
\midrule
$R_p / R_\star$ &  &  $0.0504_{-0.0012}^{+0.0013}$ & $0.0962\pm0.0055$ \\
$(R_\star + R_p)/a_p$ & & $0.0383_{-0.0042}^{+0.0035}$ & $0.0455_{-0.0083}^{+0.0088}$ \\
$\cos{i_p}$ & & $0.006_{-0.015}^{+0.029}$ & $0.014_{-0.042}^{+0.020}$ \\
Orbital period,  $P$ & days &  $3.4950412\pm0.0000022$ & $3.4950404\pm0.0000028$ \\
Mid-transit time, $T_0$ & BJD$_{TDB}$-2457000 & $2525.64728\pm0.00033$ & $2525.64715_{-0.00050}^{+0.00041}$ \\
\midrule
\multicolumn{2}{c}{\it{Derived Parameters}} & & \\
\midrule
Planet Radius, $R_p$ & $R_\oplus$ & $1.14\pm0.13$ & $1.36\pm0.33$ \\
Semimajor axis, $a$ & AU &  $0.0263_{-0.0036}^{+0.0040}$ & $0.0145_{-0.0037}^{+0.0045}$ \\
Inclination, $i$ & $^{\circ}$ &  $89.66_{-0.89}^{+1.6}$ & $89.2_{-1.1}^{+2.4}$ \\
Transit duration, $T_{1-4}$ & hrs & $0.893_{-0.022}^{+0.023}$ & $0.996_{-0.059}^{+0.075}$ \\
Impact parameter, $b$ & &  $0.19_{-0.77}^{+0.38}$ &  $0.41_{-1.1}^{+0.31}$ \\
$^{a}$ Equilibrium Temperature, $T_{eq}$ & K & $374_{-21}^{+23}$  & $385_{-37}^{+40}$ \\
Insolation Flux, $S$ & S$_{\oplus}$ & $4.7_{-1.4}^{+2.1}$ & $4.6_{-2.5}^{+6.8}$  \\ 
Host density from orbit, $\rho_\mathrm{\star}$ & $g\,cm^{-3}$ & $31.9_{-8.5}^{+11}$ & $21.7_{-8.8}^{+18}$ \\ 
\midrule
\multicolumn{2}{c}{\it{Shared Parameters}} & & \\
\midrule
Dilution factor \textit{TESS}, $D_\mathrm{0;\,\textit{TESS}}$ & & $0.311_{-0.030}^{+0.032}$ & $0.820_{-0.015}^{+0.013}$ \\
Dilution factor $z'$, $D_\mathrm{0;\,z'}$ & & $0.152\pm0.036$ & $0.728_{-0.021}^{+0.023}$ \\
Dilution factor $i'$, $D_\mathrm{0;\,i'}$ & & $0.219_{-0.043}^{+0.045}$ & $0.780_{-0.034}^{+0.031}$ \\
Dilution factor $I_c$, $D_\mathrm{0;\,I_c}$ & & $0.230769$ (fixed) & $0.769230$ (fixed) \\
GP hyper-parameter, $\mathrm{\ln {S_0}_{;\,\textit{TESS}}}$ & rel. flux &  $-15.211_{-0.047}^{+0.043}$ & $-15.219_{-0.049}^{+0.046}$ \\
GP hyper-parameter, $\mathrm{\ln {Q}_{;\,\textit{TESS}}}$ & rel. flux &  $0.807_{-0.050}^{+0.047}$ & $0.811_{-0.056}^{+0.061}$ \\
GP hyper-parameter, $\mathrm{\ln {\omega_0}_{;\,\textit{TESS}}}$ & rel. flux &  $2.4442_{-0.050}^{+0.047}$ & $2.4441_{-0.010}^{+0.0097}$ \\
Period ratio, $P_\mathrm{c} / P_\mathrm{b}$ & & $1.5268255_{-0.0000011}^{+0.0000012}$ & $1.5268255\pm0.0000017$ \\
Combined host density from all orbits, $\rho_\mathrm{\star}$ & $g\,cm^{-3}$ & $34.3_{-8.3}^{+6.8}$ & $31.1_{-15}^{+9.6}$ \\
\bottomrule
\end{tabular}
\tablefoot{
$^{a}$ Values derived assuming an albedo of 0.3 (Earth-like) and emissivity of 1.
}
\end{threeparttable}
}
\end{table*}
\renewcommand{\arraystretch}{1.0} % restaura el valor normal

\newpage

\begin{figure*}[ht]
\centering
\includegraphics[width=\textwidth]{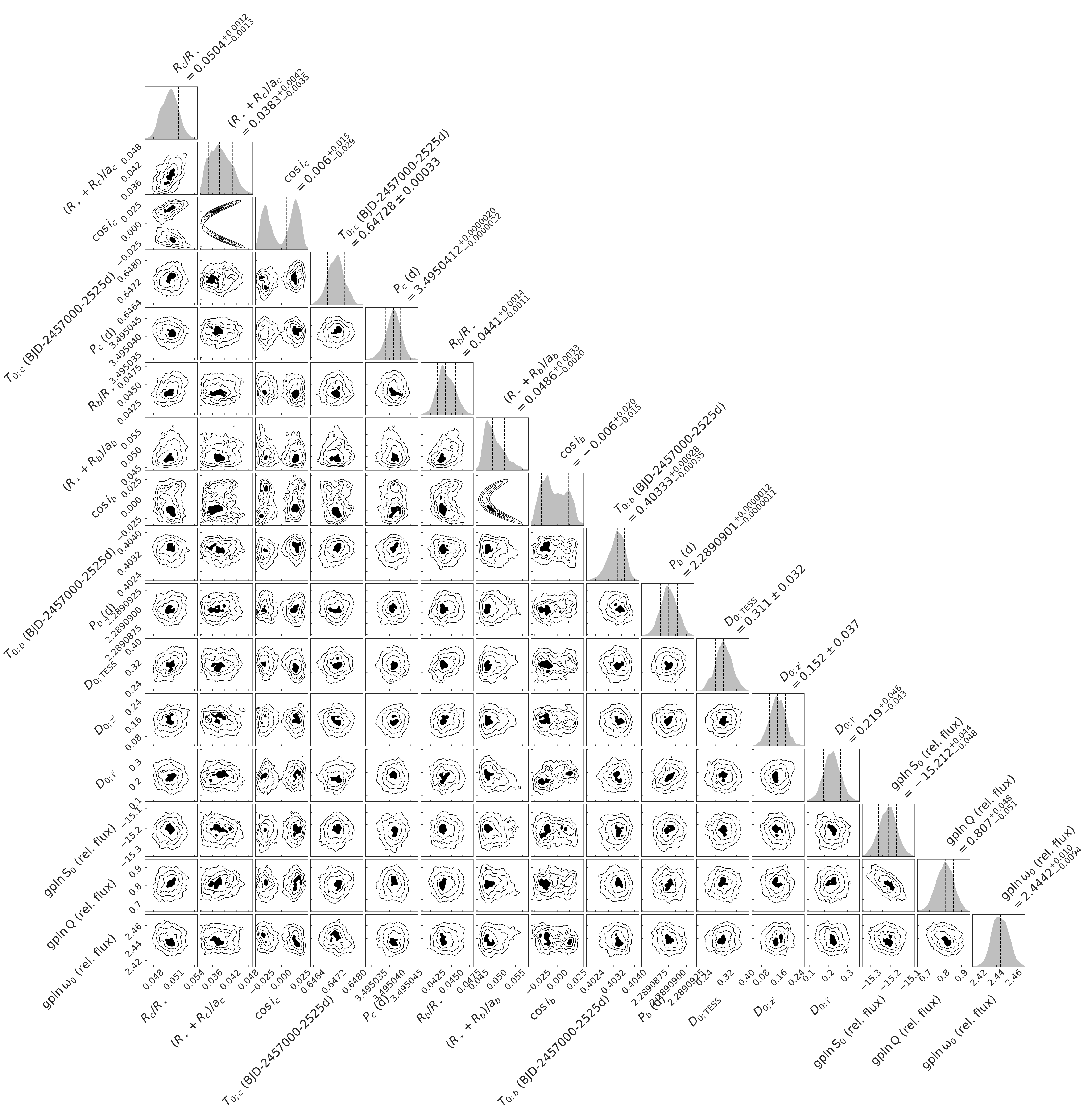}
\caption{Posterior probability distributions for all the physical parameters fitted for TOI-2267 bc systems (Primary star host) using \allesfitter nested sampling as described in Sect.~\ref{sec:global_fit}. The vertical dashed lines represent the median and the 68$\%$ credible interval. The figure highlights the correlation (or absence thereof) between all the parameters.} 
\label{fig:bc_primary_posteriors}
\end{figure*}

\newpage
\begin{figure*}[ht!]
\centering
\includegraphics[width=\textwidth]{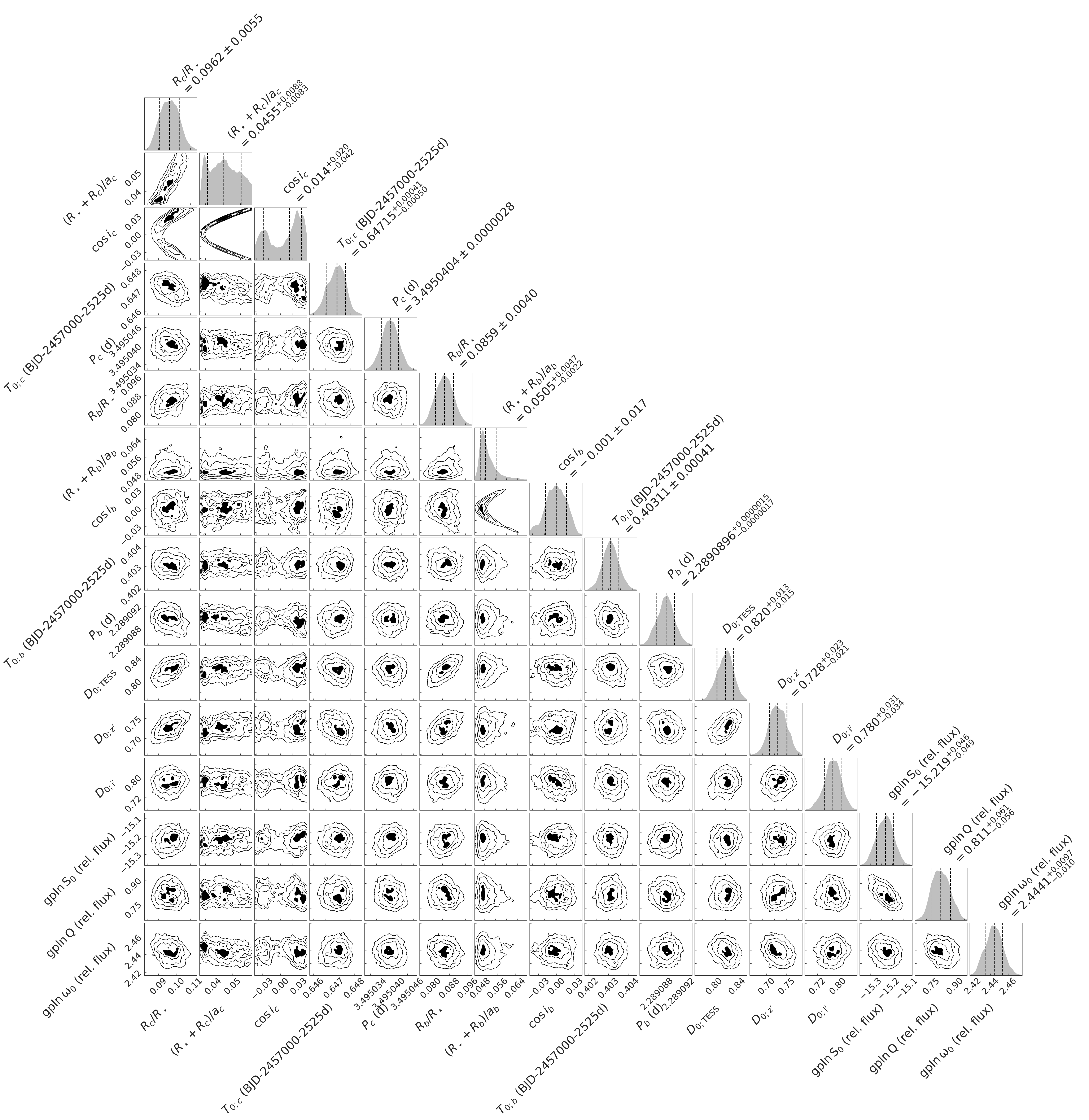}
\caption{Posterior probability distributions for all the physical parameters fitted for TOI-2267 bc systems (Secondary star host) using \allesfitter nested sampling as described in Sect.~\ref{sec:global_fit}. The vertical dashed lines represent the median and the 68$\%$ credible interval. The figure highlights the correlation (or absence thereof) between all the parameters.} 
\label{fig:bc_secondary_posteriors}
\end{figure*}

\newpage

\renewcommand{\arraystretch}{1.1} % aire entre filas
\begin{table*}[htb!]
\caption{Parameters for the TOI-2267.02 candidate.}
\centering
\resizebox{0.9\textwidth}{!}{%
\begin{threeparttable}
\begin{tabular}{l c c r}
\toprule
Parameter & Unit & Primary host & Secondary host \\ \midrule
\multicolumn{2}{c}{\it{Fitted Parameters}} & \multicolumn{2}{c}{TOI-2267\,d / TOI-2267.02}  \\
\midrule \smallskip
$R_p / R_\star$ &  &  $0.0422_{-0.0025}^{+0.0028}$ & $0.0799_{-0.0093}^{+0.010}$ \\ \smallskip
$(R_\star + R_p)/a_p$ & & $0.0486_{-0.0072}^{+0.018}$ & $0.0501_{-0.0062}^{+0.016}$ \\ \smallskip
$\cos{i_p}$ & & $-0.006_{-0.038}^{+0.033}$ & $0.012_{-0.028}^{+0.031}$ \\ \smallskip
Orbital period,  $P$ & days &  $2.0344678_{-0.0000023}^{+0.0000020}$ & $2.0344671\pm0.0000023$ \\ \smallskip
Mid-transit time, $T_0$ & BJD$_{TDB}$-2457000 & $1817.0810\pm0.0014$ & $1817.0813_{-0.0015}^{+0.0014}$  \\ \smallskip
Dilution factor \textit{TESS}, $D_\mathrm{0;\,\textit{TESS}}$ & & $0.288\pm0.049$ & $0.806_{-0.045}^{+0.040}$ \\ \smallskip
GP hyper-parameter, $\mathrm{\ln {S_0}_{;\,\textit{TESS}}}$ & rel. flux &  $-15.220\pm0.047$ & $-15.227_{-0.045}^{+0.048}$ \\ \smallskip
GP hyper-parameter, $\mathrm{\ln {Q}_{;\,\textit{TESS}}}$ & rel. flux &  $0.818\pm0.053$ & $0.817\pm0.052$ \\ \smallskip
GP hyper-parameter, $\mathrm{\ln {\omega_0}_{;\,\textit{TESS}}}$ & rel. flux &  $2.4434_{-0.0097}^{+0.0092}$ & $2.444\pm0.010$ \\
\midrule
\multicolumn{2}{c}{\it{Derived Parameters}} & & \\
\midrule \smallskip
Planet Radius, $R_p$ & $R_\oplus$ & $0.95\pm0.12$ & $1.13_{-0.28}^{+0.31}$ \\ \smallskip
Semimajor axis, $a$ & AU & $0.0203_{-0.0053}^{+0.0046}$ & $0.0124_{-0.0035}^{+0.0040}$  \\ \smallskip
Inclination, $i$ & $^{\circ}$ &  $90.3_{-1.9}^{+2.2}$  & $89.3_{-1.8}^{+1.6}$ \\ \smallskip
Transit duration, $T_{1-4}$ & hrs &  $0.654_{-0.036}^{+0.051}$ & $0.708_{-0.048}^{+0.060}$ \\ \smallskip
Impact parameter, $b$ & & $-0.15_{-0.59}^{+0.72}$ &   $0.29_{-0.65}^{+0.44}$ \\ \smallskip
$^{a}$ Equilibrium Temperature, $T_{eq}$ & K & $424_{-35}^{+70}$ & $412_{-36}^{+58}$ \\ \smallskip
Insolation Flux, $S$ & S$_{\oplus}$ & $8.6_{-3.3}^{+6.4}$ & $7.6_{-4.1}^{+10.5}$  \\ \smallskip
Transit Depth (\textit{TESS}, undiluted), $\delta$ & ppt &  $1.95_{-0.20}^{+0.22}$ & $7.1_{-1.4}^{+2.0}$ \\ \smallskip
Transit Depth (\textit{TESS}, diluted), $\delta$ & ppt &  $1.39\pm0.12$ &  $1.37_{-0.11}^{+0.12}$ \\ \smallskip
Host density from orbit, $\rho_\mathrm{\star}$ & $g\,cm^{-3}$ & $45\pm28$ & $46_{-25}^{+22}$ \\ 
\bottomrule
\end{tabular}
\tablefoot{
$^{a}$ Values derived assuming an albedo of 0.3 (Earth-like) and emissivity of 1.}
\end{threeparttable}
}
\label{table:planet_params_d}
\end{table*}
\renewcommand{\arraystretch}{1.0} % restaura el valor por defecto

\newpage
\begin{figure*}[h!]
\centering
\includegraphics[width=0.6\textwidth]{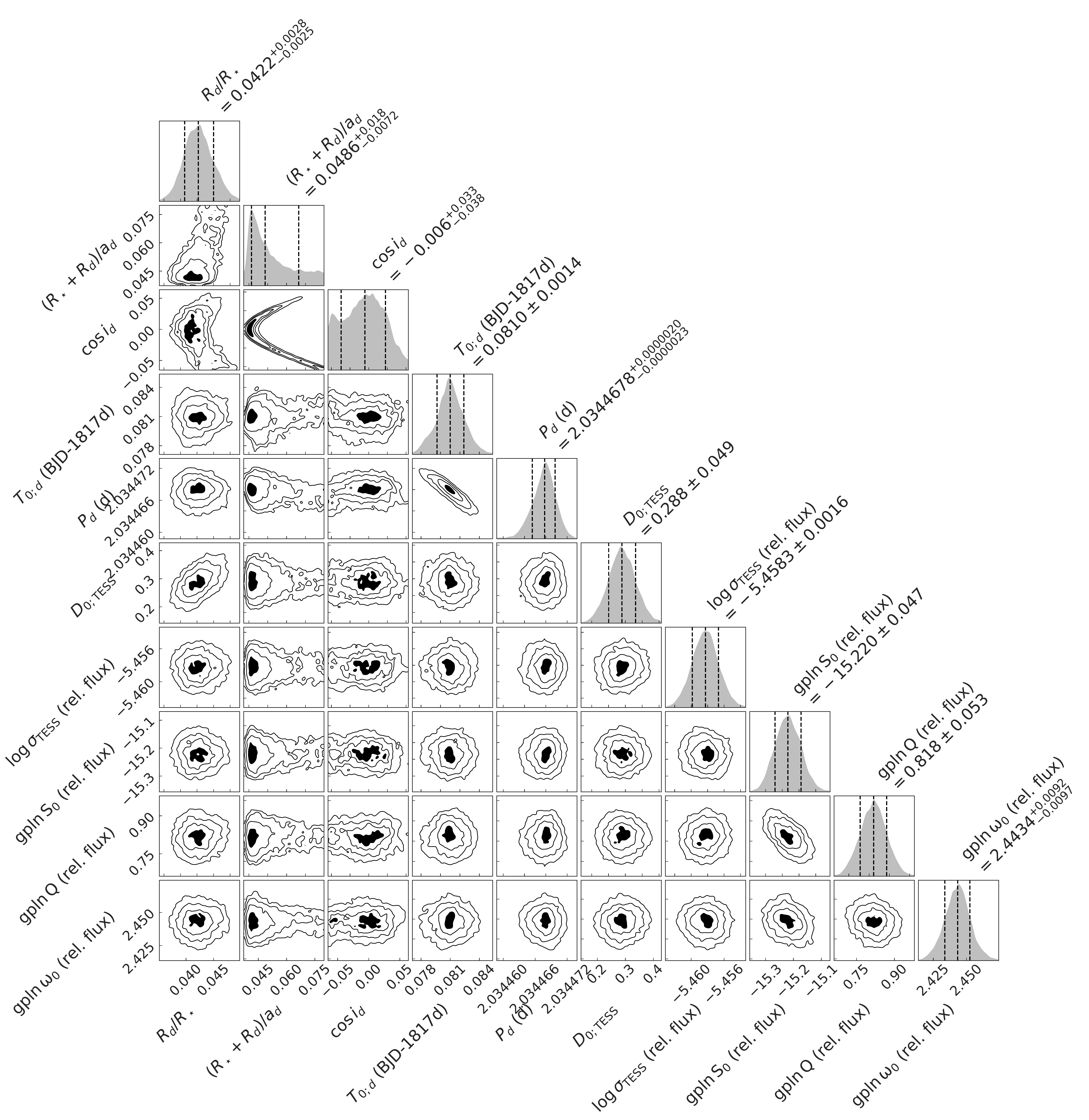}
\includegraphics[width=0.6\textwidth]{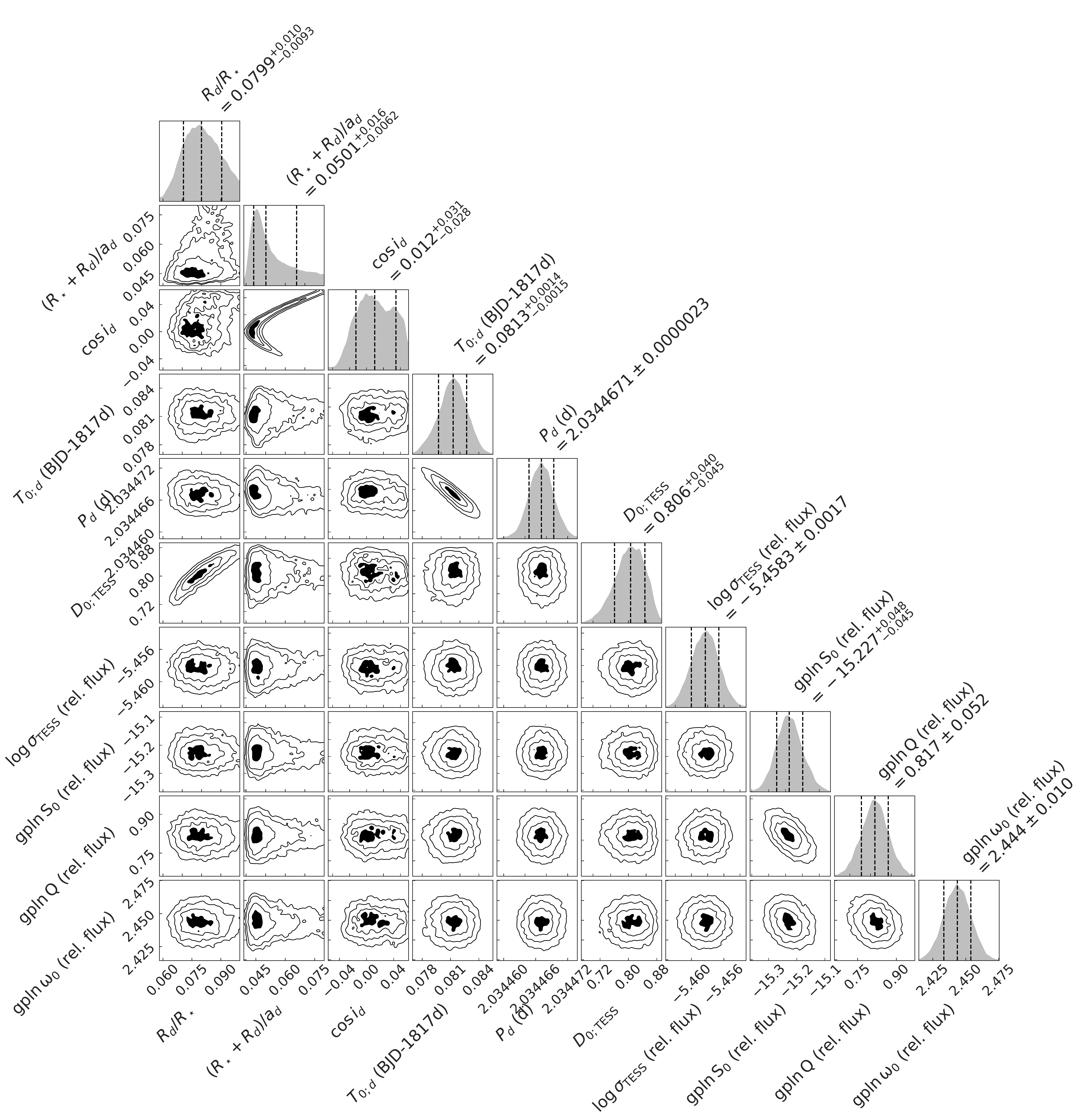}
\caption{Posterior probability distributions for all the physical parameters fitted for TOI-2267.02 using \allesfitter nested sampling as described in Sect.~\ref{sec:global_fit}. Upper panel for the primary star host and lower panel for the secondary star host case. The vertical dashed lines represent the median and the 68$\%$ credible interval. The figure highlights the correlation (or absence thereof) between all the parameters.} 
\label{fig:bc_secondary_posteriors}
\end{figure*}

\newpage
\twocolumn
\section{Host density from orbital fit}
\begin{figure}[htb!]
    \centering
    \includegraphics[width=\linewidth]{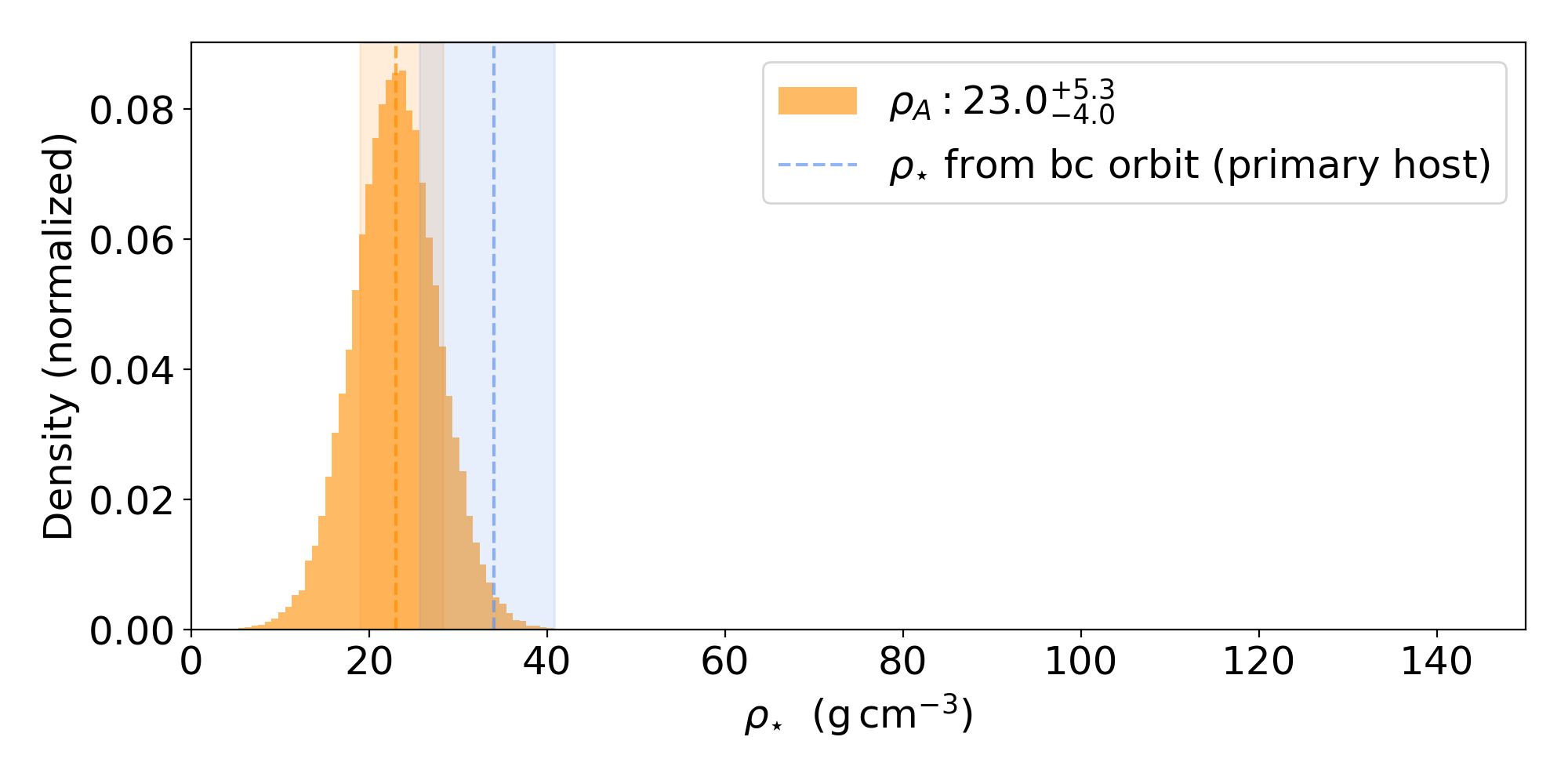}
    \includegraphics[width=\linewidth]{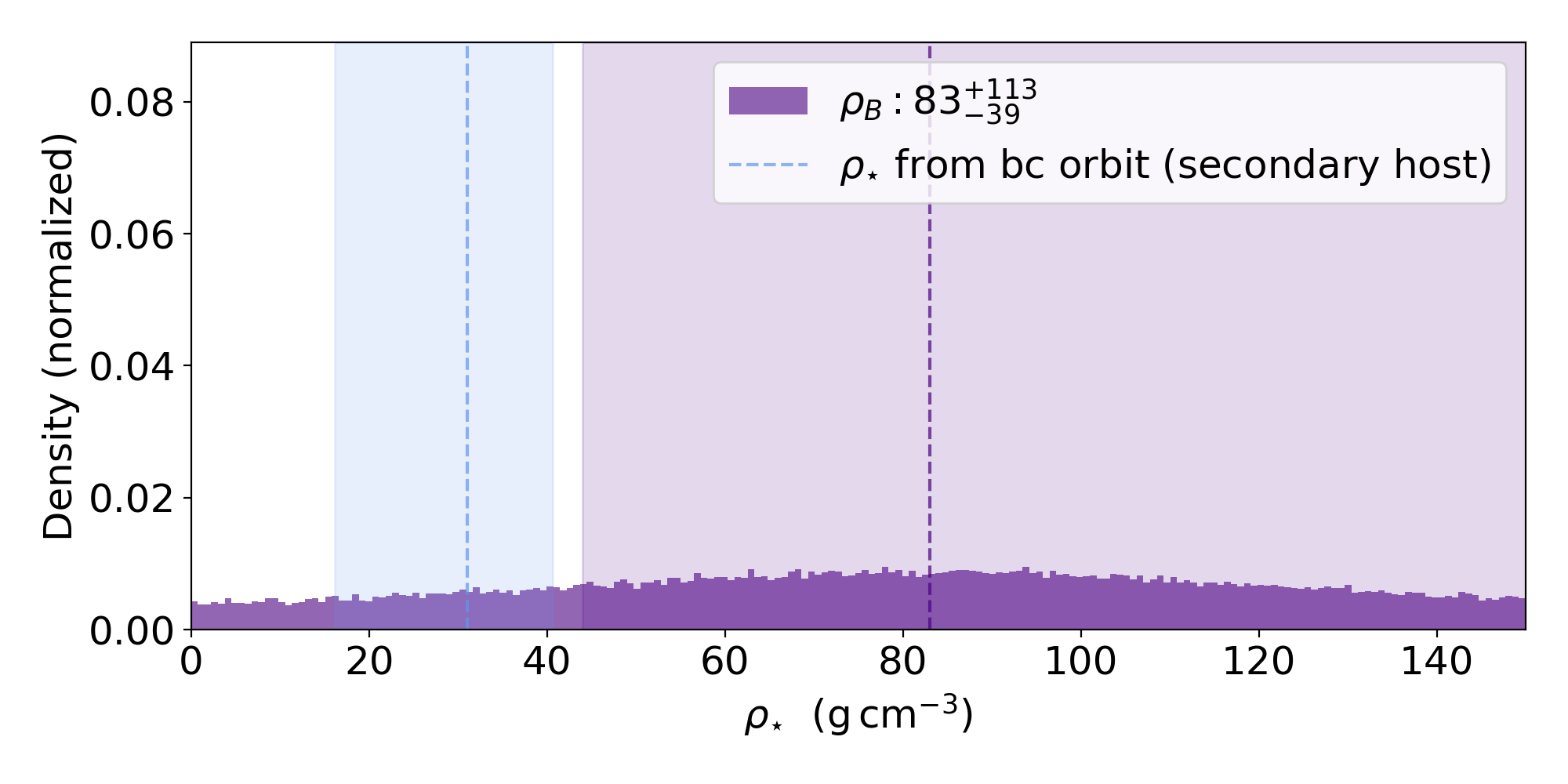}
    \caption{Comparison between host density obtained from stellar model and orbital fitting for TOI-2267 bc for primary (Top panel) and secondary (Bottom panel) host.}
    \label{fig:bc_density}
\end{figure}

\vspace{0.5cm}

\begin{figure}[htb!]
    \centering
    \includegraphics[width=\linewidth]{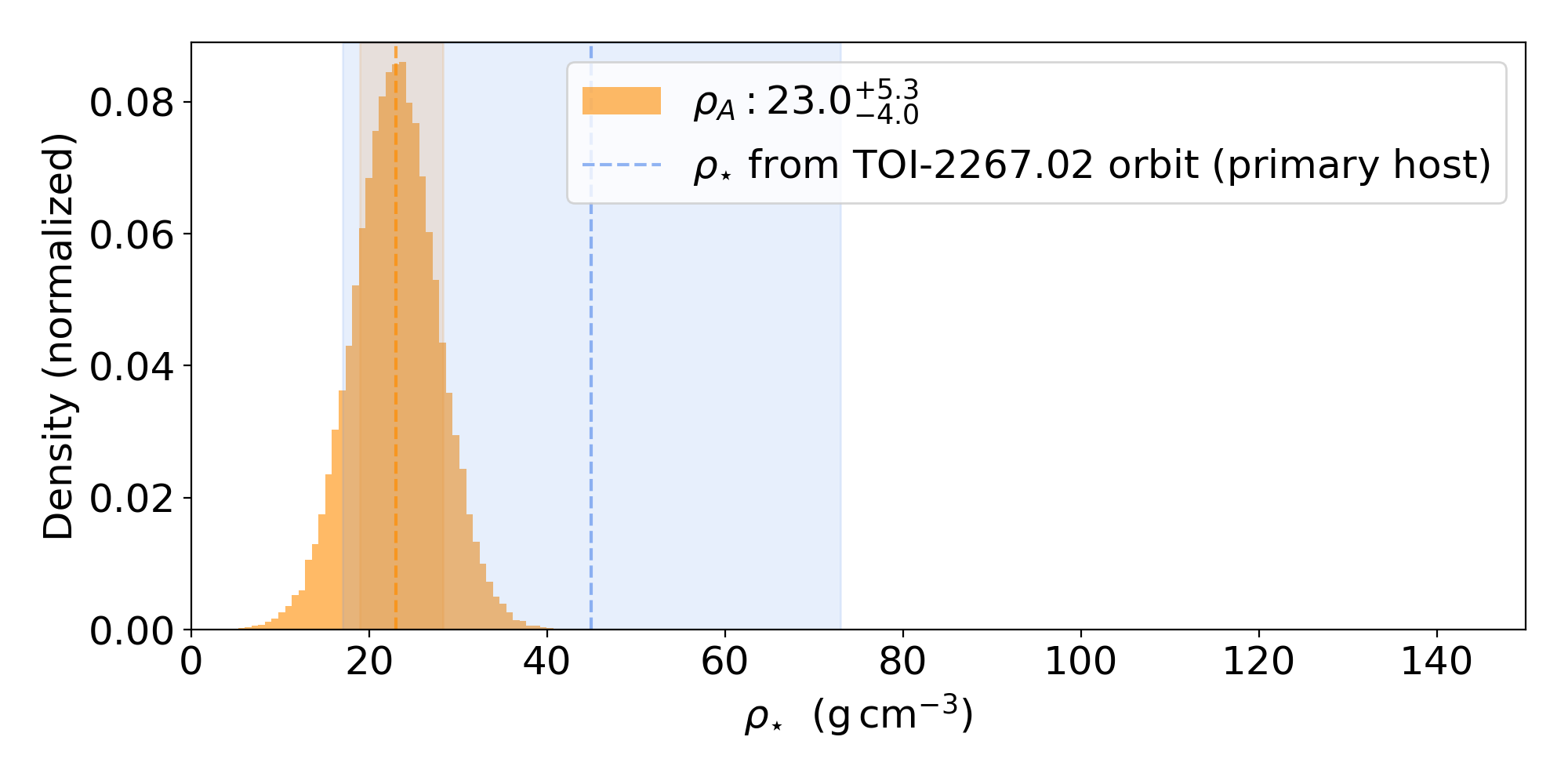}
    \includegraphics[width=\linewidth]{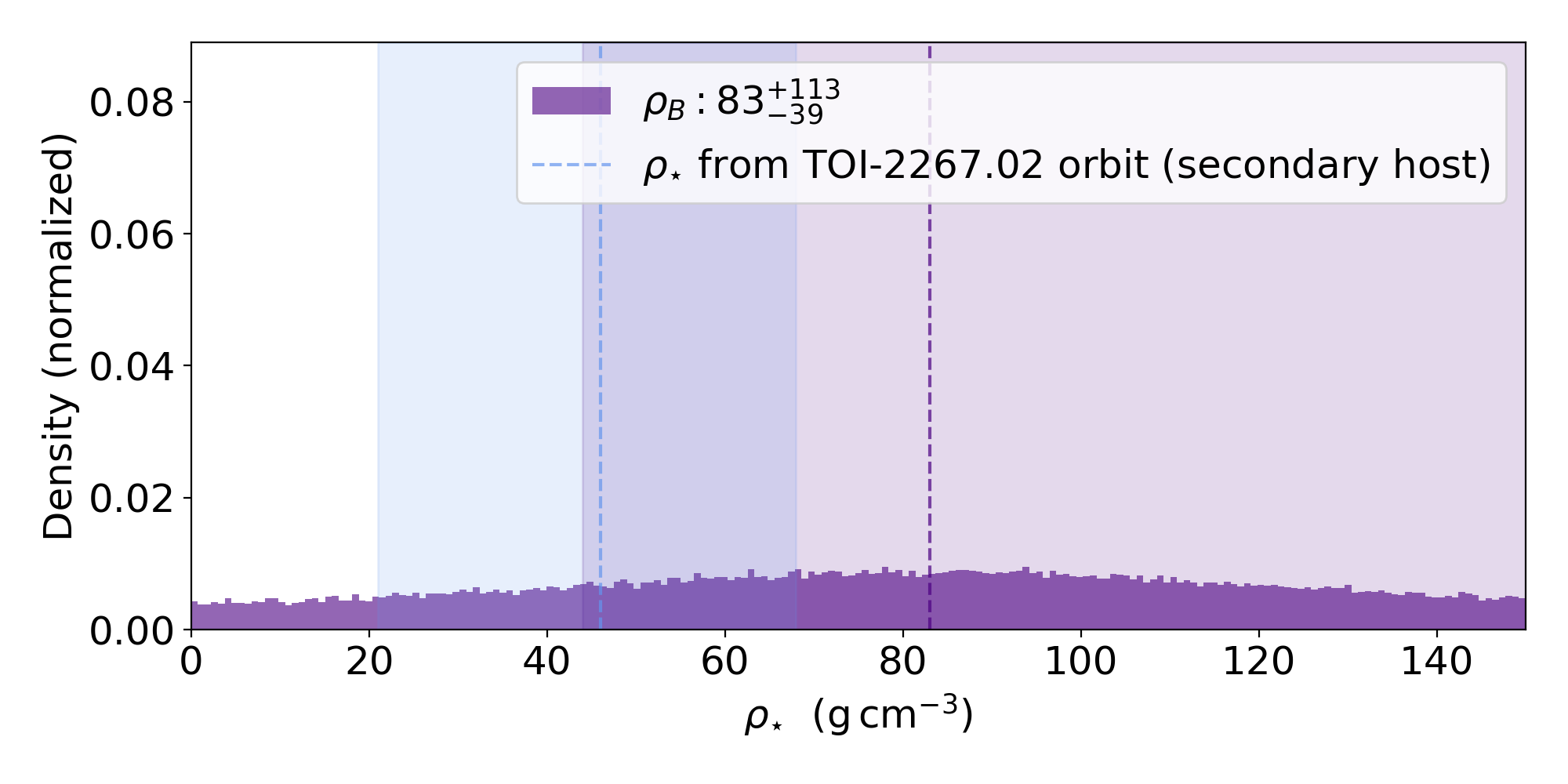}
    \caption{Comparison between host density obtained from stellar model and orbital fitting for TOI-2267.02 for primary (Top panel) and secondary (Bottom panel) host}
    \label{fig:d_density}
\end{figure}

\section{Transit depths}

\begin{figure}[htb!]
    \centering
    \includegraphics[width=0.9\linewidth]{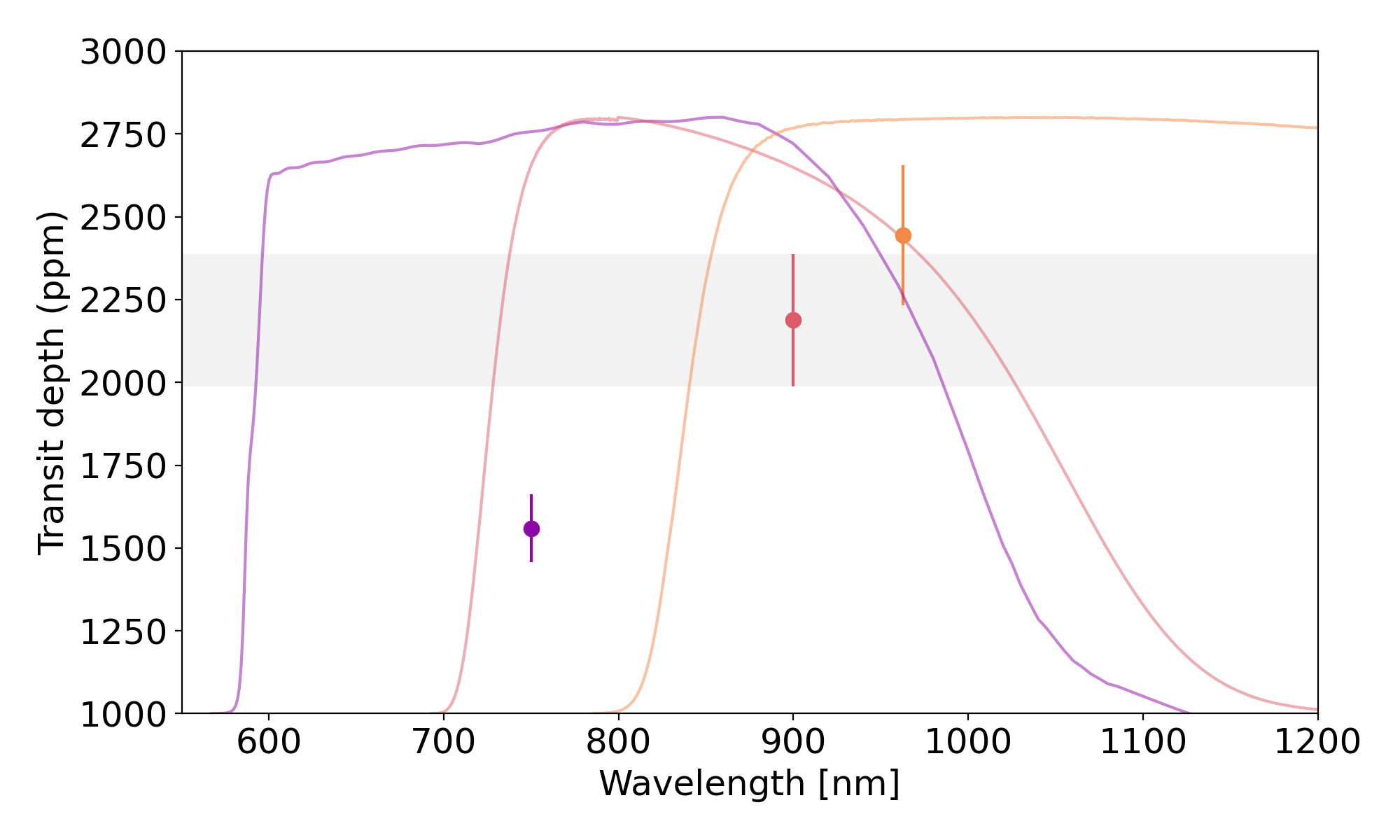}
    \includegraphics[width=0.9\linewidth]{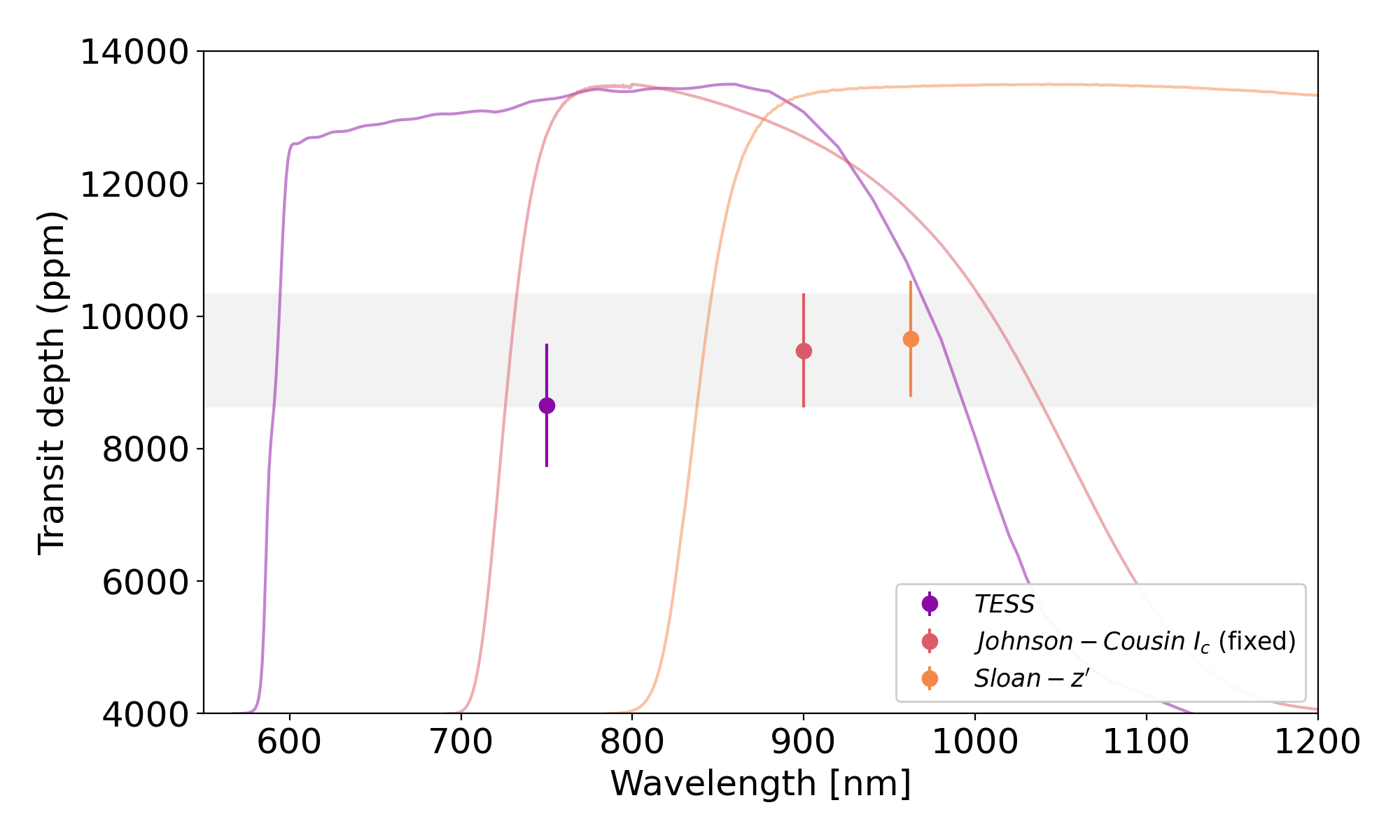}
    \caption{Measured transit depth vs wavelength for TOI-2267 b (secondary host). The dark grey horizontal line indicates the depth of the Johnson-Cousin $I_c$ filter (dilution factor fixed in the global fit, see Sect. \ref{sec:global_fit}); circles highlight other filters for comparison. \textit{Top}: Diluted transit depths. \textit{Bottom}: Undiluted transit depths corrected by the dilution factor.}
    \label{fig:B_transit_dephts_b}
\end{figure}

\begin{figure}[htb!]
    \centering
    \includegraphics[width=0.9\linewidth]{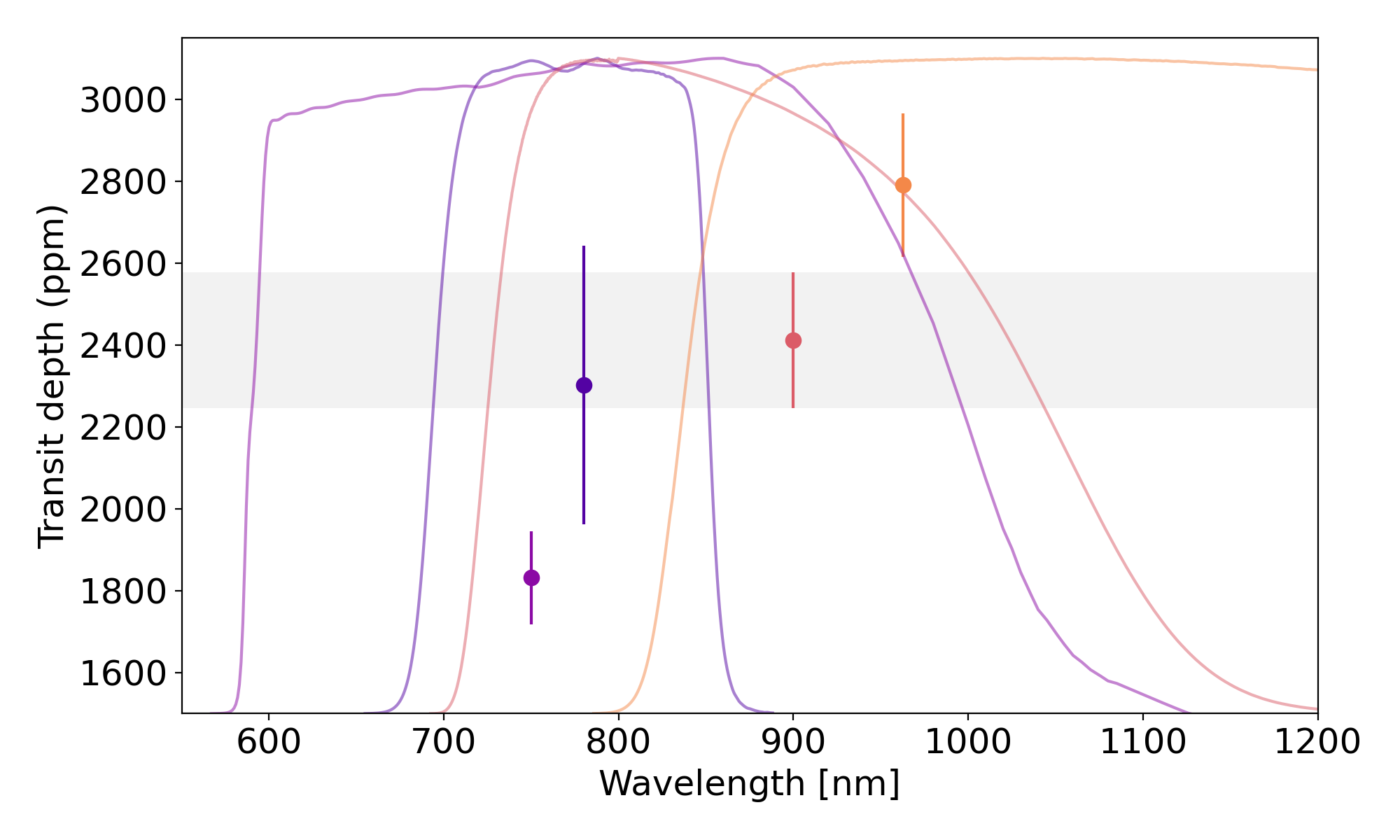}
    \includegraphics[width=0.9\linewidth]{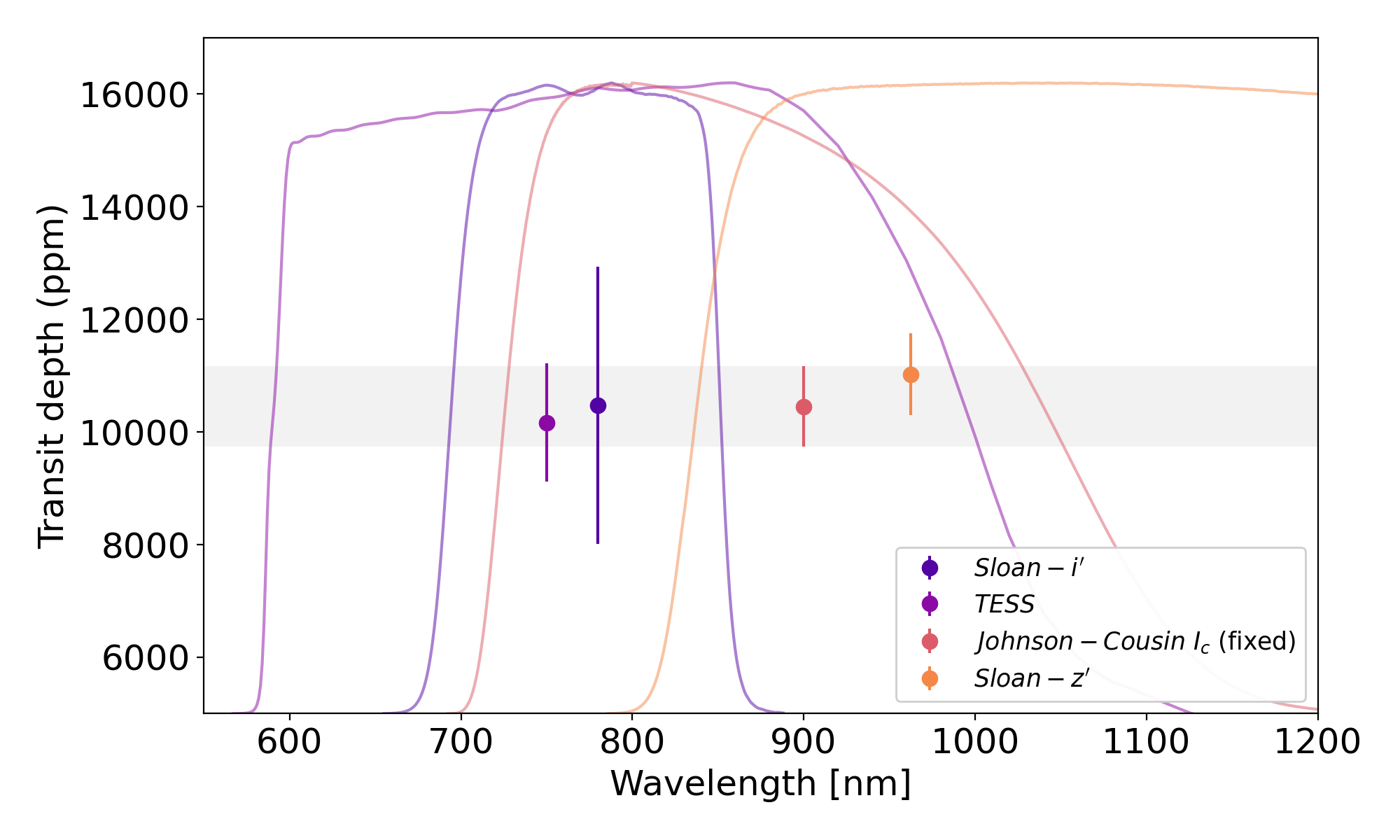}
    \caption{Measured transit depth vs wavelength for TOI-2267 c (secondary host). The dark grey horizontal line indicates the depth of the Johnson-Cousin $I_c$ filter (dilution factor fixed in the global fit, see Sect. \ref{sec:global_fit}); other filters are highlighted by circles for comparison. \textit{Top}: Diluted transit depths. \textit{Bottom}: Undiluted transit depths corrected by the dilution factor.}
    \label{fig:B_transit_dephts_c}
\end{figure}

\renewcommand{\arraystretch}{1.2} % más espacio entre filas
\begin{table}[htb!]
\caption{Measured transit depths at each bandpass for the global fit assuming the secondary star as the host.}
\centering
\resizebox{0.4\textwidth}{!}{%
\begin{threeparttable}
\begin{tabular}{lcc}
\toprule
Transit depth, $\delta$ & Diluted (ppm) & Undiluted (ppm)\\
& & \\
\midrule
Bandpass & \multicolumn{2}{c}{TOI-2267B b}\\
\midrule
\textit{TESS} & $1560 \pm 100$ & $8650_{-840}^{+930}$ \\
$Ic$ & $2190_{-180}^{+190}$ & $9470_{-770}^{+840}$ \\
$z'$ & $2470_{-220}^{+190}$ & $9750_{-780}^{+840}$ \\
\midrule
 & \multicolumn{2}{c}{TOI-2267B c}\\
\midrule
$i'$ & $2300 \pm 340$  & $10500_{-2000}^{+2500}$ \\
\textit{TESS} & $1830_{-190}^{+170}$ & $10170_{-960}^{+1100}$ \\
$Ic$ & $2420_{-160}^{+170}$  & $10480_{-670}^{+720}$ \\
$z'$ & $2790_{-170}^{+160}$ & $11140_{-670}^{+630}$ \\
\bottomrule
\end{tabular}
\tablefoot{
Undiluted transit depths are corrected by the dilution factor (see Sect.~\ref{sec:global_fit}).
}
\end{threeparttable}
}
\label{tab:transit_depths_b}
\end{table}
\renewcommand{\arraystretch}{1.0} % restaurar valor por defecto

\end{appendix}

\end{document}